\pdfoutput=1
\documentclass[cernpreprint,texlive=2016,txfonts,UKenglish]{latex/atlasdoc}

\usepackage[subfigure=true,block=none]{latex/atlaspackage}
\usepackage{latex/atlasphysics}

\usepackage{amssymb}

\nolinenumbers
\usepackage[centering,scale=0.75]{geometry}
\usepackage{graphicx}

\usepackage{adjustbox,lipsum}
\usepackage{rotating}
\usepackage{pdflscape}

\usepackage{latex/atlascover}

\usepackage{hyperref}
\usepackage{latex/atlascontribute}

\usepackage{latex/atlasbiblatex}
\usepackage{mathrsfs}
\usepackage{xspace}
\usepackage{float}
\usepackage{color, colortbl}
\definecolor{Gray}{gray}{0.9}

\newcommand{\ttb}{\ensuremath{{\ttbar}}}

\newcommand{\absyttbar}{\ensuremath{\left|y^{\ttbar}\right|}\xspace}

\newcommand{\absythad}{\ensuremath{\left|y^{t,\textrm{had}}\right|}\xspace}

\newcommand{\mttbar}{\ensuremath{m^{\ttbar}}}

\newcommand{\ptthad}{\ensuremath{\pt^{t,\textrm{had}}}}

\newcommand{\ptttbar}{\ensuremath{\pt^{\ttbar}}}
\newcommand{\Etmiss}{\ensuremath{\ET^{\mathrm{miss}}}}

\newcommand{\ejets}{{$e+$jets}\xspace}
\newcommand{\mujets}{{$\mu+$jets}\xspace}
\newcommand{\ljets}{{$\ell+$jets}\xspace}

\newcommand{\mtw}{\ensuremath{m_{\mathrm{T}}^W}}

\newcommand{\lumitot}{\mbox{3.2\,fb$^{-1}$}}

\newcommand{\Powheg}{\textsc{Powheg}\xspace}
\newcommand{\PowHeg}{\textsc{Powheg}\xspace}
\newcommand{\PowhegBox}{\textsc{Powheg-Box}\xspace}
\newcommand{\PowHegBox}{\textsc{Powheg-Box}\xspace}

\newcommand{\Pythia}{\textsc{Pythia}\xspace}
\newcommand{\PythiaSix}{\textsc{Pythia6}\xspace}
\newcommand{\PythiaEight}{\textsc{Pythia8}\xspace}
\renewcommand{\Herwig}{\textsc{Herwig}\xspace}

\newcommand{\aMCatNLO}{a\textsc{MC{@}NLO}\xspace}

\newcommand{\Sherpa}{\textsc{Sherpa}\xspace}
\newcommand\herwigpp{\textsc{Herwig\raise0.2ex\hbox{\scriptsize\kern-0.5pt+\kern-0.5pt+}}}
\newcommand\mgamcatnlo{\textsc{MadGraph5}\_a\textsc{MC@NLO}}
\newcommand{\HerwigSeven}{\textsc{Herwig7}\xspace}

\newcommand{\hdamp}{\ensuremath{h_{\mathrm{damp}}}}

\newcommand{\HDampMT}{\ensuremath{h_{\mathrm{damp}}\!=\!m_{t}}}

\newcommand{\NNLL}{\textsc{NNLL}}
\renewcommand{\NNLO}{\textsc{NNLO}}
\renewcommand{\NLO}{\textsc{NLO}}

\AtlasTitle{Measurements of top-quark pair differential cross-sections in the lepton+jets channel in $pp$ collisions at $\sqrt{s}=13$ \TeV{} using the ATLAS detector}
\author{The ATLAS Collaboration}

\AtlasAbstract{
  Measurements of differential cross-sections of top-quark pair production 
in fiducial phase-spaces 
are presented as a~function of
top-quark and $t\bar{t}$ system kinematic observables in proton--proton collisions at a~centre-of-mass energy of $\sqrt{s}=13$ \TeV{}. 
The data set corresponds to an integrated luminosity of $3.2$~\ifb, recorded in 2015 with the ATLAS detector at the CERN Large Hadron Collider. Events with exactly one
electron or muon and at least two jets in the final state are used for the measurement. Two separate selections are applied that each focus on different top-quark momentum regions, referred to as resolved and boosted topologies of the $t\bar{t}$ final state. The measured spectra are corrected for detector effects and are compared to several Monte Carlo simulations by means of calculated $\chi^2$ and $p$-values.
} 

\PreprintIdNumber{CERN-EP-2017-058}

\author{The ATLAS Collaboration}

\AtlasJournal{JHEP}
\date{\today}

\begin{document}

\title{Measurements of top-quark pair differential cross-sections in the lepton+jets channel in $pp$ collisions at $\sqrt{s}=13$ \TeV{} using the ATLAS detector}
\author{The ATLAS Collaboration}
\maketitle{}

\section{Introduction}\label{sec:Introduction}
The large top-quark pair production cross-section at the Large Hadron Collider (LHC) allows detailed studies of the characteristics of $\ttbar{}$ production to be performed with respect to different kinematic variables, providing a~unique opportunity to test the Standard Model (SM) at the~\TeV{} scale.
Furthermore, extensions of the SM may modify the expected $\ttbar{}$ differential distributions based solely on the SM
in ways not detectable by an inclusive cross-section measurement~\cite{Frederix:2009}. In particular, such effects may distort the top-quark momentum distribution, especially at higher values~\cite{Atwood:1994vm,Englert:2012by}. Therefore, a~precise measurement of the $\ttbar{}$ differential cross-section has the potential to enhance the sensitivity to possible effects beyond the SM, as well as to challenge theoretical predictions.

The ATLAS and CMS experiments have published measurements of the $\ttbar{}$ differential cross-sections in $pp$ collisions at centre-of-mass energies of $\sqrt{s}=7$~\TeV{} (ATLAS~\cite{atlasDiff1,atlasDiff2, atlasDiff3}, CMS~\cite{cmsDiff1}) and $\sqrt{s}=8$~\TeV{} (ATLAS~\cite{atlasDiff4}, CMS~\cite{cmsDiff2}), both in the full phase-space using parton-level variables and in fiducial phase-space regions using observables constructed from final-state particles (particle level). In addition, both experiments published measurements of the top-quark transverse momentum ($\pt$) spectrum which focused on the highest momentum region using the $\sqrt{s}=8$~\TeV{} data set~\cite{CON-2014-057,cmsBoosted}.  
The results presented in this paper probe the top-quark kinematic properties at a~centre-of-mass energy of $\sqrt{s}=13$~\TeV{}
and complement recent measurements involving leptonic final states (ATLAS~\cite{atlasDiff13Dilep}, CMS~\cite{Khachatryan:2016mnb}). At this energy, the prediction for the inclusive cross-section is increased by a~factor of 3.3 compared to 8~\TeV{}, and the top quarks are produced at higher transverse momenta. This allows the top-quark $\pt$ reach to be extended
up to 1.5~\TeV{} in order to explore both the low- and the high-momentum top-quark kinematic regimes.

In the SM, the top quark decays almost exclusively into a~\Wboson{} boson and a~$b$-quark. The signature of a~\ttbar{} decay is therefore determined by the \Wboson{} boson decay modes. This analysis makes use of the lepton$+$jets \ttbar{} decay mode, where one \Wboson{} boson decays into an electron or a~muon and a~neutrino, and the other \Wboson{} boson decays into a~pair of quarks, with the two decay modes referred to as the $e$+jets and $\mu$+jets channels, respectively. Events in which the \Wboson{} boson decays into an electron or muon through a~$\tau$ lepton decay may also meet the selection criteria.

Two complementary topologies of the \ttbar{} final state in the lepton+jets channel are exploited, dubbed "resolved" and "boosted", where the decay products of the hadronically decaying top quark are either angularly well separated or collimated into a~single large jet reconstructed in the calorimeter, respectively.
Where the jet selection efficiency of the resolved analysis decreases with the increasing top-quark transverse momentum, the boosted selection takes over to efficiently select events at higher momenta of the hadronically decaying top quarks.

This paper presents a~set of measurements of the \ttbar{} production cross-section as a~function of different properties of the reconstructed top quark (transverse momentum and rapidity) and of the \ttbar{} system (transverse momentum, rapidity and invariant mass). The results, unfolded to a~fiducial particle-level phase-space, are presented as both
absolute and relative differential cross-sections and are compared to the predictions of Monte Carlo (MC) event generators.
The goal of unfolding to a~fiducial particle-level phase-space and of using variables directly related to detector observables is to allow precision tests of quantum chromodynamics (QCD), avoiding uncertainties due to model-dependent extrapolations both to parton-level objects and to phase-space regions outside the detector sensitivity.

\section{ATLAS detector}\label{sec:Detector}

ATLAS is a multipurpose detector \cite{Aad:2008zzm} that provides nearly
full solid angle\footnote{ATLAS uses a
right-handed coordinate system with its origin at the nominal
interaction point (IP) in the centre of the detector and the $z$-axis
along the beam pipe. The $x$-axis points from the IP to the centre of
the LHC ring, and the $y$-axis points upward. Cylindrical coordinates
($r$,$\phi$) are used in the transverse plane, $\phi$ being the azimuthal angle
around the beam pipe. The pseudorapidity is defined in terms of the
polar angle $\theta$ as $\eta = - \ln \tan(\theta/2)$ and the angular separation between particles is
defined as $\Delta R = \sqrt{(\Delta \phi)^2 + (\Delta \eta)^2}$.} coverage around the interaction point.
This analysis
exploits all major components of the detector. Charged-particle trajectories
with pseudorapidity
$|\eta| <2.5$ are reconstructed
in the inner detector, which comprises a~silicon pixel detector, a~silicon 
microstrip detector and a~transition radiation tracker (TRT). The innermost pixel layer, the insertable B-layer~\cite{IBL}, was added before the start of the 13 \TeV{} LHC operation, at a~radius of 33~mm around a~new, thinner beam pipe. 
The inner detector is embedded in a~2~T axial magnetic field, allowing
precise measurement of charged-particle momenta. 
Sampling calorimeters with several different
designs span the pseudorapidity range up to $|\eta| = 4.9$. High-granularity 
liquid argon (LAr) electromagnetic (EM) calorimeters are used up
to $|\eta| = 3.2$. Hadronic calorimeters based on scintillator-tile
active material cover $|\eta| < 1.7$ while LAr technology is used
for hadronic calorimetry in the region $1.5 < |\eta| < 4.9$. The
calorimeters are surrounded by a~muon spectrometer within a~magnetic field  
provided by air-core toroid magnets with a~bending integral of about 2.5~Tm 
in the barrel and up to 6~Tm in the end-caps.
Three layers of precision drift tubes and cathode-strip chambers provide 
an accurate measurement of the muon track curvature in the region  $|\eta| < 2.7$. 
Resistive-plate and thin-gap chambers provide muon triggering capability up to $|\eta| = 2.4$.

Data are selected from inclusive $pp$ interactions using a~two-level trigger system~\cite{Aaboud:2016leb}. 
 A~hardware-based trigger uses custom-made hardware and coarser-granularity detector data to initially reduce the trigger rate to approximately $75\,$kHz from the original 40 MHz LHC collision bunch rate. 
Next, a~software-based high-level trigger, which has access to full detector granularity, is applied to further reduce the event rate to $1\,$kHz.

\section{Data and simulation samples} \label{sec:DataSimSamples}
The differential cross-sections are measured using a data set collected
during the 2015 LHC $pp$ run at $\rts =13$~\TeV{} and with 25 ns bunch spacing. The average number of proton--proton interactions per bunch crossing ranged from approximately 5 to 25, with a~mean of~14. 
After applying data-quality assessment criteria based on beam, detector and data-taking quality, the available data correspond to a~total integrated luminosity of $3.2$~\ifb. The uncertainty in the integrated luminosity is 2.1\% and is derived, following techniques similar to those described in Ref.~\cite{Aaboud:2016hhf}, from the luminosity scale calibration using a~pair of $x$--$y$ beam-separation scans performed in August 2015. 

The data sample is collected using single-muon and single-electron triggers. For each lepton type, multiple trigger conditions are combined in order to maintain good efficiency in the full momentum range, while controlling the trigger rate.
For electrons the \pt{} thresholds are 24~\GeV{}, 60~\GeV{} and 120~\GeV{}, while for muons the thresholds are 20~\GeV{} and 50~\GeV{}.
In the case of the lowest-$\pt{}$ thresholds, isolation requirements are also applied.

The signal and background processes are modelled with various Monte Carlo event generators. Multiple overlaid proton--proton collisions are simulated with the soft QCD processes of \Pythia 8.186~\cite{Sjostrand:2007gs} using parameter values from tune A2~\cite{ATLAS:2012uec} and the MSTW2008LO~\cite{Martin:2009iq} set of parton distribution functions (PDFs). The detector response is simulated~\cite{ATLASsim} in \GEANT4~\cite{GEANT4}. The data and MC events are reconstructed with the same software algorithms. Simulation samples are reweighted so that the distribution of the number of proton--proton interactions per event (pile-up) matches the one observed in data.

For the generation of \ttb{} samples and those with a single top quark from the $Wt$ and $s$-channel samples, the \PowhegBox v2~\cite{Frixione:2007vw} event generator with the CT10 PDF set~\cite{CT10} in the matrix element calculations is used~\cite{PUBSimtop}.  Events where both top quarks decay into hadronically decaying $W$ bosons are not included. The overlap between the $Wt$ and \ttb{}~samples is handled using the diagram removal scheme \cite{SingleTopWt}.

The top-quark mass is set to 172.5 \GeV{}. The EvtGen v1.2.0 program~\cite{EvtGen} is used to simulate the decay of bottom and charm hadrons. 
The $h_\textrm{damp}$ parameter, which controls the \pT\ of the first additional emission beyond the Born configuration in \Powheg, is set to the mass of the top quark. The main effect of this is to regulate the high-\pT\ emission against which the \ttbar{} system recoils.  Signal \ttb{} events generated with these settings are referred to as the nominal signal MC sample.

Electroweak $t$-channel single-top-quark events are generated using the \PowhegBox v1 event generator which uses the four-flavour scheme for the next-to-leading-order (\NLO{}) matrix element calculations together with the fixed four-flavour PDF set CT10f4. For this process, the top quarks are decayed using MadSpin~\cite{Artoisenet:2012st} to preserve all spin correlations. For all processes, the parton shower, fragmentation and underlying event are simulated using \Pythia 6.428~\cite{Sjostrand:2006za} with the CTEQ6L1 PDF sets~\cite{CTEQ6L1} and the corresponding Perugia2012 tune~\cite{perugia}. The single-top cross-sections for the $t$- and $s$-channels are normalised using their \NLO{} predictions, while for the $Wt$ channel it is normalised using its \NLO+\NNLL{} prediction~\cite{Kidonakis:2011wy,Aliev:2010zk,Kant:2014oha}. 

To estimate the effect of the parton shower (PS) algorithm, a~\Powheg{}+\herwigpp{} sample is generated using the same set-up for \Powheg{} as for the \Powheg{}+\PythiaSix{} sample. For alternative choices of PS, hadronisation and underlying event (UE) simulation, samples are produced with \herwigpp v2.7.1~\cite{Bahr:2008pv} using the UE-EE-5 tune~\cite{A14tune} and the CTEQ6L1 PDFs. The impact of the matrix element (ME) generator choice is evaluated using events generated with \mgamcatnlo{} v2.1.1~\cite{Alwall:2014hca}  at NLO and the CT10 PDF set, interfaced with \herwigpp{} and the UE-EE-5 tune.

The factorisation and hadronisation scales, as well as the $h_\textrm{damp}$ parameter, are varied in signal samples used to study the effect of possible mismodelling of QCD radiation. The following two samples are produced and compared to the nominal sample, where, in the first sample,
the factorisation and hadronisation scales are varied downward by a~factor of~0.5, the \hdamp{} parameter is increased to $2 m_{\textrm top}$ and the `radHi' tune variation from the Perugia2012 tune set is used.
In the second sample the factorisation and hadronisation scales are varied upward by a~factor of 2.0, the \hdamp{} parameter is unchanged and the `radLo' tune variation from the Perugia2012 tune set is used.

The unfolded data are compared to three additional \ttbar{} simulated samples~\cite{PUBSimtop} which use the NNPDF3.0NLO PDF set~\cite{Ball:2014uwa} for the ME: a~\mgamcatnlo{}+\PythiaEight{} sample using the A14 tune, a~\Powheg{}+\PythiaEight{} sample simulated with the  $h_\textrm{damp}$ parameter set to the top-quark mass, also using the A14 tune and a~\Powheg{}+\HerwigSeven{} sample generated with the $h_\textrm{damp}$ parameter set to 1.5 times the top-quark mass, using the H7-UE-MMHT tune.

The $t\bar{t}$ samples are normalised
using $\sigma_{t\bar{t}} = 832^{+46}_{-51}$~pb where the uncertainty includes effects due to scale, PDF and $\alpha_\textrm{S}$ variations,
evaluated using the Top++2.0 program~\cite{Czakon:2011xx}. The calculation includes next-to-next-to-leading-order (NNLO) QCD corrections and resums next-to-next-to-leading logarithmic (NNLL) soft gluon terms~\cite{Cacciari:2011hy,Beneke:2011mq,Baernreuther:2012ws,Czakon:2012zr,Czakon:2012pz,Czakon:2013goa}.

For the simulation of background events, inclusive samples containing single $W$ or $Z$ bosons in association with jets are simulated using the \Sherpa v2.1.1~\cite{Gleisberg:2008ta} event generator. Matrix elements are calculated for up to two partons at \NLO{} and four partons at \LO~using the Comix~\cite{Gleisberg:2008fv} and OpenLoop~\cite{Cascioli:2011va} matrix element event generators and merged with the \Sherpa parton shower~\cite{Schumann:2007mg} using the ME+PS@NLO prescription~\cite{Hoeche:2012yf}. The CT10 PDF set is used in conjunction with dedicated parton shower tuning developed by the authors of \Sherpa{}. The $W/Z+$jets events are normalised using the \NNLO~cross-sections~\cite{PUBSimWJets}.

Diboson processes with one of the bosons decaying hadronically and the other leptonically are simulated using the \Sherpa v2.1.1 event generator~\cite{Gleisberg:2008ta, PUBSimDiboson}. They are calculated for up to one ($ZZ$) or zero ($WW$, $WZ$) additional partons at \NLO{} and up to three additional partons at LO using the Comix  and OpenLoops matrix element event generators and merged with the \Sherpa parton shower using the ME+PS@NLO prescription. The CT10 PDF set is used in conjunction with dedicated parton shower tuning developed by the authors of \Sherpa{}. The event-generator cross-sections, already evaluated at NLO accuracy, are used in this case.

The $t\bar{t}$ state produced in association with weak bosons ($t\bar{t}$ + $W/Z/WW$, denoted as $t\bar{t}V$) are simulated using the \mgamcatnlo{} event generator at LO interfaced to the \Pythia 8.186 parton shower model~\cite{PUBSimttV}. The matrix elements are simulated with up to two ($t\bar{t}$ + $W$), one ($t\bar{t}$ + $Z$) or no ($t\bar{t}$ + $WW$) extra partons. The ATLAS underlying-event tune A14 is used together with the NNPDF2.3LO PDF set. The events are normalised using their respective NLO cross-sections~\cite{ttVSim}. 

A summary of the MC samples used in this analysis is shown in Table~\ref{tab:MC}.

\begin{table*} 
\scriptsize
\centering
\begin{tabular}{|l|l|c|c|c|c|} \hline
Physics process & Event generator & Cross-section & PDF set for & Parton shower & Tune \\
& & normalisation & hard process & & \\ \hline
$\ttbar$ Nominal & \PowHegBox v2 & \NNLO{}+\NNLL & CT10 &  \Pythia 6.428 & Perugia2012 \\ 
$\ttbar$ PS syst. & \PowHegBox v2 & \NNLO{}+\NNLL & CT10 & \herwigpp v2.7.1 & UE-EE-5 \\
$\ttbar$ ME syst. & \textsc{MadGraph5}\_ & \NNLO{}+\NNLL & CT10 & \herwigpp v2.7.1 & UE-EE-5 \\
                            & a\textsc{MC@NLO} & & & & \\
$\ttbar$ rad. syst. & \PowHegBox v2 & \NNLO{}+\NNLL & CT10 &  \Pythia 6.428 & `radHi/Lo'  \\ 
Extra $\ttbar$ model & \PowHegBox v2 & \NNLO{}+\NNLL &  NNPDF3.0NLO &  \Pythia 8.210 & A14 \\ 
Extra $\ttbar$ model & \PowHegBox v2 & \NNLO{}+\NNLL &  NNPDF3.0NLO &  \Herwig v7.0.1 & H7-UE-MMHT \\
Extra $\ttbar$ model &  \textsc{MadGraph5}\_  & \NNLO{}+\NNLL &  NNPDF3.0NLO &  \Pythia 8.210 & A14 \\  
                            & a\textsc{MC@NLO} & & & & \\

Single top  $t$-channel &\PowHegBox v1 & \NLO{} & CT10f4 &  \Pythia 6.428 & Perugia2012 \\ 
Single top  $s$-channel & \PowHegBox v2 & \NLO{} & CT10 &  \Pythia 6.428 & Perugia2012 \\ 
Single top $Wt$-channel & \PowHegBox v2 & \NLO{}+\NNLL & CT10 &  \Pythia 6.428 & Perugia2012 \\ 
$\ttbar$+$W/Z/WW$ &  \textsc{MadGraph5}\_ & \NLO{} & NNPDF2.3LO & \Pythia 8.186 & A14 \\ 
                            & a\textsc{MC@NLO} & & & & \\
$W(\to \ell \nu) $+ jets & \Sherpa v2.1.1 & \NNLO{} & CT10 & \Sherpa & \Sherpa  \\ 
$Z(\to \ell {\bar \ell}) $+ jets & \Sherpa v2.1.1 & \NNLO{} & CT10 & \Sherpa & \Sherpa  \\ 
$WW, WZ, ZZ$ & \Sherpa v2.1.1 & \NLO{} & CT10 & \Sherpa & \Sherpa  \\ \hline
\end{tabular}
\caption{Summary of MC samples, showing the event generator for the hard-scattering process, cross-section normalisation precision, PDF choice as well as the parton shower and the corresponding tune used in the analysis. The \PythiaSix{} and \herwigpp{} parton-shower models use the CTEQ6L1 PDF set, while \PythiaEight{} uses the NNPDF2.3LO PDF set and \HerwigSeven{} uses the MMHT2014lo68cl PDF set.}
\label{tab:MC}
\end{table*}

\section{Event reconstruction and selection}\label{sec:EventReco}
The lepton+jets \ttbar{} decay mode is characterised by the presence of a~high-\pt{} lepton, missing transverse momentum due to the neutrino from the semileptonic top-quark decay, and two jets originating from $b$-quarks.
Furthermore, in the resolved topology, two jets from the hadronic decay of the \Wboson{} boson are expected, while in the boosted topology, the presence of a~large-$R$ jet is required, in order to select events with a~high-\pt{} (boosted) hadronically decaying top quark.

The following sections describe the detector-level and particle-level objects used to characterise
the final-state event topology and to define a~fiducial phase-space region for the measurements.

\subsection{Detector-level objects}\label{sec:ObjectDef}

Primary vertices are formed from reconstructed tracks spatially compatible with the interaction region. The hard-scatter primary vertex is chosen to be the vertex with the highest $\sum \pt^2$ where the sum extends over all associated tracks with $\pt > 0.4\,\mathrm{\GeV{}}$.

Electron candidates are reconstructed by matching tracks in the inner detector to energy deposits in the EM calorimeter. They must satisfy a~``tight'' likelihood-based identification criterion based on shower shapes in the EM calorimeter, track quality and detection of transition radiation produced in the TRT detector~\cite{PUB2011006:elecperf}. The EM clusters are required to have a~transverse energy $\ET>$ 25~\GeV{} and be in the pseudorapidity region $|\eta| < 2.47$, excluding the transition region between the barrel and the end-cap calorimeters ($1.37 < |\eta| < 1.52$). The associated track must have a~longitudinal impact parameter $|z_0\, \sin \theta|<0.5$~mm and a~transverse impact parameter significance $|d_0|/\sigma(d_0)<5$ where $d_0$ is measured with respect to the beam line.
Isolation requirements based on calorimeter and tracking quantities are used to reduce the background from non-prompt and fake (mimicked by a~photon or a jet) electrons~\cite{atlasInclusive13}. The isolation criteria are \pT{}- and $\eta$-dependent and ensure an efficiency of 90\% for electrons with \pT{} of 25~\GeV{} and 99\% for electrons at 60~\GeV{}. These efficiencies are measured using electrons from $Z$ boson decays~\cite{ElecID}. 

Muon candidates~\cite{Aad:2016jkr} are identified by matching tracks in the muon spectrometer to tracks in the inner detector. The track $\pt$ is determined through a~global fit of the hits which takes into account the energy loss in the calorimeters.  Muons are required to have \pt $>$ 25~\GeV{} and to be within $|\eta|<2.5$.
To reduce the background from muons originating from heavy-flavour decays inside jets, muons are required to be separated by $\Delta R>0.4$ from the nearest jet and to be isolated using track quality and isolation criteria similar those applied for the electrons. If a~ muon shares a~track with an electron, it is likely to have undergone bremsstrahlung and hence the electron is not selected.

Jets are reconstructed using the anti-$k_{t}$ algorithm~\cite{akt1} implemented in the \textsc{FastJet} package \cite{Fastjet}. Two types of anti-$k_t$ jets are considered: so-called \textit{small-R} jets with radius parameter $R = 0.4$ and \textit{large-R} jets with radius parameter $R=1.0$. Jet reconstruction in the calorimeter starts from topological clusters calibrated to be consistent with expected electromagnetic or hadronic cluster shapes using corrections determined in simulation and inferred from test beam data. Jet four-momenta are then corrected for pile-up effects using the jet-area method \cite{Cacciari:2008gn}. In order to reduce the number of small-$R$ jets originating from pile-up, an additional selection criterion based on a~jet-vertex tagging (JVT) technique is applied. The JVT is a~likelihood discriminant that combines information from several track-based variables~\cite{Aad:2015ina}
and the criterion is only applied to small-$R$ jets with \pT $<$ 60~\GeV{} and $|\eta|$ < 2.4.

Small-$R$ jets 
are calibrated using an energy- and $\eta$-dependent simulation-based calibration scheme with \textsl{in situ} corrections based on data~\cite{Objects_jet_calibration, PERF-2016-04}, and are accepted if they have \pT $>$ 25~\GeV{} and $|\eta| < 2.5$. 

Objects can satisfy both the jets and leptons selection criteria and as such a procedure called "overlap removal" is applied in order to associate objects to a unique hypothesis. To prevent double-counting of electron energy deposits as jets, the closest small-$R$ jet lying  \mbox{$\Delta R < 0.2$} from a~reconstructed electron is discarded.
Subsequently, to reduce the impact of non-prompt leptons, if an electron is  \mbox{$\Delta R < 0.4$} from a~small-$R$ jet, then that electron is removed. If a~small-$R$ jet has fewer than three tracks and is  \mbox{$\Delta R < 0.4$} from a~muon, the small-$R$ jet is removed. Finally, the muon is removed if it is  \mbox{$\Delta R < 0.4$} from a~small-$R$ jet which has at least three tracks.

The purity of the selected \ttbar{} sample is improved by identifying small-$R$ jets containing $b$-hadrons. This identification exploits the long decay time of $b$-hadrons and the mass of the corresponding reconstructed secondary vertex, which is several~\GeV{} larger than that in jets originating from gluons or light-flavour quarks. Information from
the track impact parameters, secondary vertex location and 
decay topology are combined in a~multivariate algorithm (MV2c20).
The operating point used corresponds to an overall 77\% $b$-tagging
efficiency in \ttbar{} events, with a corresponding rejection of charm-quark jets (light-flavour and gluon jets) by a factor of 4.5 (140), respectively \cite{btagRun2}.

Large-$R$ jets associated with hadronically decaying top quarks are selected over jets originating from the fragmentation of other quarks or gluons by requiring that they contain several high-\pt objects and have a~mass compatible with the top-quark mass. A trimming algorithm~\cite{trim} is applied to large-$R$ jets to mitigate the impact of initial-state radiation, underlying-event activity and pile-up, with the goal of improving the mass resolution. Trimmed large-$R$ jets are considered if they fulfill $|\eta| <$ 2.0 and \pt $>$ 300~\GeV{}. Since large-$R$ jets with $m < 50$~\GeV{} or \pt $>$ 1500~\GeV{} are outside of a~well-calibrated region of phase-space, they are excluded from the selection.

Sub-jets, with radius $R_{\textrm{sub}}=0.2$, are clustered starting from the large-$R$ jet constituents by means of a $k_{t}$ algorithm. A sub-jet is selected only if it contains at least 5\% of the total large-$R$ jet transverse momentum, thereby removing the soft constituents from the large-$R$ jet. The $N$-subjettiness $\tau_N$~\cite{Thaler:2010tr} measures the consistency of the large-$R$ jet with its $N$ sub-jets when the jet constituents are reclustered with a~smaller-$R$ jet algorithm. 
A top-tagging algorithm~\cite{ATL-PHYS-PUB-2015-053} is applied that depends on the calibrated jet mass and the $N$-subjettiness ratio $\tau_{32} \equiv \tau_3 / \tau_2$: going from \pt = 300~\GeV{} to 1500~\GeV{}, the  $\tau_{32}$ upper requirement varies from 0.85 to 0.70, while the lower requirement on the minimum calibrated jet mass varies from 70~\GeV{} to 120~\GeV{}. These correspond to a loose working point with an approximately flat top-tagging efficiency of 80\% above \pt of 400~\GeV{}.

The missing transverse momentum $\Etmiss$ is computed from the vector sum of the transverse momenta of 
the reconstructed calibrated physics objects (electrons, photons, semi-hadronically decaying $\tau$ leptons, jets and muons) together with the transverse energy 
deposited in the calorimeter cells, calibrated using tracking information, not associated with these objects \cite{atlasEtmisPerf}.
The contribution from muons is added using their momenta. To avoid double-counting of energy, the muon energy loss in the calorimeters is subtracted in the $\Etmiss$ calculation.

\subsection{Event selection at detector level}\label{sec:EventSelection}
The event selection comprises a~set of requirements based on the general event quality and on the reconstructed objects, defined above, that characterise the final-state event topology. The analysis applies two non-exclusive event selections: one corresponding to a~resolved topology and another targeting a boosted (collimated decay) topology. 

For both selections, events must have a~reconstructed primary vertex with two or more associated tracks and contain exactly one reconstructed
lepton candidate with \pt $>$ 25~\GeV{} geometrically matched to a~corresponding object at trigger level.

For the resolved event selection, each event must also contain at least four small-$R$ jets with \pt $>$ 25~\GeV{} and $|\eta| < 2.5$ of which at least two must be tagged as $b$-jets. 

For the boosted event selection,
at least one small-$R$ jet close to the lepton, i.e. with $\Delta R$(small-$R$ jet, lepton) $< 2.0$, and at least one large-$R$ top-tagged jet are required.
The large-$R$ jet must be well separated from the lepton, $\Delta\phi($large-$R$ jet, lepton$) >$ 1.0, and from the small-$R$ jet associated with the lepton, $\Delta R($large-$R$ jet, small-$R$ jet) $> 1.5$. In addition, it is required that at least one $b$-tagged small-$R$ jet fulfills the following requirements: it is either inside the large-$R$ jet, $\Delta R($large-$R$ jet, $b$-tagged jet)$\,< 1.0$, or it is the small-$R$ jet associated with the lepton. Finally, in order to suppress the multijet background\footnote{Also referred to as non-prompt real-leptons and fake-leptons background, as described in Section~\ref{sec:BackgroundDetermination}.} in the boosted topology the missing transverse momentum must be larger than 20~\GeV{} and the sum of \Etmiss{} and $m_{\textrm T}^W$ (transverse mass of the $W$ boson\footnote{$m_{\textrm T}^{W} = \sqrt{2 p_{\textrm T}^{\ell} E_{\textrm T}^{\textrm{miss}} (1 - \cos \Delta \phi(\ell, E_{\textrm T}^{\textrm{miss}}))}$, where $\ell$ stands for the charged lepton.}) must be larger than 60~\GeV{}.

The event selection is summarised in Table~\ref{tab:seln}.

\begin{table} [!htbp]
\resizebox{\textwidth}{!}{
 \begin{tabular}{|l|c|c|l|}
  \hline
  Level & \multicolumn{2}{c|}{Detector} & Particle \\
  \hline
   Topology         & Resolved & Boosted & \\
  \hline\hline
  Leptons &  \multicolumn{2}{c|}{\parbox[c][3em][c]{8.5cm}{ $|d_0|/\sigma(d_0)<5$ and $|z_0 \sin\theta | <$ 0.5 mm \\ Track and calorimeter isolation\\ 
 $|\eta|<1.37$ or $1.52<|\eta|<2.47$ {($e$)},  $|\eta|<2.5$ {($\mu$)} \\ $E_{\mathrm{T}} $ {($e$)}, $p_{\mathrm{T}}$ {($\mu$)} $>$ 25 \GeV{} }}  & \parbox[c][7em][c]{2.0cm}{ $|\eta|<2.5$\\ \mbox{$p_{\mathrm T} >$ 25~\GeV{}}  } \\
  \hline
  Small-$R$ jets &
  \multicolumn{2}{c|}{\parbox[c][4em][c]{8.5cm}{  $|\eta|<2.5$ \\$p_{\mathrm{T}} > $ 25~\GeV{}   \\ JVT cut (if $p_{\mathrm{T}}<$ 60~\GeV{} and $|\eta| < 2.4$)}} & \parbox[c][4em][c]{2.0cm}{$|\eta|<2.5$\\ \mbox{$p_{\mathrm T} >$ 25~\GeV{}}}\\ \hline
Num. of  small-$R$ jets &  $\geq 4$ jets &  $\geq 1$ jet & Same as detector level
\\   \hline
\parbox[l][2.5em][c]{2cm}{  \met, \mtw} & & \met $>$ 20~\GeV{}, $\met+\mtw >$ 60~\GeV{} & Same as detector level  \\
  \hline
  Leptonic top  &  \parbox[c][6em][c]{3.5cm}{ Kinematic top-quark \\ reconstruction \\ for detector \\ and particle level} & \parbox[c][4em][c]{7.5cm}{ At least one small-$R$ jet \\with $\Delta R$($\ell$, small-$R$ jet) $< 2.0$ }& \\ 
  \hline
  Hadronic top &  \parbox[c][6em][c]{3.5cm}{ Kinematic top-quark \\ reconstruction \\ for detector \\ and particle level} &
 \parbox[c][4em][c]{7.0cm}{ The
      leading-$p_{\mathrm T}$ trimmed large-$R$ jet has: \\  $|\eta|<2.0$, \\ 300~\GeV{} $< p_{\mathrm{T}}<$ 1500~\GeV{}, $m > 50$~\GeV{}, \\Top-tagging at 80\% efficiency \\ $\Delta R$(large-$R$ jet, small-$R$ jet associated with lepton) $>1.5$, \\ $\Delta \phi$($\ell$, large-$R$ jet) $ > 1.0$}
      &  \parbox[c][8.5em][c]{3.7cm}{  {\textbf{Boosted:}} \\ $|\eta|<2.0$\\ 300 $< p_{\mathrm{T}}<$ 1500~\GeV{} \\ Top-tagging:\\  $m>$ 100~\GeV{}, \\ $\tau_{32} <$ 0.75} \\
  \hline
  $b$-tagging & At least 2 $b$-tagged jets & 
   \parbox[c][4em][c]{7.5cm}{At least
      one of: \\1) the leading-$p_{\mathrm T}$ small-$R$ jet with \\ $\Delta R$($\ell$, small-$R$ jet) $ < 2.0$ is $b$-tagged \\  2) at least one small-$R$ jet with \\ $\Delta R$(large-$R$ jet, small-$R$ jet) $ < 1.0$ is $b$-tagged 
     } 
     &\parbox[c][7em][c]{3.2cm}{ Ghost-matched \\ $b$-hadron}\\
  \hline
  \end{tabular}
}
\caption{Summary of the requirements for detector-level and MC-generated particle-level events, for both the resolved and boosted event selections. The description of the particle-level selection is in Section~\ref{sec:TruthObjectDef}. The description of the kinematic top-quark reconstruction for the resolved topology is in Section~\ref{sec:PseudoTop}. Leptonic (hadronic) top refers to the top quark that decays into a leptonically (hadronically) decaying $W$ boson.}
\label{tab:seln}
\end{table}

\subsection{Particle-level objects and fiducial phase-space definition}\label{sec:TruthObjectDef}
Particle-level objects are defined for simulated events in analogy to the detector-level objects described above.
Only particles with a~mean lifetime of $\tau > 30\,$ps are considered.

The fiducial phase-space for the measurements presented in this paper is defined using a~series of requirements applied to particle-level objects analogous to those used in the selection of the detector-level objects. The procedure explained in this section is applied to the \ttbar{} signal only, since the background subtraction is performed before unfolding the data to particle level.

Electrons and muons must not originate, either directly or through a~$\tau$ decay, from a~hadron in the MC particle record. 
This ensures that the lepton is from an electroweak decay without requiring a~direct match to a~$W$ boson.
The four-momenta of leptons are modified by adding the four-momenta of all photons within $\Delta R=0.1$ and not originating from hadron decays, to take into account final-state photon radiation.
Such leptons are then required to have \pT $>$ 25~\GeV{} and $|\eta| < 2.5$. Electrons in the calorimeter's transition region (1.37 < $|\eta|$ < 1.52 ) are rejected at detector level but accepted in the fiducial selection. This difference is accounted for by the efficiency described in Section~\ref{sec:unfolding}.

Particle-level jets are clustered using the anti-$k_{t}$ algorithm with radius parameter $R = 0.4$ or $R = 1.0$, starting from all stable particles, except for selected leptons ($e$, $\mu$) and their radiated photons, as well as neutrinos.

Small-$R$ particle-level jets are required to have \pT $>$ 25~\GeV{} and $|\eta| < 2.5$.
Hadrons with \pT $>$ 5~\GeV{} containing a~$b$-quark are matched to small-$R$ jets through a~ghost-matching technique as described in Ref.~\cite{Cacciari:2008gn}.
Neutrinos and charged leptons from hadron decays are included in particle-level jets. 
The large-$R$ particle-level jets have to fulfill 300~\GeV{} $<$ \pT $<$ 1500~\GeV{}, $m > 50$~\GeV{} and $|\eta| <$ 2.0. A~top-tag requirement is applied at particle-level: if the large-$R$ jet has a~mass larger than 100~\GeV{} and $\tau_{32} < 0.75$, the large-$R$ jet is considered to be top-tagged. No overlap removal criteria are applied to particle-level objects. 

The particle-level missing transverse momentum is calculated from the four-vector sum of the neutrinos, discarding neutrinos from hadron decays, either directly or through a~$\tau$ decay.

Particle-level events in the resolved topology are required to contain exactly one lepton
and at least four small-$R$-jets passing the aforementioned requirements,
with at least two of the small-$R$ jets required to be $b$-tagged. For the boosted topology, after the same lepton requirements as in the resolved case, the events are required to contain at least one large-$R$ jet that is also top-tagged and at least one $b$-tagged small-$R$ jet fulfilling the same $\Delta R$ requirements as at detector-level as described in~Section~\ref{sec:ObjectDef}. In addition, for the boosted topology, the missing transverse momentum must be larger than 20~\GeV{} and the sum of \Etmiss$+m_{\textrm T}^W >$ 60~\GeV{}. 

Dilepton $\ttbar$ events where only one lepton satisfies the fiducial selection are by definition included in the fiducial measurement. 

Table~\ref{tab:seln} summarises the object and event selections at both detector- and particle-level for each topology.

\section{Background determination and event yields} \label{sec:BackgroundDetermination}

Following from the event selection, various backgrounds, mostly involving real leptons, will contribute to the event yields. Data-driven techniques are used to estimate backgrounds that suffer from large theoretical uncertainties like the production of $W$ bosons in association with jets, or that rely on a precise simulation of the detector for backgrounds that involve jets mimicking the signature of charged leptons. 

The single-top-quark background is the largest background contribution in both the resolved and boosted topologies, amounting to 4--6\% of the total event yield and $35$\% of the total background estimate. Shapes of all distributions of this background are modelled with MC simulation, and the event yields are normalised using calculations of its cross-section, as described in Section~\ref{sec:DataSimSamples}.

Multijet production processes, including all-hadronic \ttbar{} production, have a~large cross-section and mimic the lepton+jets signature due to jets misidentified as prompt leptons (fake leptons) or to semileptonic decays of heavy-flavour hadrons (non-prompt real leptons). The multijet background is estimated directly from data by using a~matrix-method~\cite{atlasXsec3}. The number of background events in the signal region is evaluated by applying efficiency factors (fake-lepton and real-lepton efficiencies) to the number of events satisfying a~tight (signal) as well as a~looser lepton selection. The fake-lepton efficiency is measured using data in control regions dominated by the multijet background with the real-lepton contribution subtracted using MC simulation. The real-lepton efficiency is extracted from a~tag-and-probe technique using leptons from $Z$ boson decays.
The multijet background contributes to the total event yield at the level of approximately $3$--$4$\%, corresponding to approximately $20$--$31$\% of the total background estimate. 

The $W$+jets background represents the third-largest background in both topologies, amounting to approximately $1$--$4$\% of the total event yield and $20$--$36$\% of the total background estimate. 
The estimation of this background is performed using a~combination of
MC simulation and data-driven techniques.
The \Sherpa $W$+jets samples, normalised using
the inclusive $W$ boson NNLO cross-section, are used as a~starting point while
the absolute normalisation and the heavy-flavour fractions of this process, which are affected by large theoretical uncertainties, are determined from data.

The overall $W$+jets normalisation is obtained by exploiting the expected charge asymmetry in the production of $W^+$ and $W^-$ bosons in $pp$ collisions. This asymmetry is predicted by theory~\cite{Halzen:2013bqa} and evaluated using MC simulation, assuming other processes are symmetric in charge except for a~small contamination from single-top-quark, $t\bar{t}V$ and $WZ$ events, which is subtracted using MC simulation.
The total number of $W$+jets events with a positively and negatively charged $W$ boson ($N_{W^+} + N_{W^-}$) in the sample can thus be estimated with the following equation

\begin{equation}
N_{W^+} + N_{W^-} = \left(\frac{r_{\textrm MC} + 1}{r_{\textrm MC} - 1}\right)(D_{\textrm+} - D_{\textrm-})\,,
\label{eq:Wchargeasymm}
\end{equation}

where $r_{\textrm MC}$ is the ratio of the number of events with positive leptons to the number of events with negative leptons in the MC simulation, and $D_{\textrm+}$ and $D_{\textrm-}$ are the numbers of events with positive and negative leptons in the data, respectively, corrected for the aforementioned non-$W+$jets charge-asymmetric contributions from simulation.

The corrections due to generator mismodelling of $W$ boson production in association with jets of different flavour ($W+b\bar{b}$, $W+c\bar{c}$, $W+c$, $W+\textrm{light}$ flavours)
are estimated in a~dedicated control sample in data which uses the same lepton and $\met$ selections as for the signal, but requiring exactly two small-$R$ jets. In the determination of the corrections, the overall normalisation scaling factor obtained using Eq.~(\ref{eq:Wchargeasymm}) is applied first. Then heavy-flavour scaling factors obtained in the two-jet control region are extrapolated to the signal region using MC simulation, assuming constant relative rates for the signal and control regions. Taking into account the heavy-flavour scale factors, the overall normalisation factor is calculated again using Eq.~(\ref{eq:Wchargeasymm}).
This iterative procedure is repeated until the total predicted $W+$jets yield in the two-jet control region agrees with the data yield at the per-mille level.
The detailed procedure can be found in Ref.~\cite{Aad:2015fna}.

The background contributions from $Z$+jets, $t\bar{t}V$ and diboson events are obtained from MC generators, and the event yields are normalised
as described in Section \ref{sec:DataSimSamples}. The total contribution from these processes is 1--2\% of the total event yield or 11--14\% of the total background.

Dilepton top-quark pair events (including decays to $\tau$ leptons) can satisfy the event selection, contributing approximately 5\% to the total event yield, and are considered in the analysis at both the detector and particle levels.
In the fiducial phase-space definition, semileptonic $\ttbar$ decays to $\tau$ leptons in lepton+jets $\ttbar$ events are
considered as signal only if the $\tau$ lepton decays leptonically.
Cases where both top quarks decay semileptonically to a~$\tau$ lepton, and where subsequently the $\tau$ leptons decay semihadronically, are accounted for in the multijet background.

\begin{table}[t]
\centering
 \begin{tabular}{lrr}
\hline\hline 
Process & \multicolumn{2}{c}{\ \ \ Expected events\ \ \ } \\ 
\hline
& \hbox{Resolved} & \hbox{Boosted} \\ 
 \hline
$t\bar{t}$ & 123800  $\pm$ 10600 & 7000 $\pm$ 1100 \\
 Single top & 6300  $\pm$ 800 & 500 $\pm$ 80 \\
Multijets & 5700  $\pm$ 3000 & 300 $\pm$ 80 \\ 
$W$+jets & 3600 ${}^{+2000}_{-2400}$ & $500 \pm 200$ \\
$Z$+jets & 1300  $\pm$    700 & 60 $\pm$ 40 \\ 
$t\bar{t}V$ & 400  $\pm$ 100 & 70 $\pm$ 10 \\ 
Diboson & 300  $\pm$    200 & 60 $\pm$ 10 \\
\hline
\\[-1.1em]
Total prediction & 142000 ${}^{+11000}_{-12000}$  & 8300  $\pm$ 1300 \\[2.3pt]
\hline 
 Data & 155593 & 7368  \\ 
\hline 
\hline
\end{tabular}
\caption{Event yields 
after the resolved and boosted selections. The signal model, denoted $t\bar{t}$ in the table, is generated using \PowHeg{}+\PythiaSix{}, normalised to NNLO calculations.  The uncertainties include the combined statistical and systematic uncertainties, excluding the systematic uncertainties related to the modelling of the \ttbar{} system, as described in Section~\ref{sec:Uncertainties}.
}
\label{tab:yields}
\end{table}

As the individual \ejets{} and \mujets{} channels have very similar corrections (as described in~Section~\ref{sec:unfolding}) and give consistent results at detector level, they are combined by summing the distributions. 
The event yields are displayed in Table~\ref{tab:yields} for data, simulated signal, and backgrounds.
Figures~\ref{fig:controls_4j2b_detector}--\ref{Fig:app_sig_CP15_sum} show,\footnote{All data as well as theory points are plotted at the bin centre of the $x$-axis throughout this paper.} for different distributions, the comparison between data and predictions. The selection produces a sample with an expected background of $13$\%  and $17$\% for the resolved and boosted topology, respectively. The overall difference between data and prediction is $10$\%  and $-9$\% in the resolved and boosted topology, respectively. This is in fair agreement within the combined experimental systematic and theoretical uncertainties of the \ttbar{} total cross-section used to normalise the signal MC sample (see Section~\ref{sec:DataSimSamples}), although in opposite directions between the resolved and boosted selections. This is due to the fact that each selection covers a~very different kinematic region, as described in Section~\ref{sec:TruthObjectDef}.

\begin{figure*}[p]
\centering
\subfigure[]{ \includegraphics[width=0.45\textwidth]{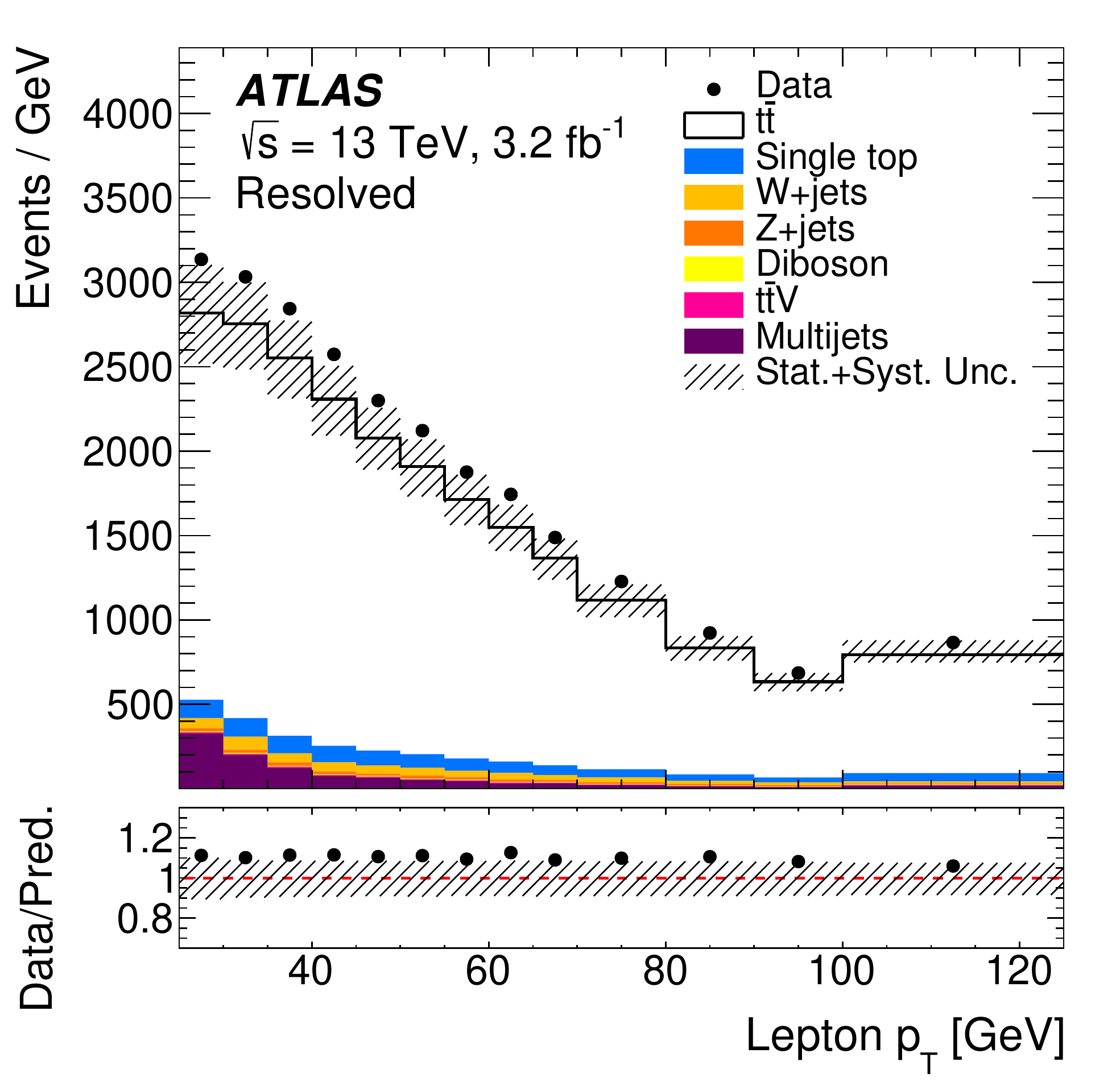}\label{fig:lep_pt_co}}
\subfigure[]{ \includegraphics[width=0.45\textwidth]{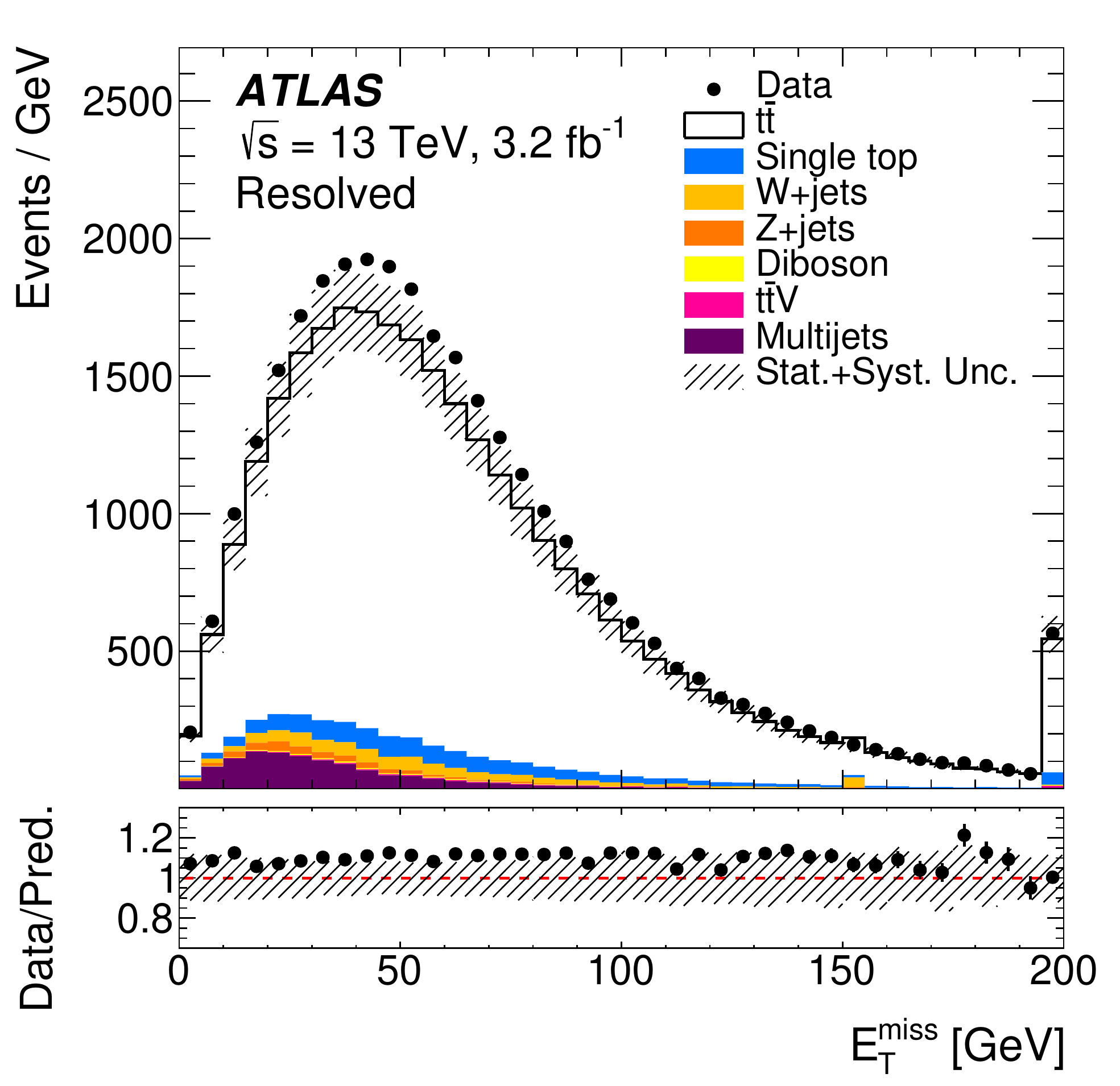}\label{fig:met_co}}
\subfigure[]{ \includegraphics[width=0.45\textwidth]{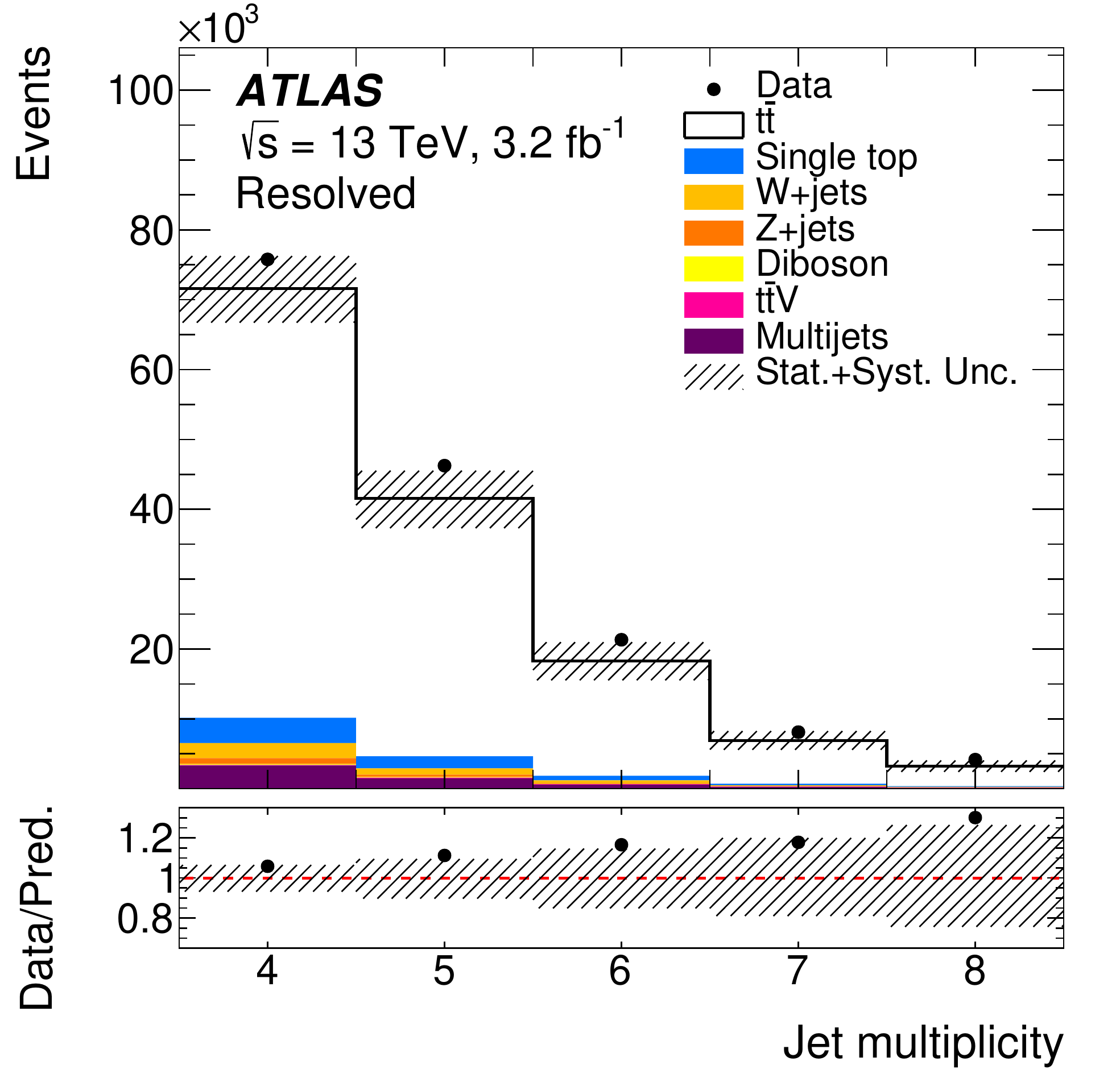}\label{fig:jet_n_co}}
\subfigure[]{ \includegraphics[width=0.45\textwidth]{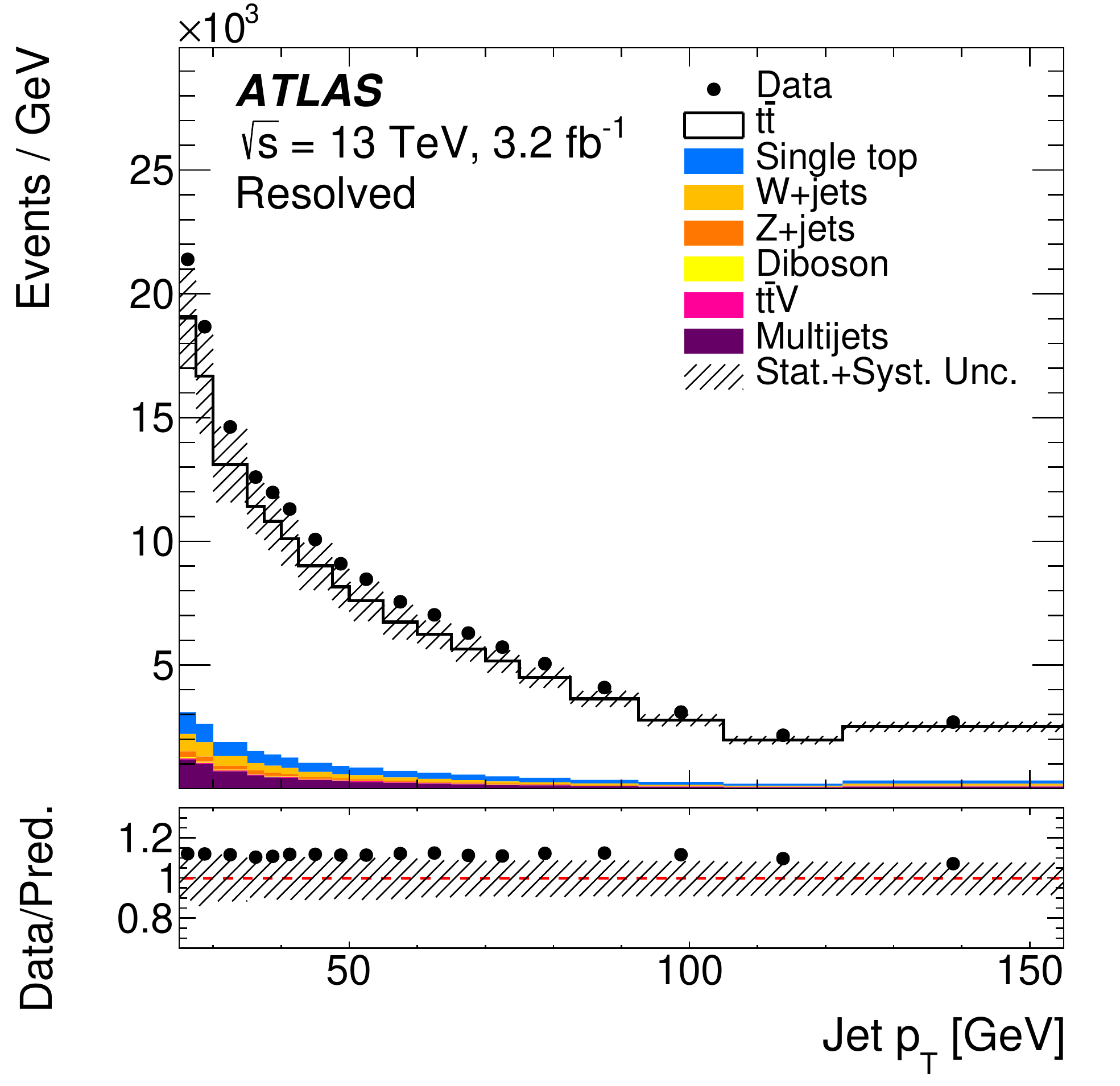}\label{fig:jet_pt_co}}
\caption{Kinematic distributions in the combined \ljets{} channel in the resolved topology at detector level: \subref{fig:lep_pt_co}~lepton transverse momentum and \subref{fig:met_co}~missing transverse momentum \Etmiss{}, \subref{fig:jet_n_co}~jet multiplicity and \subref{fig:jet_pt_co}~transverse momenta of selected jets. Data distributions are compared to predictions using \Powheg{}+\PythiaSix{} as the \ttbar{} signal model. The hatched area indicates the combined statistical and systematic uncertainties in the total prediction, excluding systematic uncertainties related to the modelling of the \ttbar{} system. Events beyond the range of the horizontal axis are included in the last bin.}
\label{fig:controls_4j2b_detector}
\end{figure*}

\begin{figure*}[p]
\centering
\subfigure[]{ \includegraphics[width=0.45\textwidth]{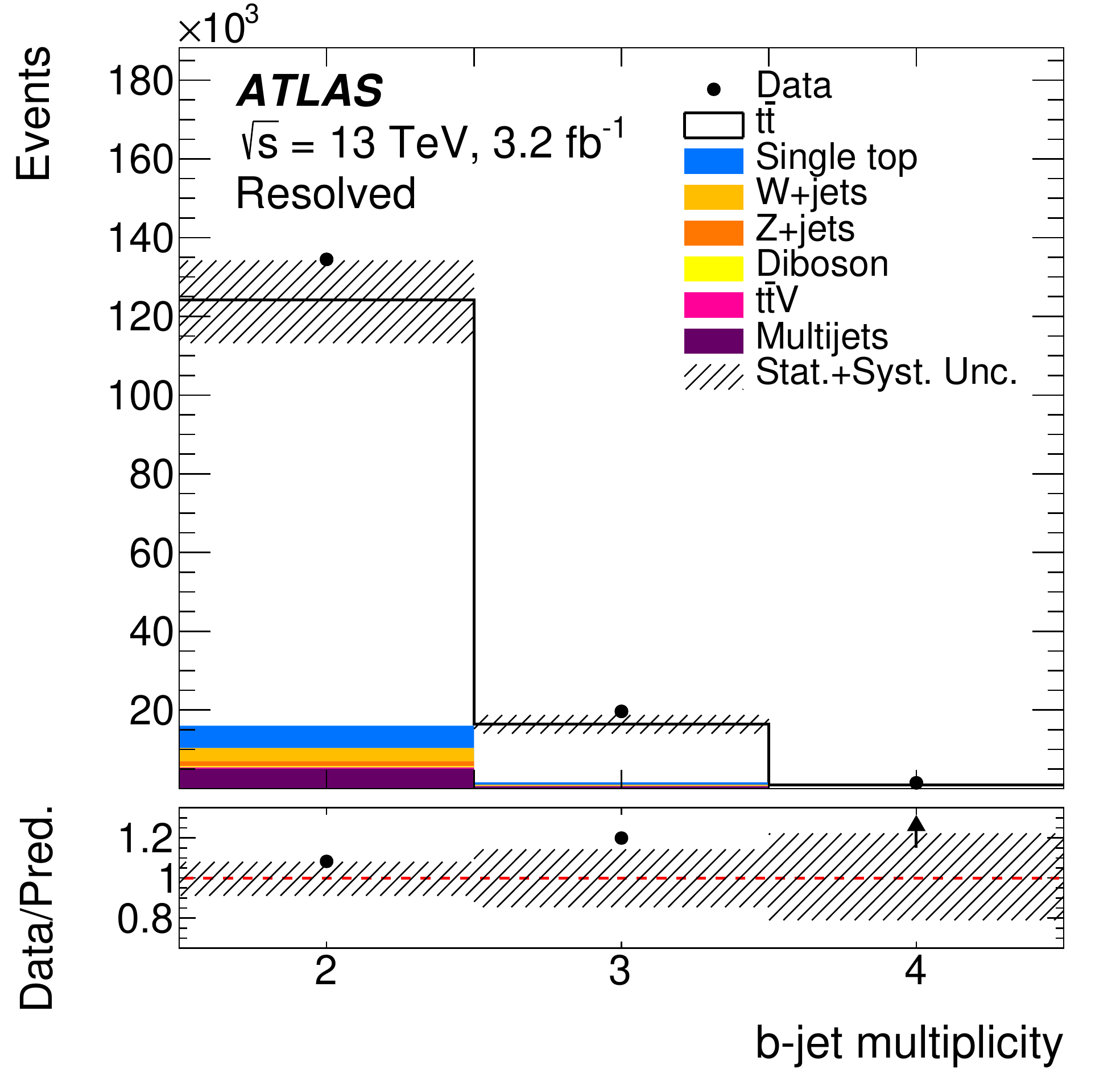}\label{fig:bjet_n_co}}
\subfigure[]{ \includegraphics[width=0.45\textwidth]{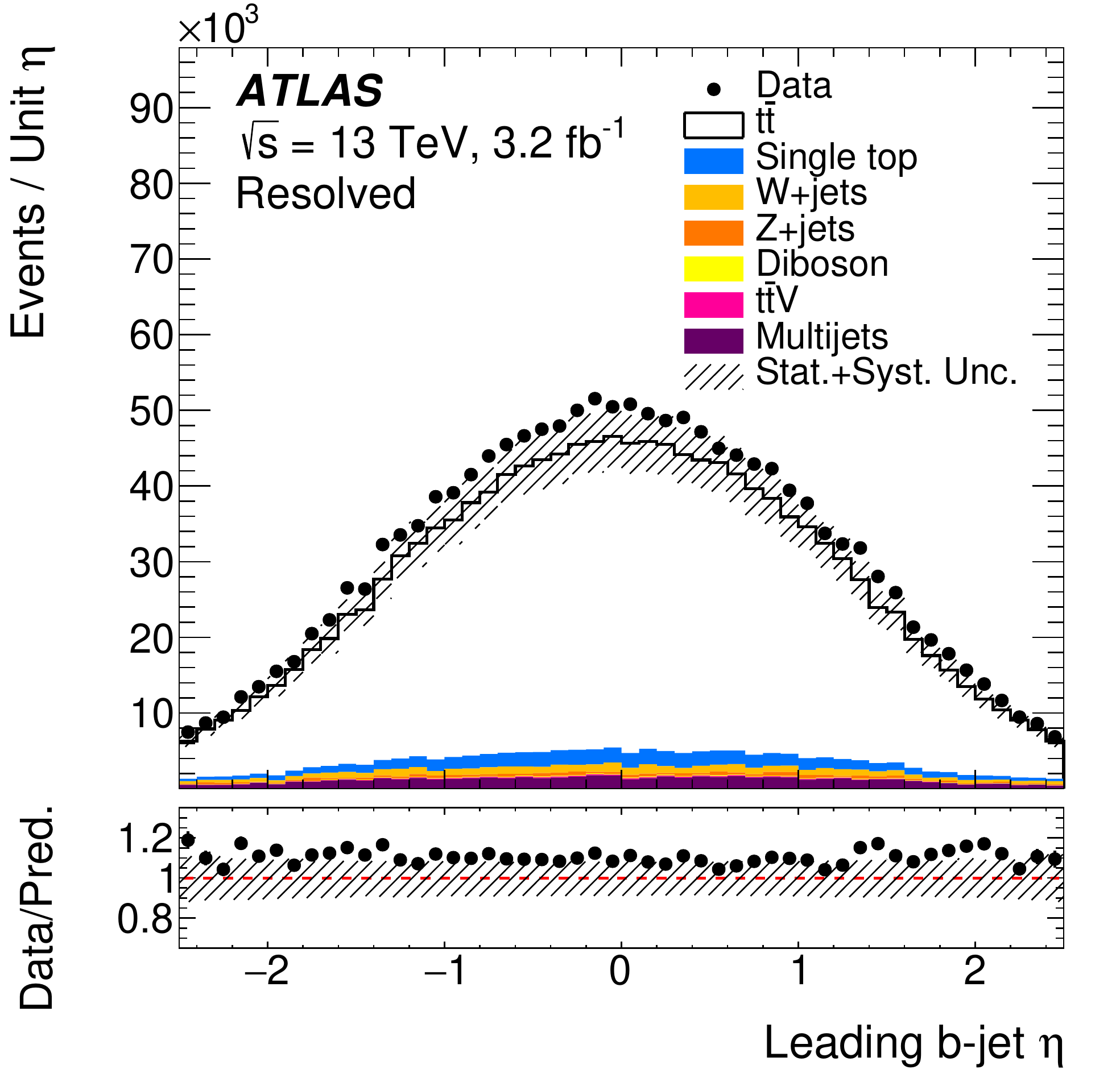}\label{fig:bjet1_eta_co}}
\caption{Kinematic distributions in the combined \ljets{} channel in the resolved topology at detector level: \subref{fig:bjet_n_co}~number of $b$-tagged jets and  \subref{fig:bjet1_eta_co}~leading $b$-tagged jet $\eta{}$. Data distributions are compared to predictions using \Powheg{}+\PythiaSix{} as the \ttbar{} signal model. The hatched area indicates the combined statistical and systematic uncertainties in the total prediction, excluding systematic uncertainties related to the modelling of the \ttbar{} system. Events (below) beyond the range of the horizontal axis are included in the (first) last bin.}
\label{fig:controls_4j2b_detector_2}
\end{figure*}

\begin{figure}[p]
\centering
\subfigure[]{\includegraphics[width=0.45\textwidth]{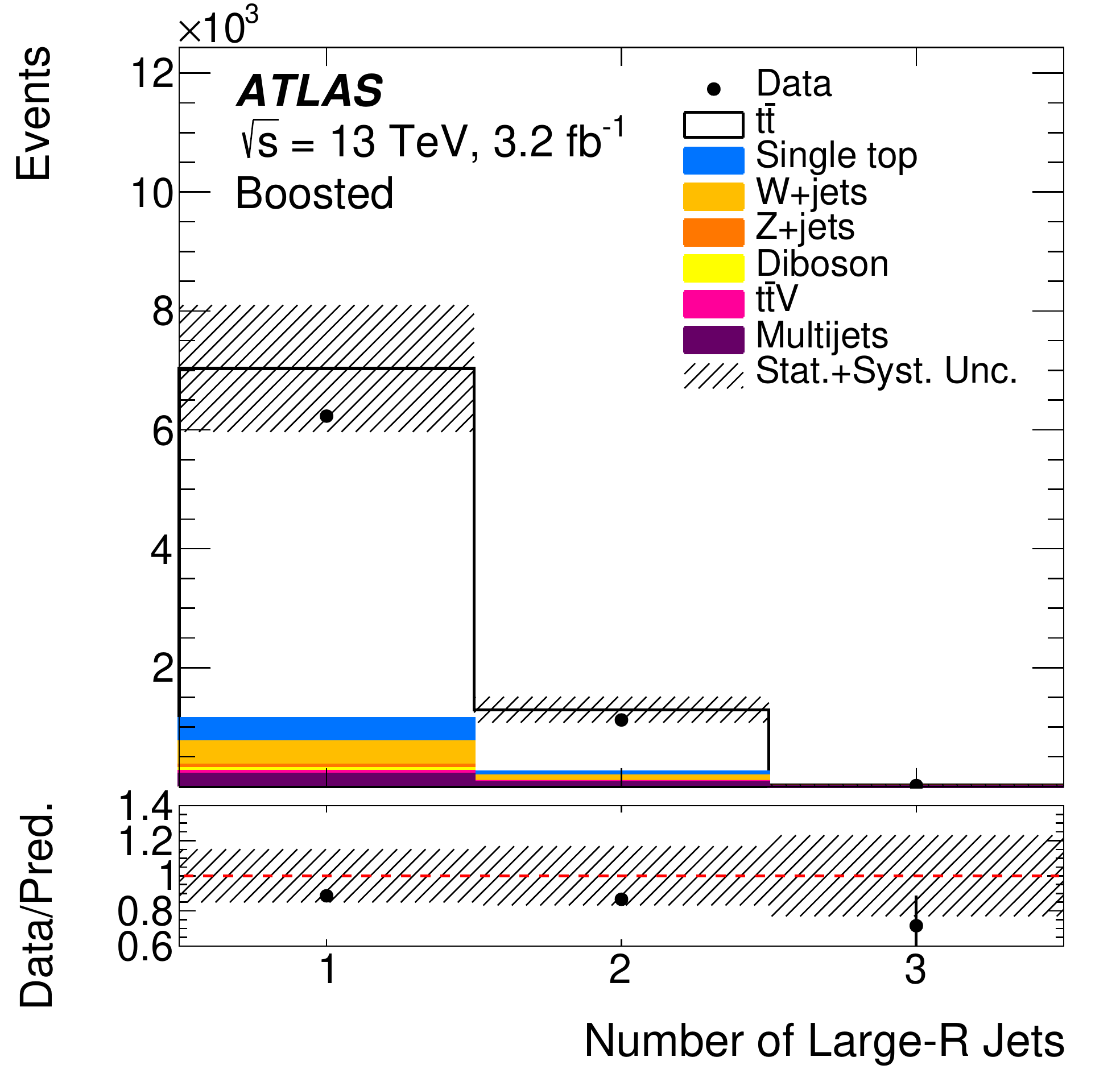} \label{fig:largejet_n}}
\subfigure[]{\includegraphics[width=0.45\textwidth]{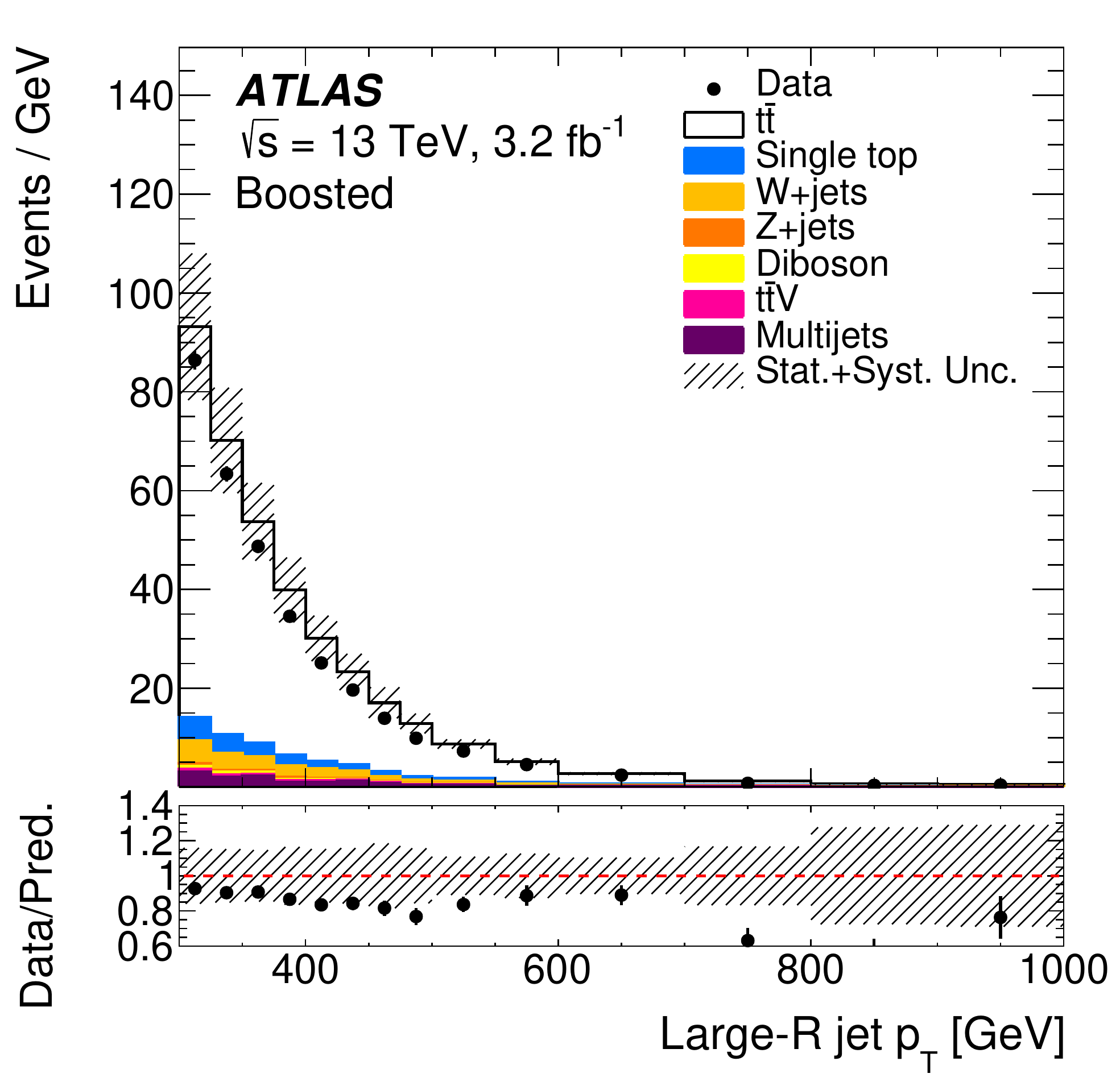}\label{fig:largejet_pt}}
\caption{Kinematic distributions in the combined \ljets{} channel in the boosted topology at detector level: \subref{fig:largejet_n} number of large-$R$ jets and \subref{fig:largejet_pt} large-$R$ jet $\pt{}$. Data distributions are compared to predictions using \Powheg{}+\PythiaSix{} as the \ttbar{} signal model. The hatched area indicates the combined statistical and systematic uncertainties in the total prediction, excluding systematic uncertainties related to the modelling of the \ttbar{} system. Events beyond the range of the horizontal axis are included in the last bin.}
\label{Fig:app_sig_CP11_sum}
\end{figure}

\begin{figure}[p]
\centering
\subfigure[]{\includegraphics[width=0.45\textwidth]{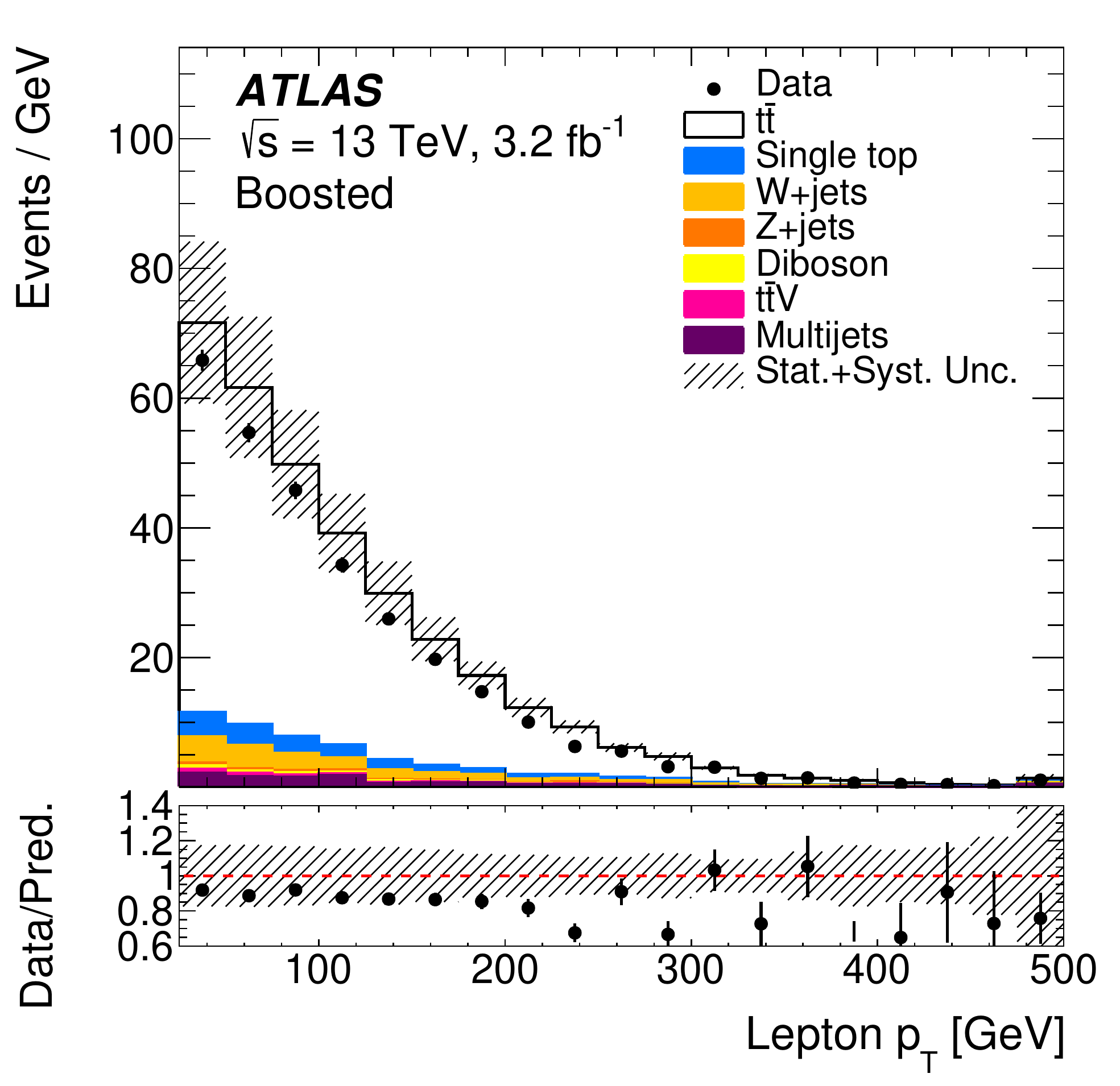} \label{fig:largejet_lep_pt}}
\subfigure[]{\includegraphics[width=0.45\textwidth]{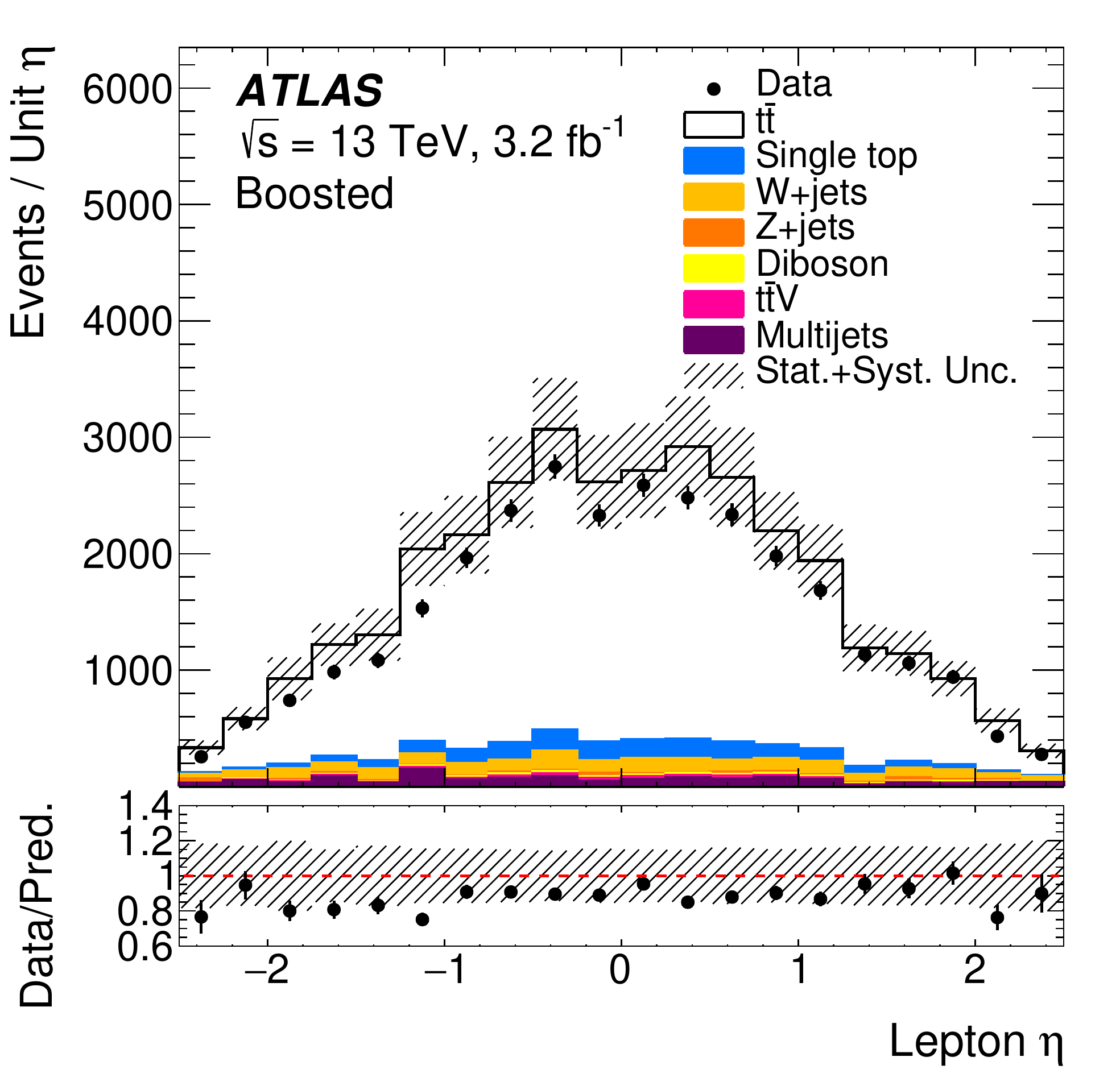} \label{fig:largejet_lep_eta}}
\subfigure[]{\includegraphics[width=0.45\textwidth]{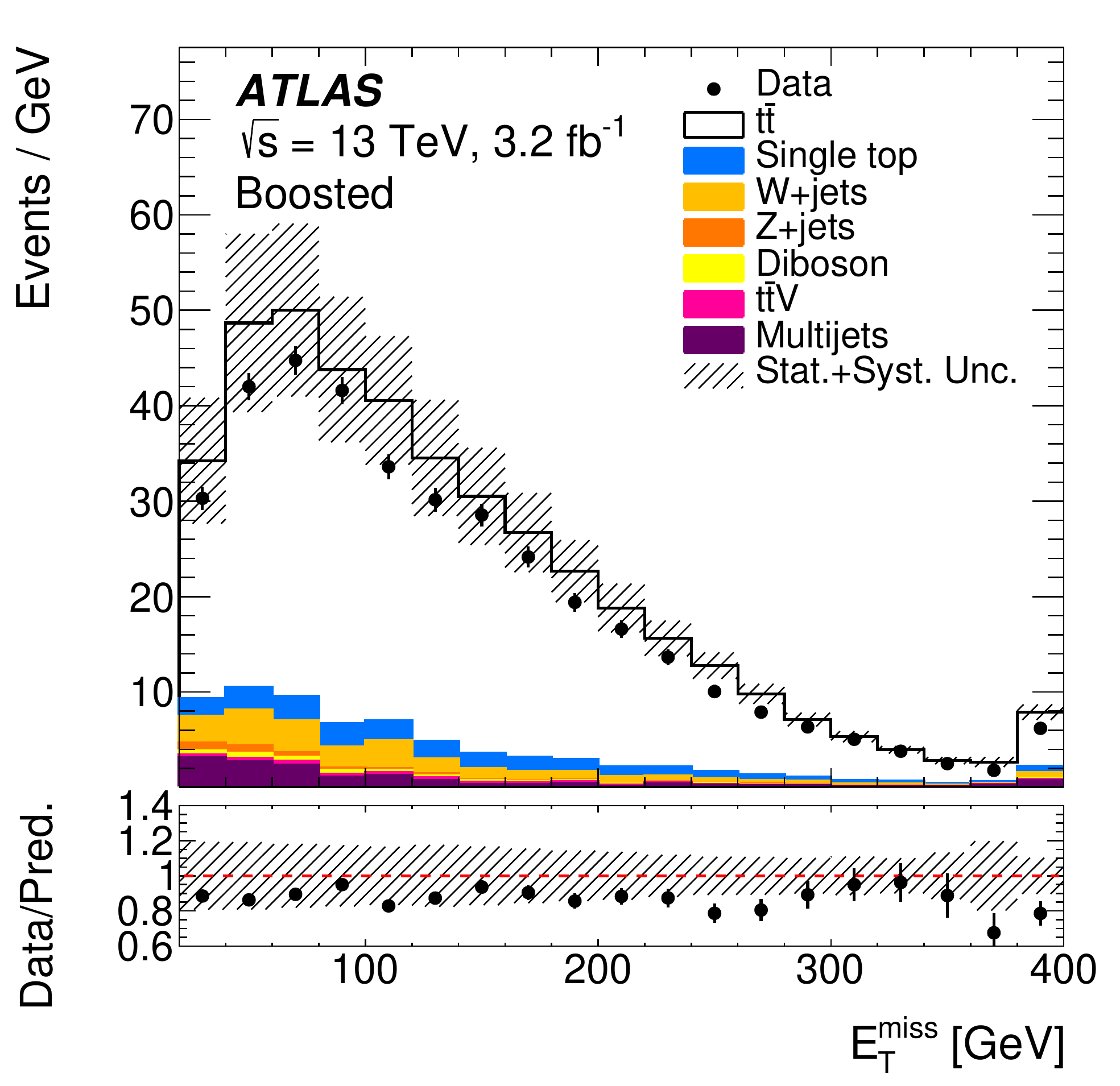} \label{fig:largejet_met}}
\subfigure[]{\includegraphics[width=0.45\textwidth]{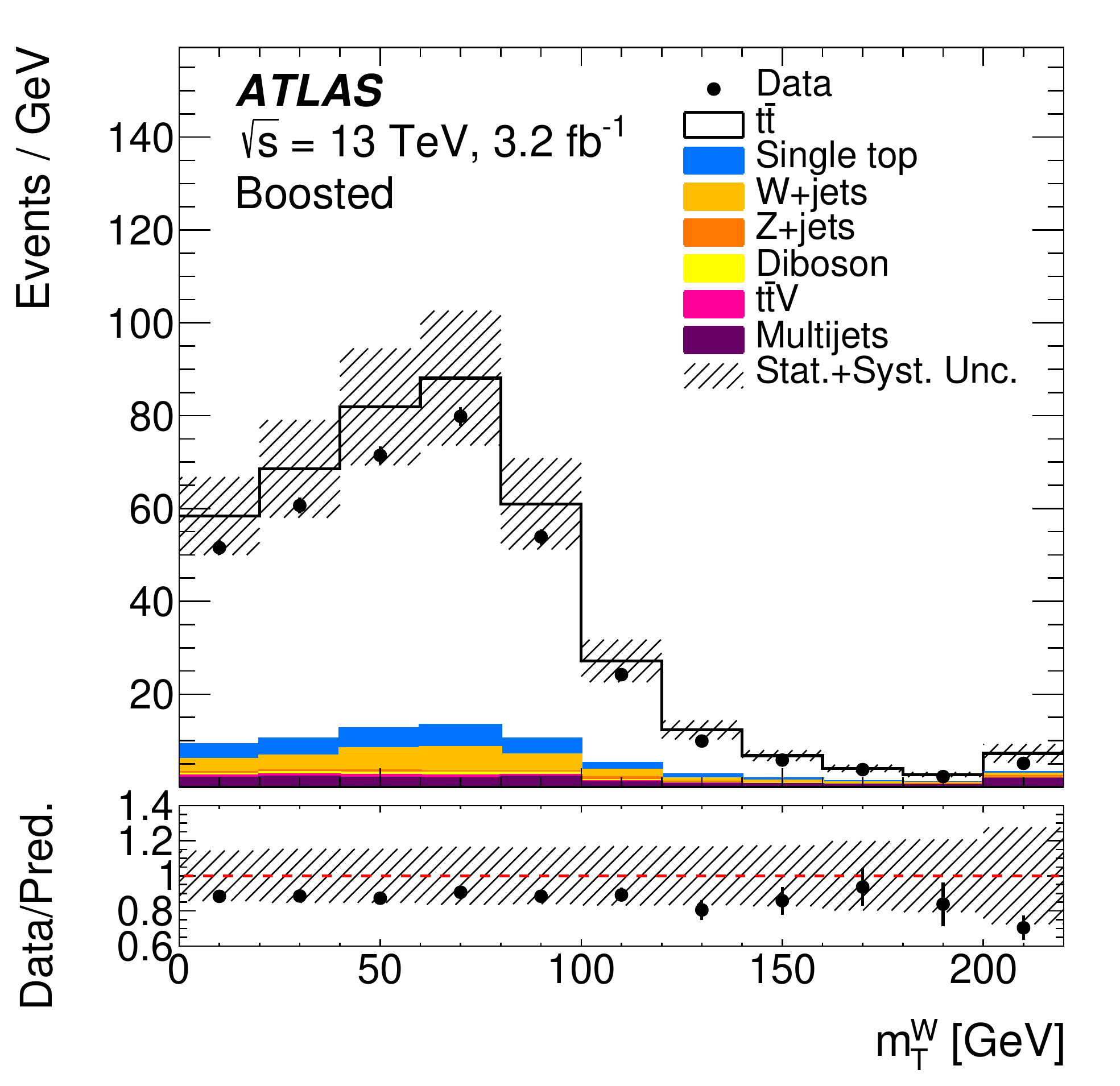} \label{fig:largejet_mtw}}
\caption{Kinematic distributions in the combined \ljets{} channel in the boosted topology at detector level: \subref{fig:largejet_lep_pt} lepton $\pt{}$ and \subref{fig:largejet_lep_eta} pseudorapidity, the \subref{fig:largejet_met} missing transverse momentum \Etmiss{} and \subref{fig:largejet_mtw} transverse mass of the $W$ boson. Data distributions are compared to predictions using \Powheg{}+\PythiaSix{} as the \ttbar{} signal model. The hatched area indicates the combined statistical and systematic uncertainties in the total prediction, excluding systematic uncertainties related to the modelling of the \ttbar{} system. Events (below) beyond the range of the horizontal axis are included in the (first) last bin.}
\label{Fig:app_sig_CP12_sum}
\end{figure}

\begin{figure}[p]
\centering
\subfigure[]{\includegraphics[width=0.45\textwidth]{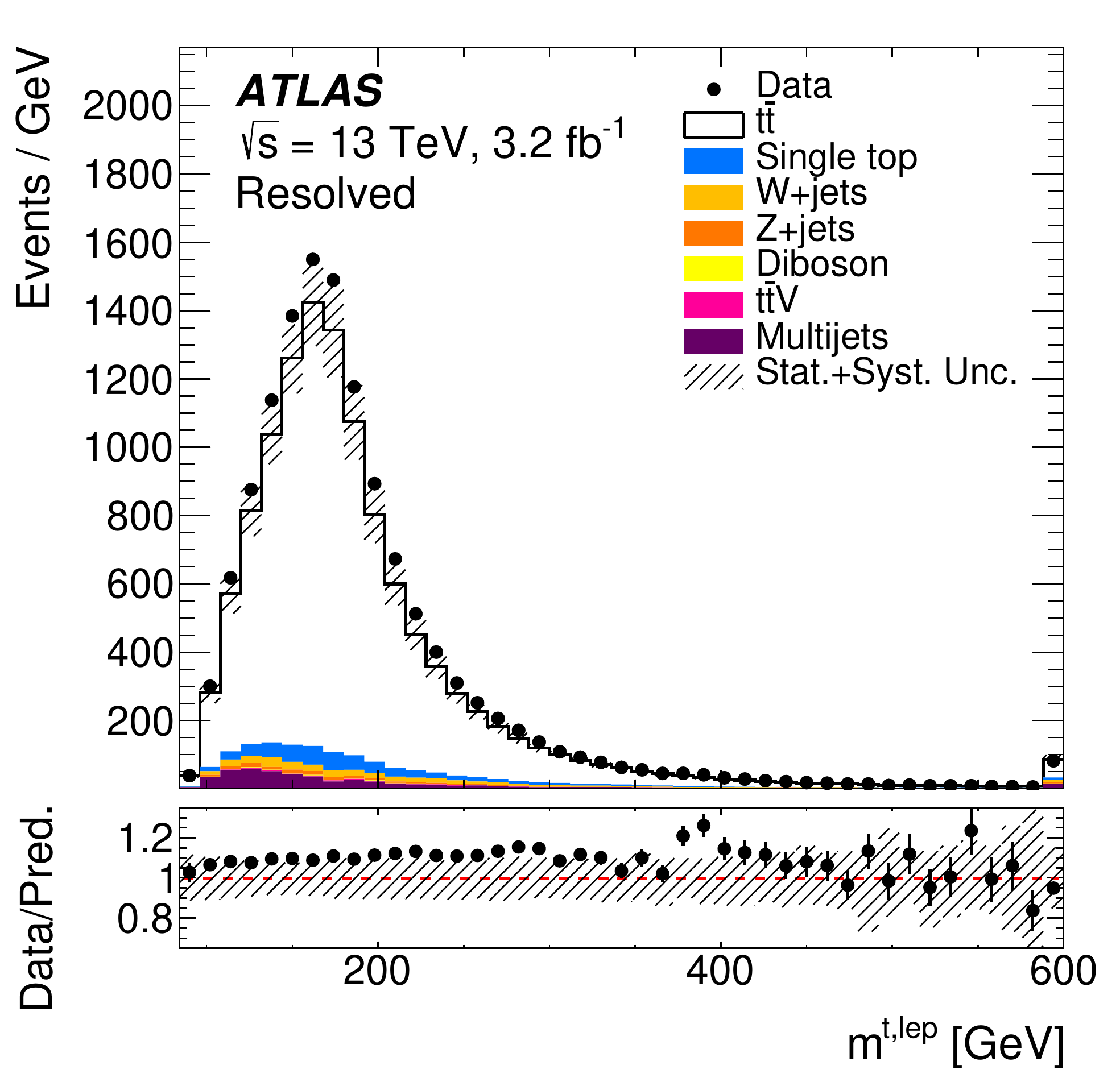} \label{fig:topL_m}}
\subfigure[]{\includegraphics[width=0.45\textwidth]{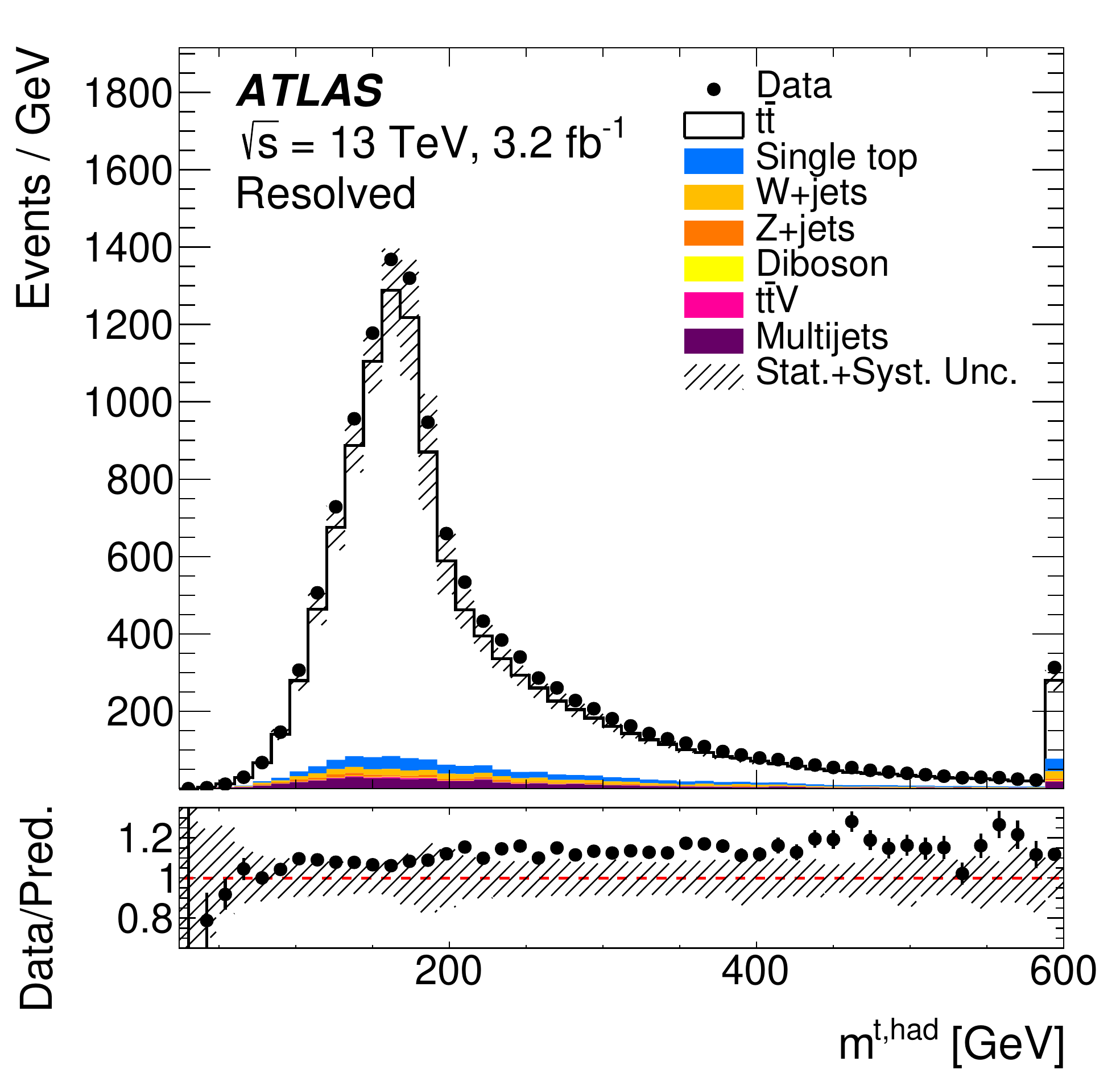} \label{fig:topH_m}}
\subfigure[]{\includegraphics[width=0.45\textwidth]{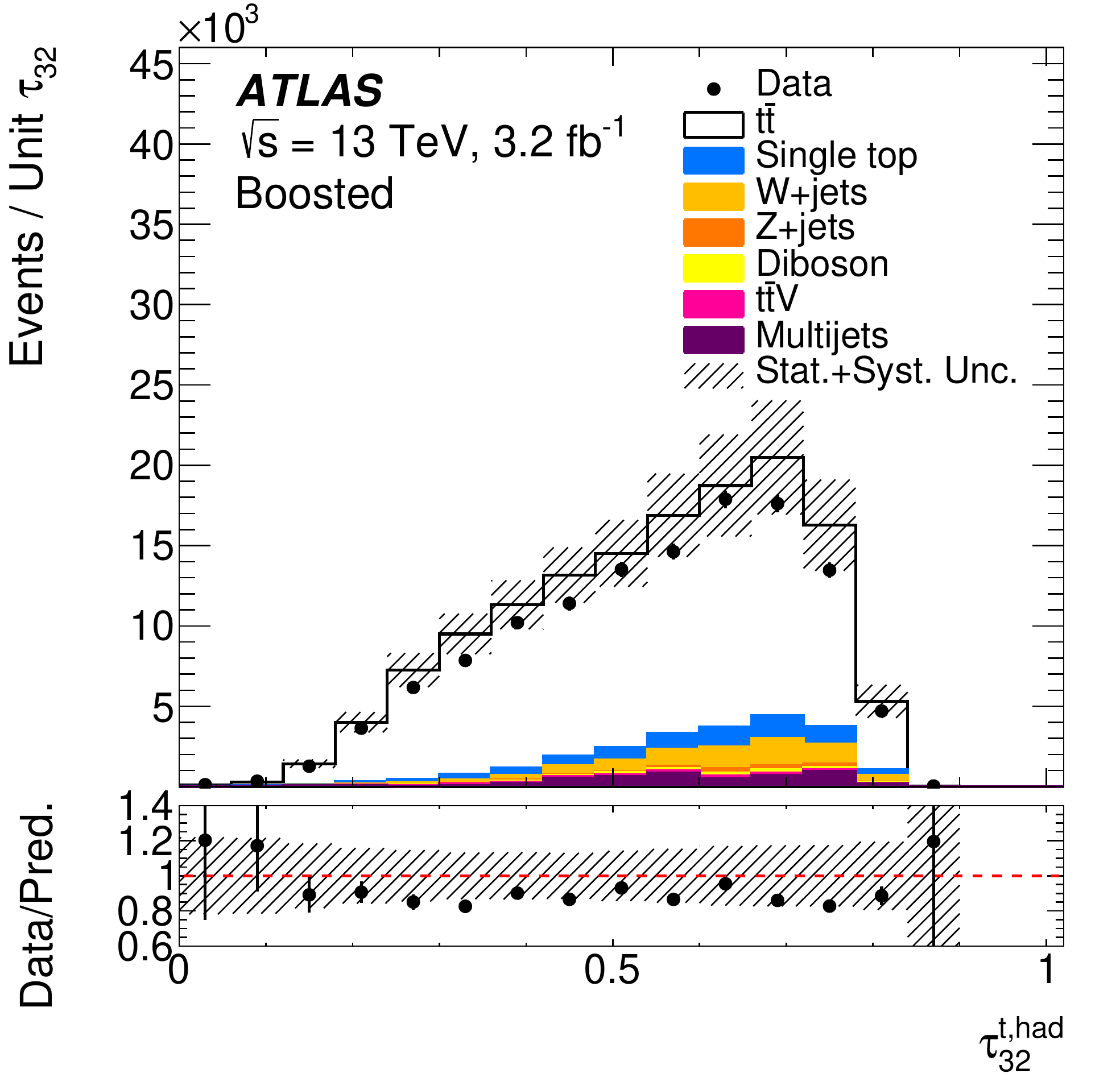} \label{fig:largejet_tau}} 
\subfigure[]{\includegraphics[width=0.45\textwidth]{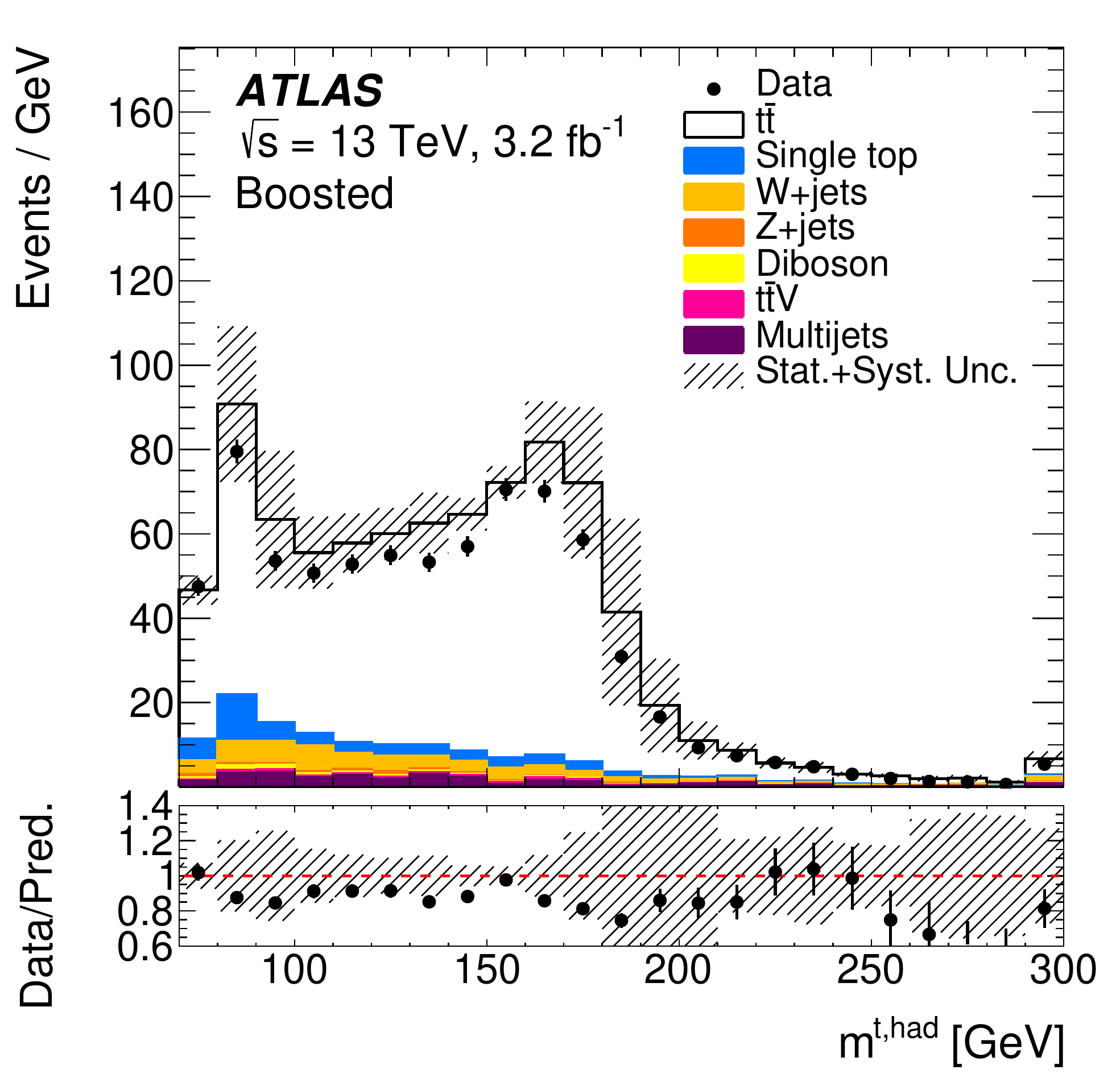} \label{fig:largejet_m}}

\caption{Kinematic distributions in the combined \ljets{} channel at detector level: reconstructed masses of the \subref{fig:topL_m} leptonic and \subref{fig:topH_m} hadronic top quark candidates in the resolved topology; \subref{fig:largejet_tau} hadronic top candidate $\tau_{32}$ and \subref{fig:largejet_m} mass in the boosted topology. Data distributions are compared to predictions using \Powheg{}+\PythiaSix{} as the \ttbar{} signal model. The hatched area indicates the combined statistical and systematic uncertainties in the total prediction, excluding systematic uncertainties related to the modelling of the \ttbar{} system. Events beyond the range of the horizontal axis are included in the last bin.}
\label{Fig:app_sig_CP15_sum}
\end{figure}

\section{Kinematic reconstruction} \label{sec:PseudoTop}
Since the \ttbar{} production differential cross-sections are measured as a function of observables involving the top quark and the \ttbar{} system, an event reconstruction is performed in each topology. In the following, the leptonic (hadronic) top quark refers to the top quark that decays into a~leptonically (hadronically) decaying $W$ boson.

In the boosted topology, the highest-$\pt{}$ large-$R$ jet that satisfies the top-tagging requirements is identified as the hadronic top-quark candidate. As shown in Figure~\ref{Fig:app_sig_CP15_sum}, the reconstructed invariant mass of the hadronic top quark has a~peak at the $W$ boson mass, indicating that not all of the top-quark decay products are always contained within the jet. However, the binning is chosen such that the correspondence of the hadronic-top-quark $\pt{}$ between detector level and particle level (where the large-$R$ jet mass is required to be greater than $100$~\GeV{}) is still very good, with more than 55\% of the events staying on the diagonal of the response matrix as shown in~Figure~\ref{Fig:corrsParticleBoosted2}.

For the resolved topology, the pseudo-top algorithm~\cite{atlasDiff3} reconstructs the four-momenta of the top quarks and their complete decay chain from final-state objects, namely the charged lepton (electron or muon), missing transverse momentum, and four jets, two of which are $b$-tagged. In events with more than two $b$-tagged jets, only the two with the highest transverse momentum are considered as $b$-jets. The same algorithm is used to reconstruct the kinematic properties of top quarks as detector- and  particle-level objects. 

The algorithm starts with the reconstruction of the neutrino four-momentum. While the $x$ and $y$ components of the neutrino momentum are set to the corresponding components of the missing transverse momentum, the $z$ component is calculated by imposing the $W$ boson mass constraint on the invariant mass of the charged-lepton--neutrino system. If the resulting quadratic equation has two real solutions, the one with the smaller  value of $|p_z|$ is chosen. If the discriminant is negative, only the real part is considered. 
The leptonically decaying $W$ boson is reconstructed from the charged lepton and the neutrino. The leptonic top quark is reconstructed from the leptonic $W$ and the \btagged jet closest in $\Delta R$ to the charged lepton. The hadronic $W$ boson is reconstructed from the two non-$b$-tagged jets whose invariant mass is closest to the mass of the $W$ boson. This choice yields the best performance of the algorithm in terms of the
correspondence between the detector and particle levels.
Finally, the hadronic top quark is reconstructed from the hadronic $W$ boson and the other \bjet. 

\section{Measured observables}\label{sec:YieldsAndPlots}

A set of measurements of the \ttbar{} production differential cross-sections are presented as a function of different kinematic observables. These include the transverse momentum of the hadronically decaying top quark (\ptthad) and absolute value of its rapidity (\absythad) for both the resolved and boosted topologies, as well as the absolute value of the rapidity (\absyttbar), invariant mass (\mttbar) and transverse momentum (\ptttbar) of the $\ttbar$ system  in the resolved topology only.
The hadronic top quark is chosen in the resolved topology over the leptonic top quark due to better resolution and correspondence to the particle level. The $\ttbar$ system is not reconstructed in the boosted topology as the leptonic top quark reconstruction would necessitate some optimisation in order to ensure good correspondence between detector level and particle level for the $\ttbar$ system.
These observables, shown in Figures~\ref{Fig:reco_level_variables_topH} and \ref{fig:controls_4j2b_tt} for the top quark and the $\ttbar{}$ system, respectively, 
were measured previously by the ATLAS experiment using the 7 and 8~\TeV{} data sets~\cite{atlasDiff2,atlasDiff3,CON-2014-057,atlasDiff4}, except for \absythad{} in the boosted topology, which is presented here for the first time.
The level of agreement between data and prediction is within the quoted uncertainties for \absythad, \mttbar{}, \ptttbar{} and \absyttbar{}, while for the \ptthad~distribution, a linear mismodelling of the data by the prediction is observed.

\begin{figure*}[p]
\centering
\subfigure[]{ \includegraphics[width=0.45\textwidth]{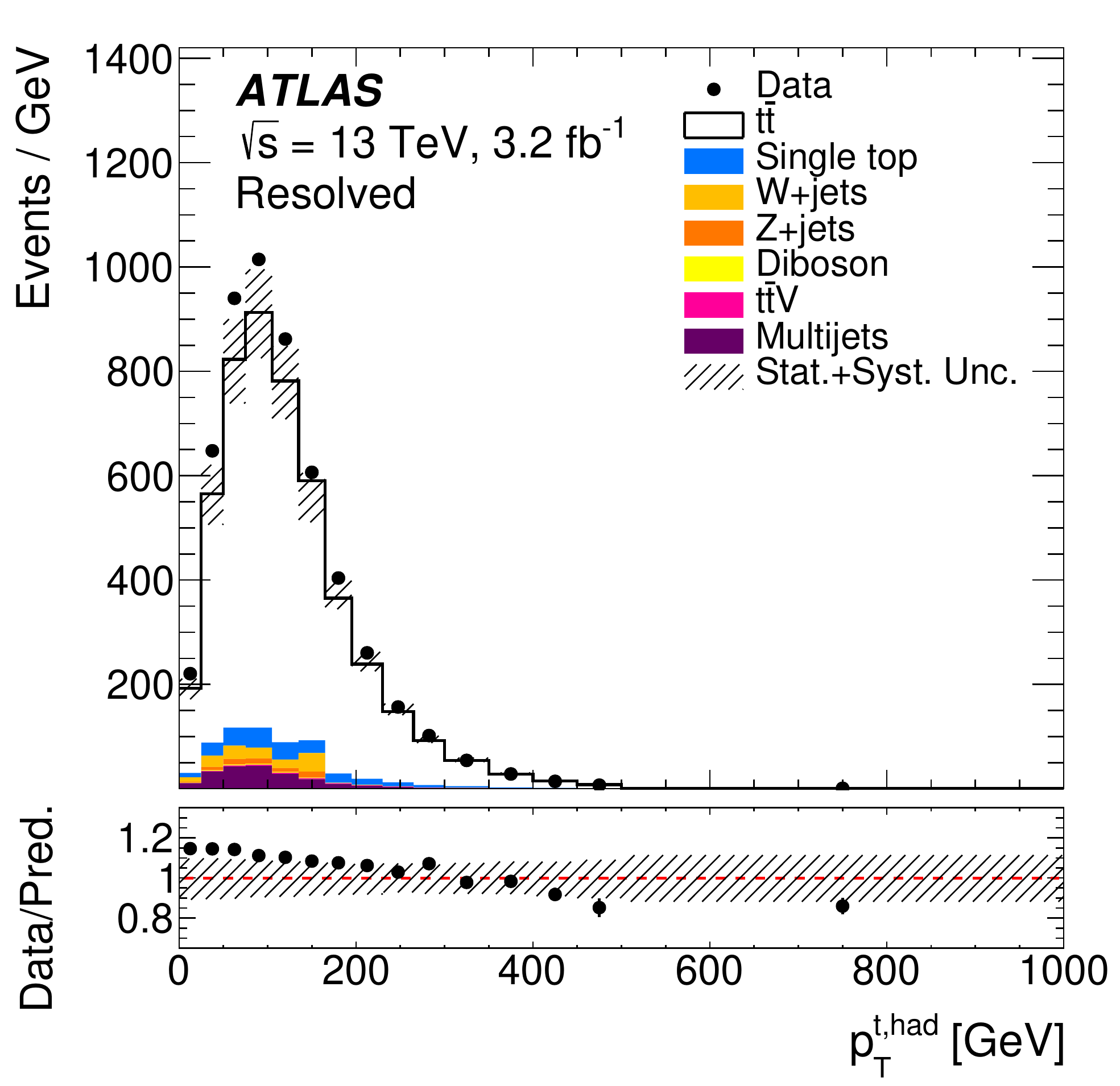}\label{fig:topH_pt_co}}
\subfigure[]{ \includegraphics[width=0.45\textwidth]{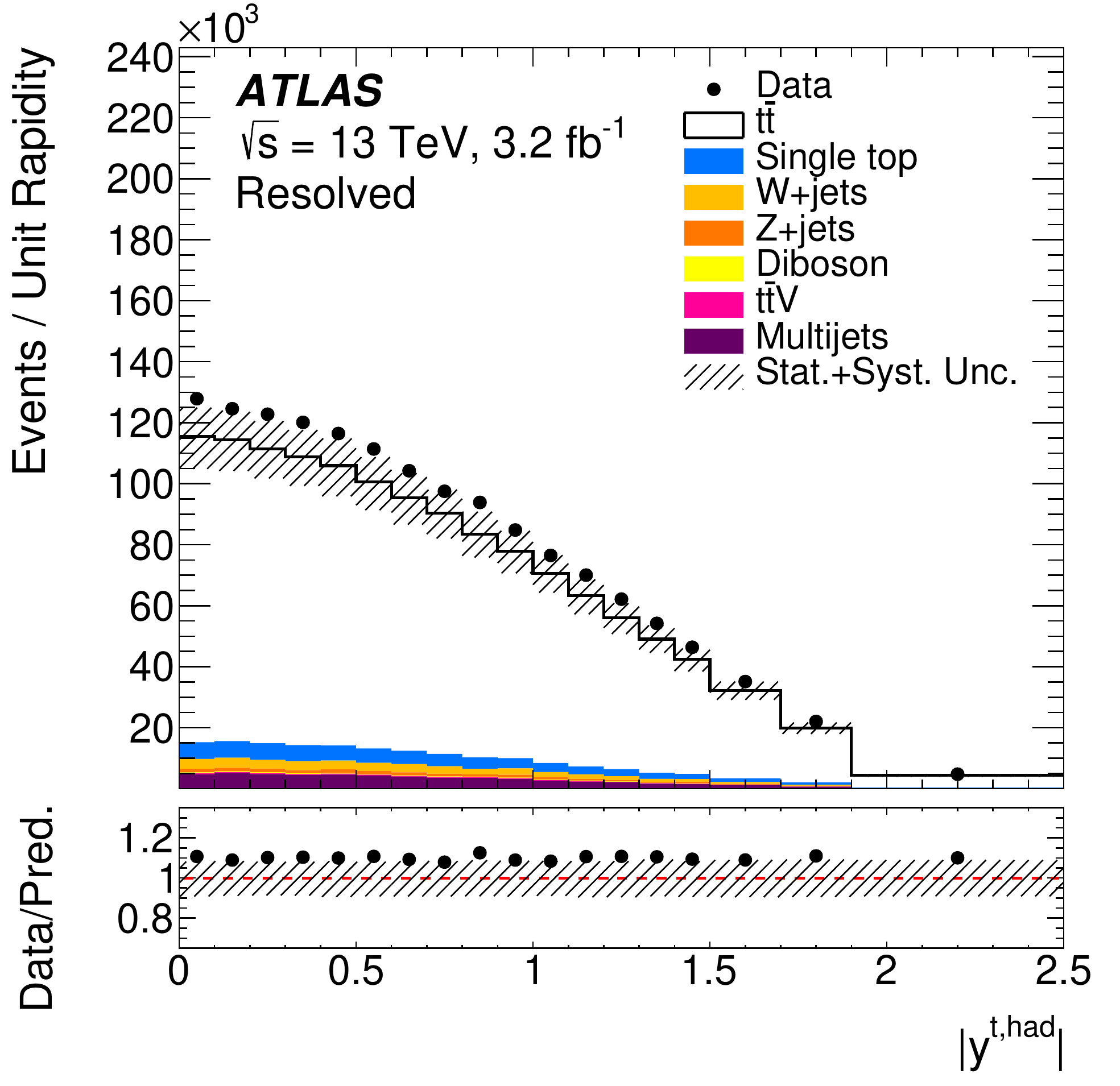}\label{fig:topH_absrap_co}}
\subfigure[]{ \includegraphics[width=0.45\textwidth]{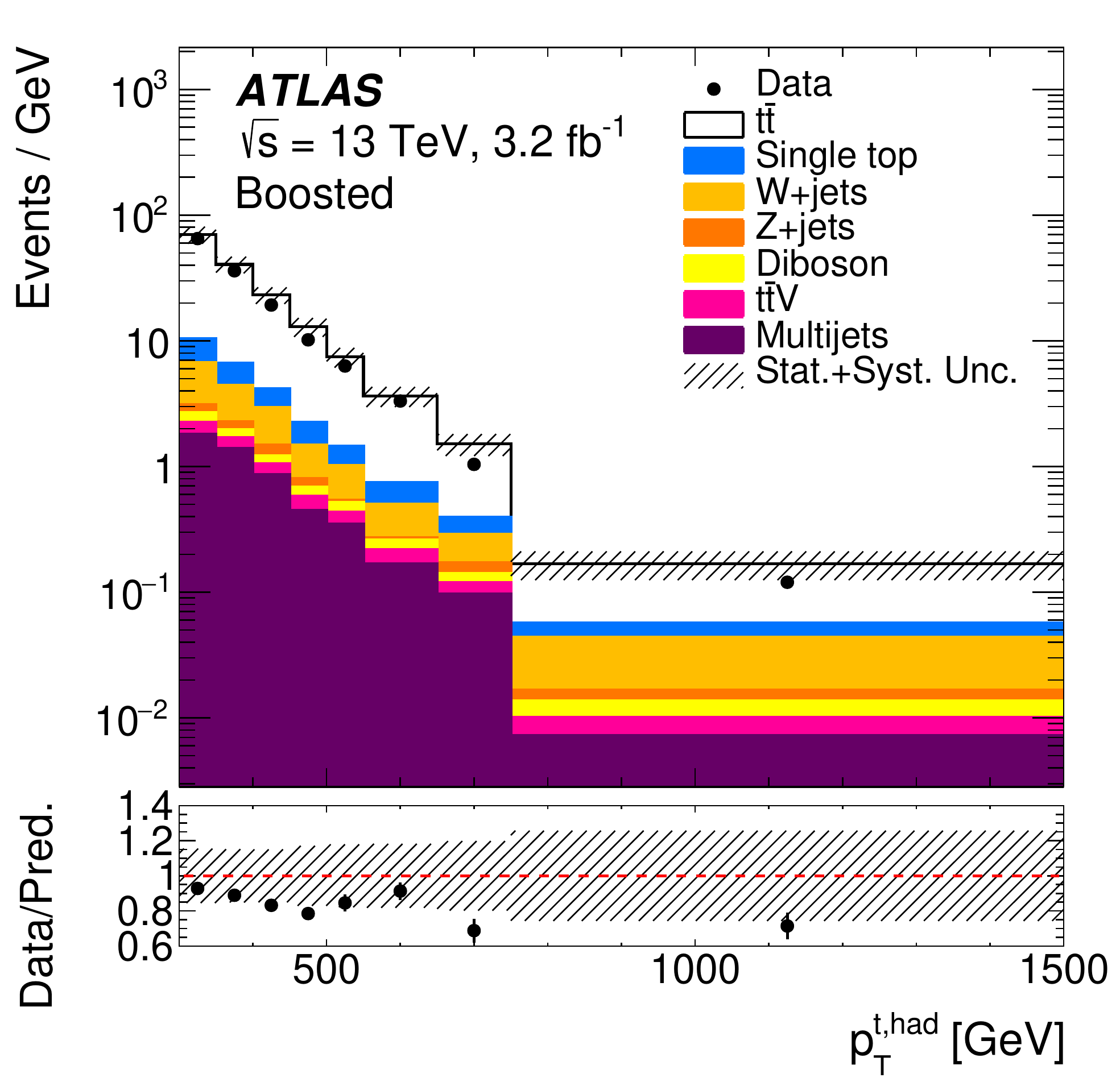}\label{fig:topH_pt_boosted}}
\subfigure[]{ \includegraphics[width=0.45\textwidth]{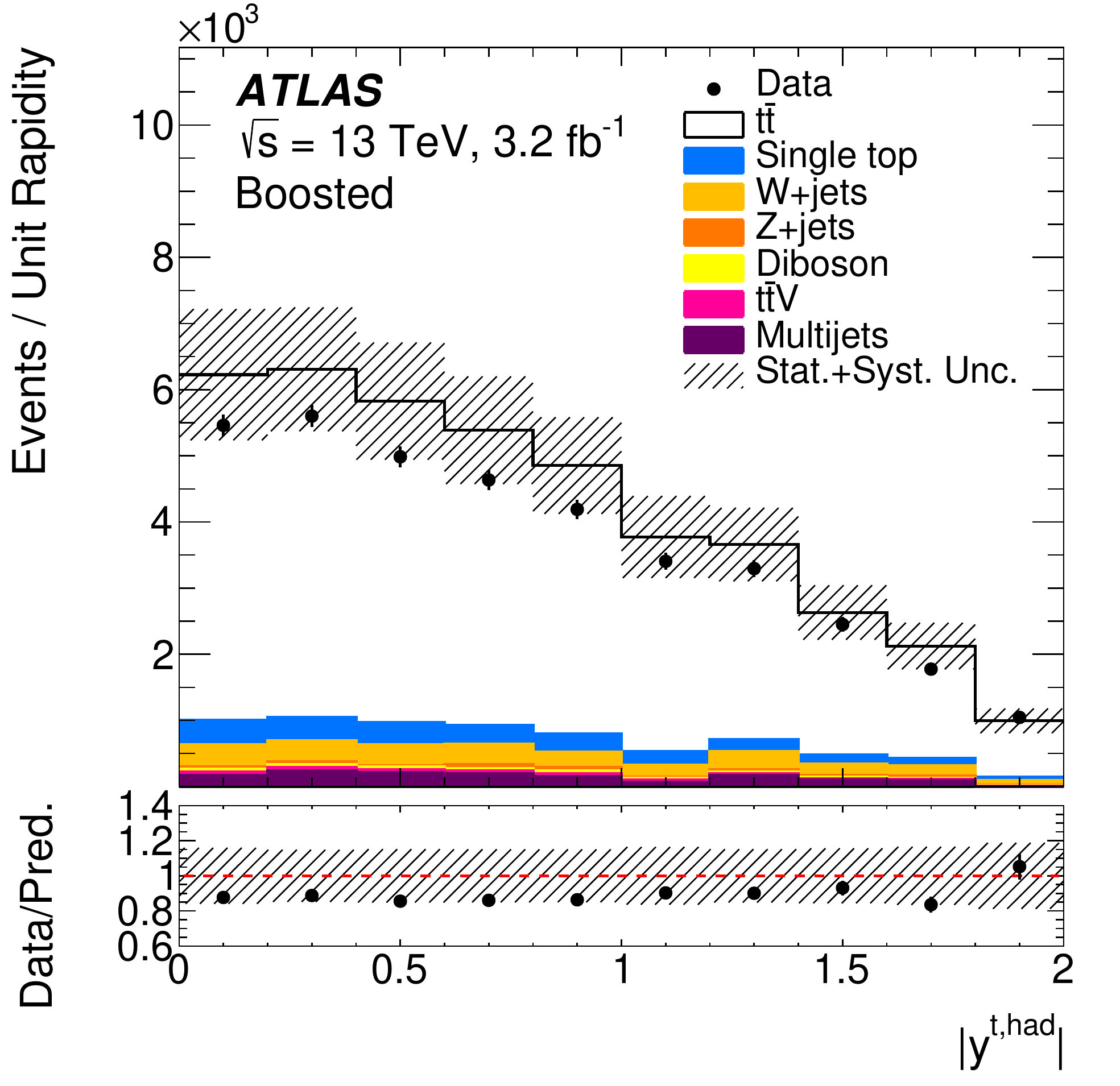}\label{fig:topH_absrap_boosted}}

\caption{Distributions of observables in the combined \ljets{} channel at detector level:
  \subref{fig:topH_pt_co}~hadronic top-quark transverse momentum \ptthad{} and \subref{fig:topH_absrap_co}~absolute value of the rapidity \absythad{} in the resolved topology, and the same variables in the boosted topology \subref{fig:topH_pt_boosted}, \subref{fig:topH_absrap_boosted}.  Data distributions are compared to predictions, using \Powheg{}+\PythiaSix{} as the \ttbar{} signal model. The hatched area indicates the combined statistical and systematic uncertainties (described in Section~\ref{sec:Uncertainties}) in the total prediction, excluding systematic uncertainties related to the modelling of the $\ttbar$ system.
}
\label{Fig:reco_level_variables_topH}
\end{figure*}

\begin{figure*}[p]
\centering
\subfigure[]{ \includegraphics[width=0.45\textwidth]{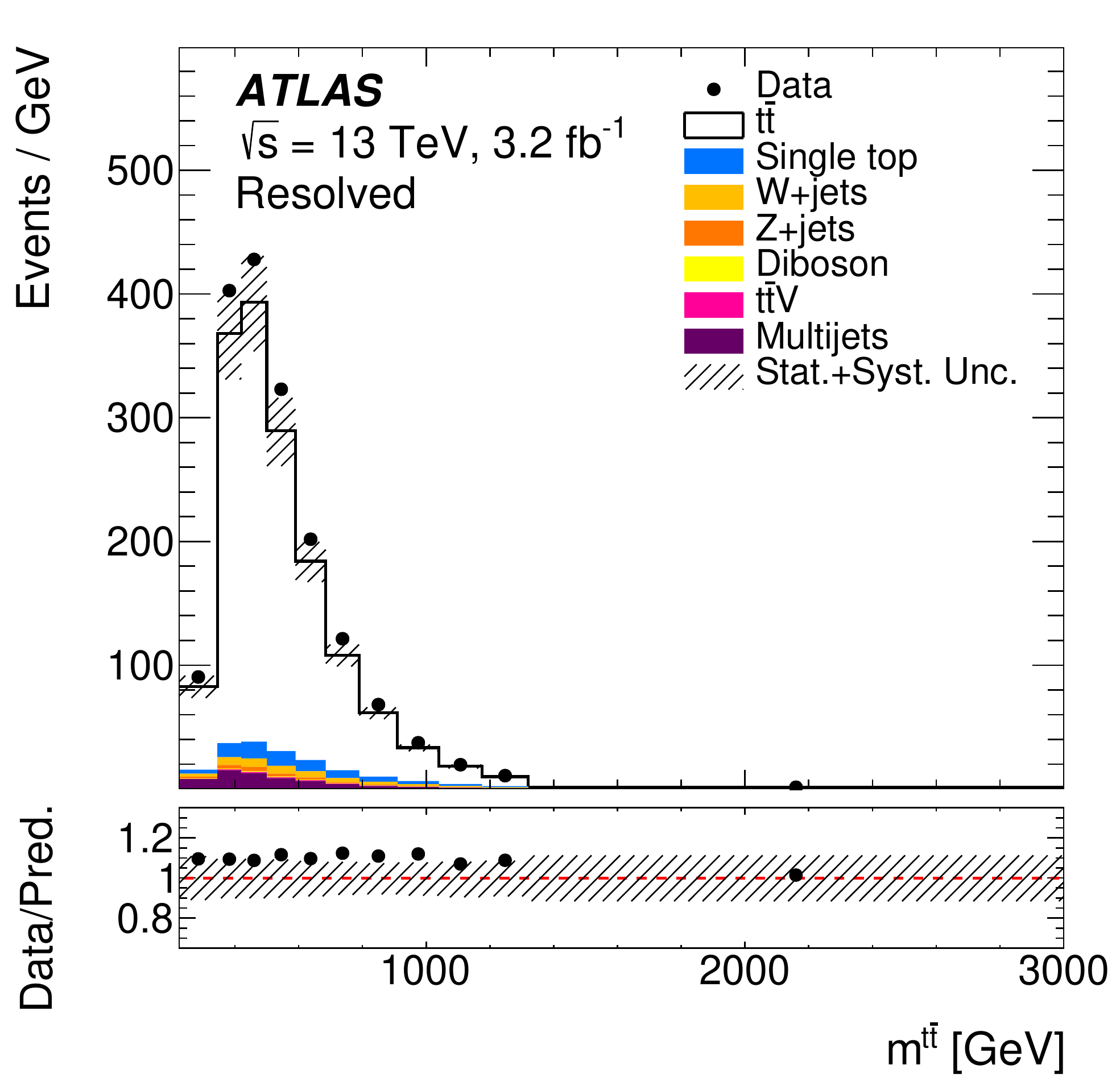}\label{fig:tt_m_co}}
\subfigure[]{ \includegraphics[width=0.45\textwidth]{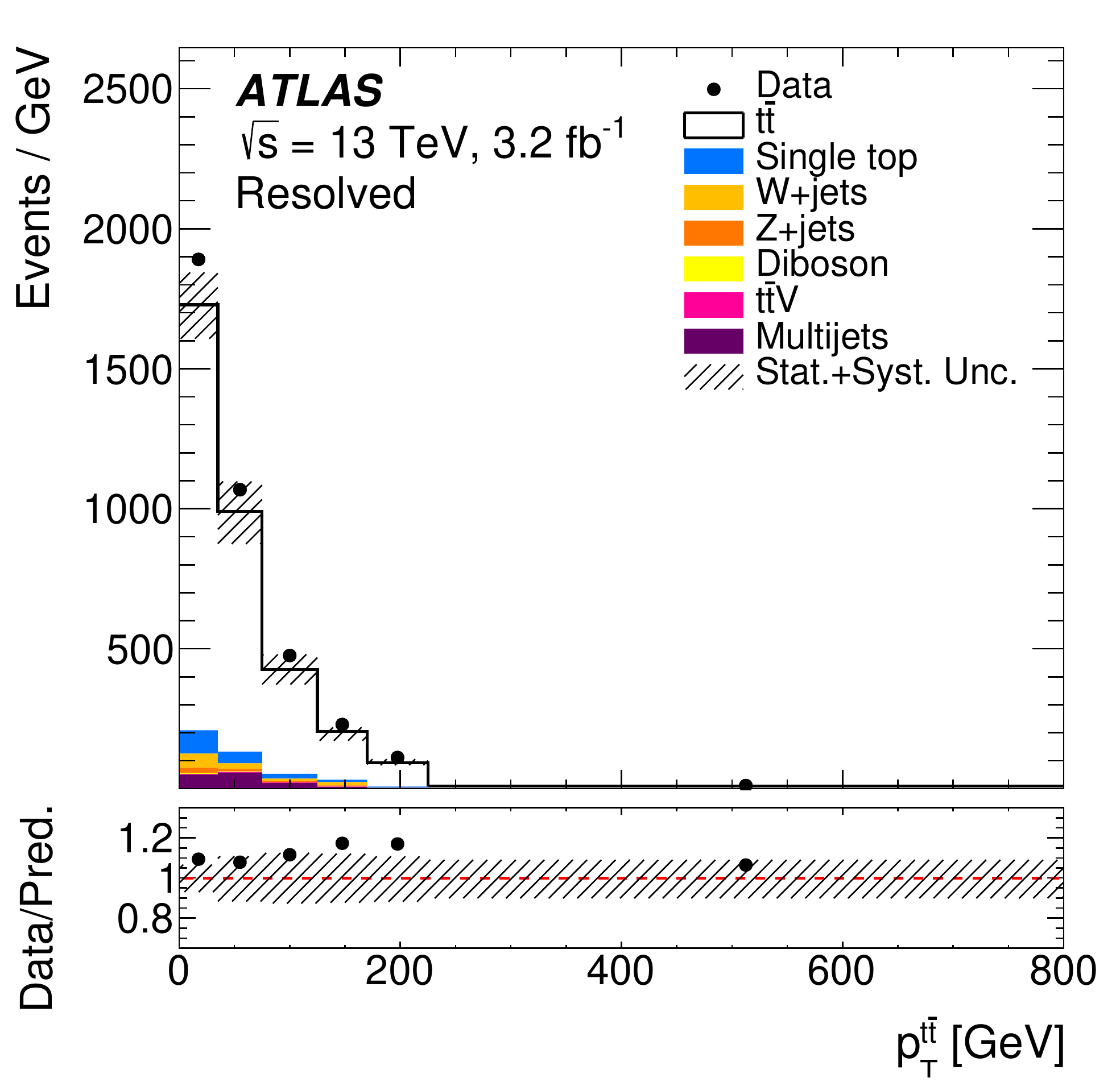}\label{fig:tt_pt_co}}
\subfigure[]{ \includegraphics[width=0.45\textwidth]{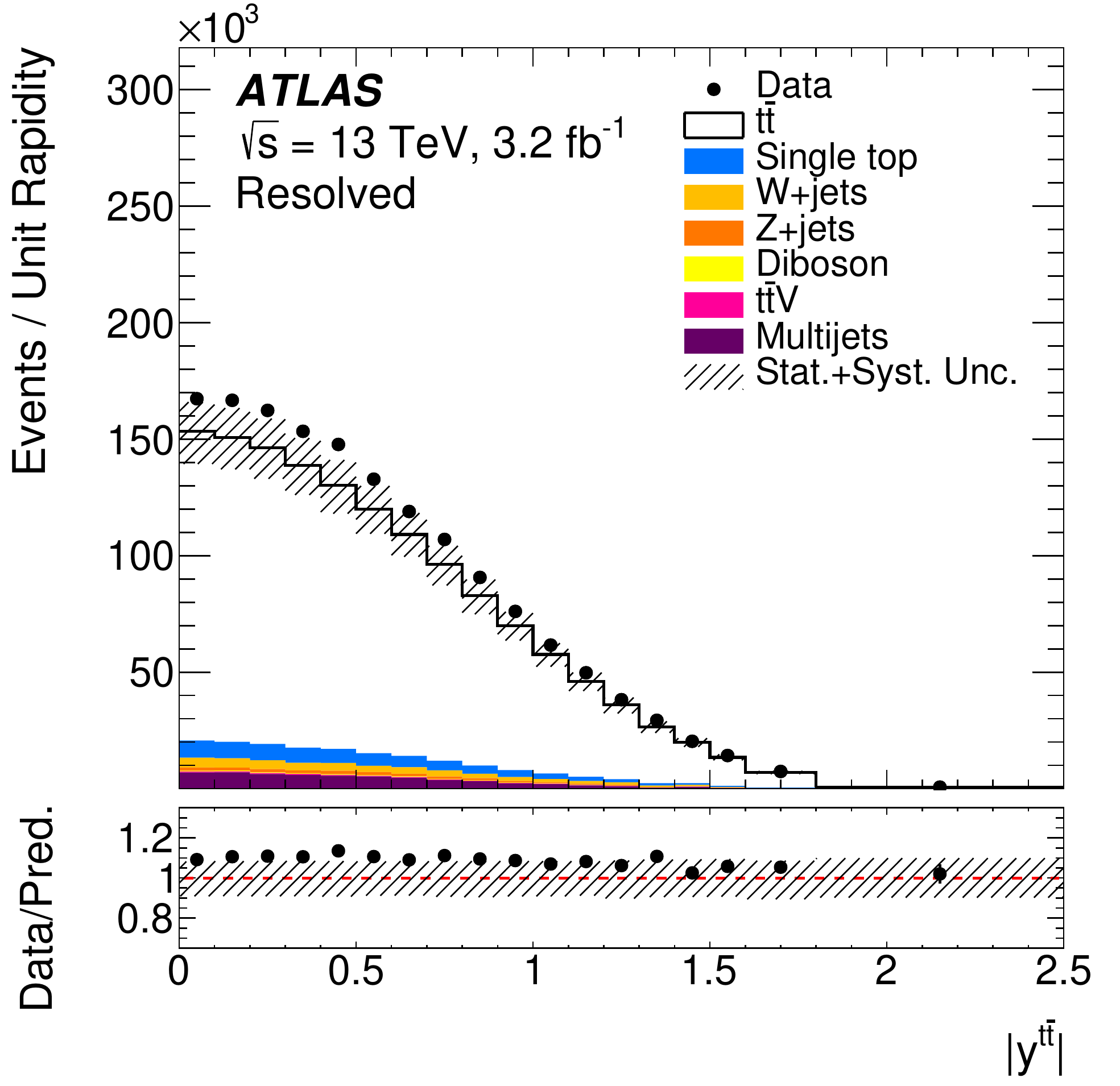}\label{fig:tt_absrap_co}}
\caption{Distributions of observables in the resolved topology in the combined \ljets{} channel at detector level: \subref{fig:tt_m_co}~\ttb{} invariant mass \mttbar{}, \subref{fig:tt_pt_co}~transverse momentum \ptttbar{} and \subref{fig:tt_absrap_co}~absolute value of the rapidity \absyttbar{}. Data distributions are compared to predictions, using \Powheg{}+\PythiaSix{} as the \ttbar{} signal model. The hatched area indicates the combined statistical and systematic uncertainties (described in Section~\ref{sec:Uncertainties}) in the total prediction, excluding systematic uncertainties related to the modelling of the $\ttbar$ system.
}
\label{fig:controls_4j2b_tt}
\end{figure*}

\section{Unfolding procedure}\label{sec:unfolding}

The measured differential cross-sections are obtained from the detector-level distributions using an unfolding technique which corrects for detector effects. The iterative Bayesian method~\cite{unfold:bayes} as implemented in RooUnfold~\cite{Adye:2011gm} is used.

For each observable, the unfolding starts from the detector-level distribution ($N_\textrm{reco}$), after subtracting the backgrounds ($N_\textrm{bg}$).
Next, the acceptance correction $f_\textrm{acc}$ corrects for 
events that are generated outside the fiducial phase-space but pass the detector-level selection.

In the resolved topology, in order to separate resolution and combinatorial effects leading to events migrating from a~particle- to various detector-level bins, distributions are corrected such that detector- and particle-level objects forming the pseudo-top quarks are angularly well matched, leading to a better correspondence between the particle and detector levels.
The matching correction $f_\textrm{match}$, evaluated in the simulation, accounts for the corresponding efficiency.
The matching is performed using geometrical criteria based on the distance  $\Delta R$. Each particle $e$ ($\mu$) is matched to the closest detector-level $e$ ($\mu$) within $\Delta R < 0.02$.
Particle-level jets forming the pseudo-top quark candidates at the particle level are then required to be geometrically matched to the corresponding jets (respecting their
assignment to the pseudo-top candidates) at the detector level within $\Delta R < 0.35$, allowing for a swap of light jets forming the hadronically decaying $W$-boson candidate.

The unfolding step uses a~migration matrix ($\mathcal{M}$) derived from simulated \ttbar{} events which maps the binned generated particle-level events to the binned detector-level events.
The probability for particle-level events to remain in the same bin is therefore represented by the elements on the diagonal, and the off-diagonal elements describe the fraction of particle-level events that migrate into other bins. Therefore, the elements of each row add up to unity (within rounding) as shown in Figures~\ref{fig:particle:migra:topH_pt} and~\ref{Fig:corrsParticleBoosted2}.
The binning is optimised to minimise off-diagonal elements in the migration matrix, have a sufficient number of data events in each bin and have stability in systematic uncertainties propagation, taking into account detector resolution and reconstruction effects.
The unfolding is performed using four iterations to balance the unfolding stability with respect to the previous iteration (below 0.1\%) and the growth of the statistical uncertainty. The effect of varying the number of iterations by one is negligible. Finally, the efficiency $\epsilon$ corrects for events which pass the particle-level selection but are not reconstructed at detector level.

All corrections are evaluated with simulation and are presented in~Figure~\ref{fig:corrs:fiducial:topH} for the case of the $\pt$ of the top quark decaying hadronically in the resolved topology. Similar corrections in the boosted topology for the hadronic top quark $\pt$ and $\absythad$ are shown in~Figures~\ref{Fig:corrsParticleBoosted1} and \ref{Fig:corrsParticleBoosted2}. 

The top-quark transverse momentum is chosen as an example to show how the corrections vary in size since the kinematic properties of the decay products of the top quark change substantially in the observed range of this observable.
The efficiency decreases in the resolved topology at high values primarily due to the increasingly large fraction of non-isolated leptons and close or merged jets in events with high top-quark $\pt$. Consequently, the boosted topology is included in this paper where jets with large radius are used, resulting in an improved efficiency at high $\pt$, as shown in~Figure~\ref{Fig:corrsParticleBoosted1:topH_pt_eff}. The progressive decrease in efficiency seen in Figure~\ref{Fig:corrsParticleBoosted1:topH_pt_eff} is caused by the lepton isolation requirements becoming too stringent as the top-quark momentum increases, as well as a~decrease in efficiency of the $b$-tagging requirements at very high jet momentum \cite{btagRun2}. The acceptance in the boosted topology decreases at low $\pt$ due to a~simpler definition of top-tagging at particle level than at detector level, where $\pt$-dependent mass and $\tau_{32}$ requirements are used.

The unfolding procedure for an observable $X$ at particle level is summarised by the expression for the absolute differential cross-section
\begin{equation*}
\frac{\textrm{d}\sigma^\textrm{fid}}{\textrm{d}X^i} \equiv \frac{1}{\mathcal{L} \cdot \Delta X^i} \cdot  \frac{1}{\epsilon^i} \cdot \sum_j \mathcal{M}_{ij}^{-1} \cdot f_\textrm{match}^j \cdot  f_\textrm{acc}^j \cdot \left(N_\textrm{reco}^j - N_\textrm{bg}^j\right)\hbox{,}
\end{equation*}
where the index $j$ iterates over bins of $X$ at detector level while the $i$ index labels bins at particle level; $\Delta X^i$ is the bin width while $\mathcal{L}$ is the integrated luminosity and the Bayesian unfolding is symbolised by $\mathcal{M}_{ij}^{-1}$. No matching correction is applied in the boosted case ($f_\textrm{match}$ =1).
The integrated fiducial cross-section is obtained by integrating the unfolded differential cross-section over the kinematic bins, and its value is used to compute the relative differential cross-section $1/\sigma^\textrm{fid}\cdot\textrm{d}\sigma^\textrm{fid} / \textrm{d}X^i$.

\begin{figure*}[htbp]
\centering
\subfigure[]{  \includegraphics[width=0.417\textwidth]{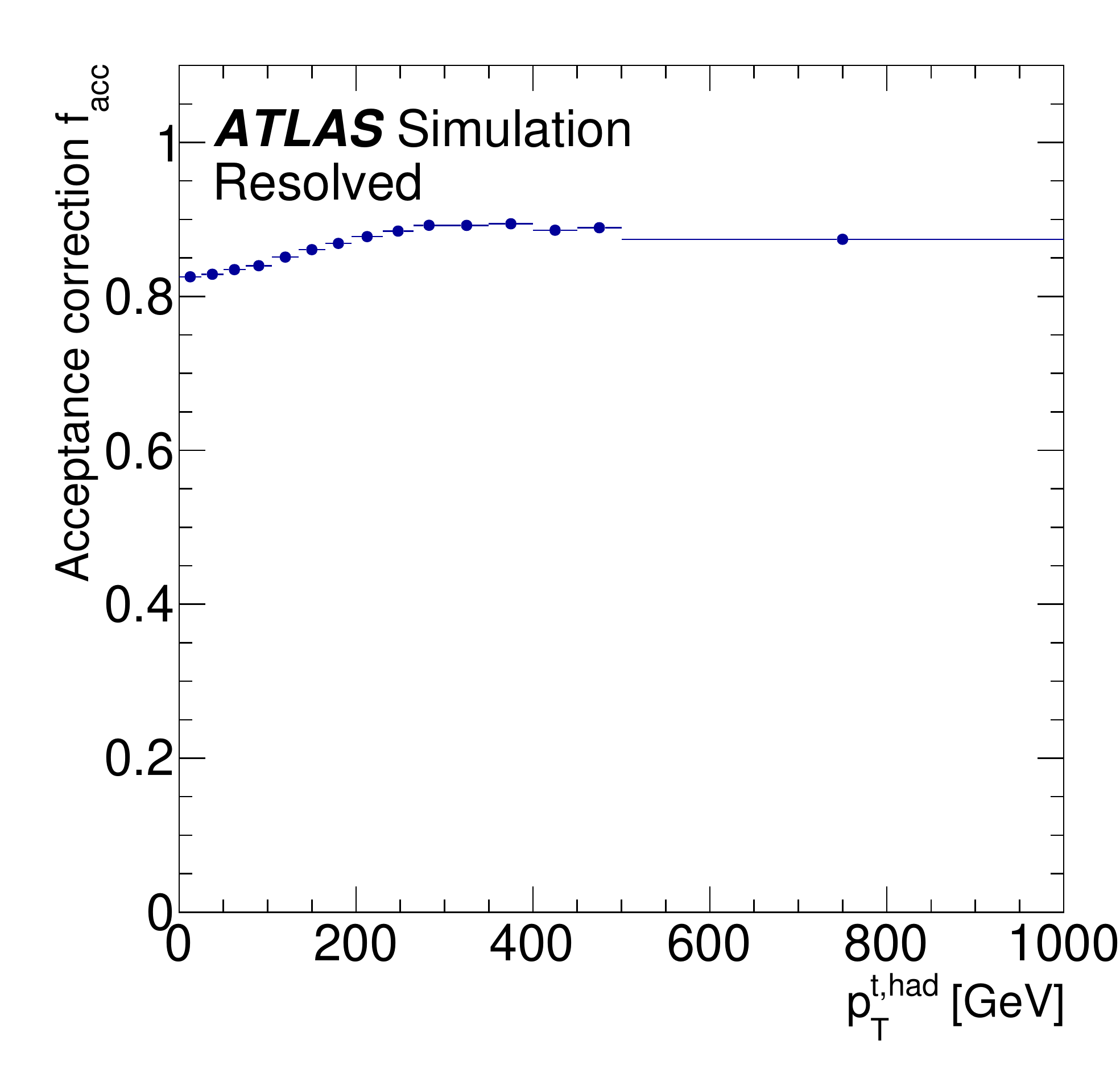}\label{fig:particle:acc:topH_pt} }
\subfigure[]{  \includegraphics[width=0.417\textwidth]{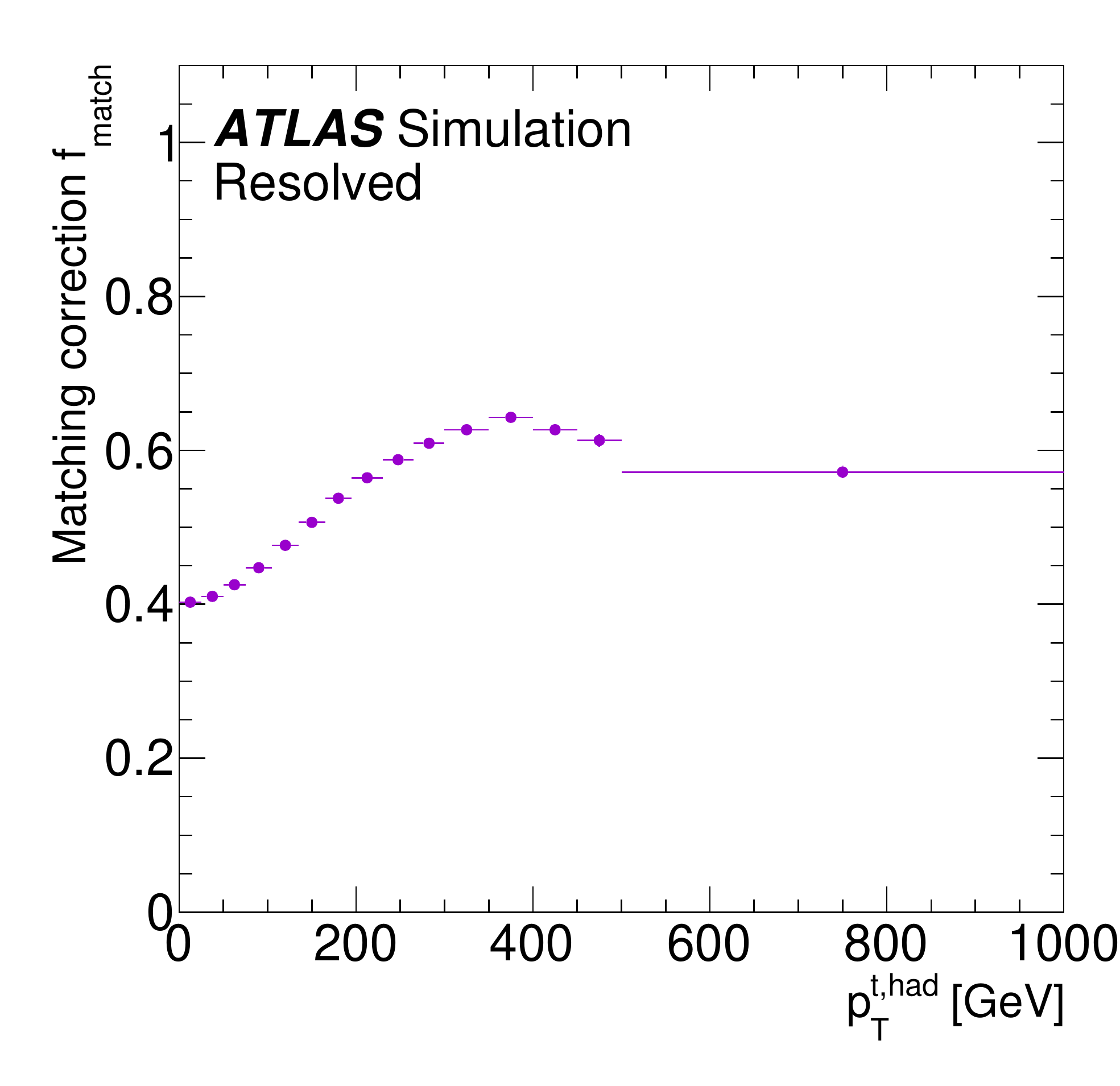}\label{fig:particle:match:topH_pt} }
\subfigure[]{  \includegraphics[width=0.417\textwidth]{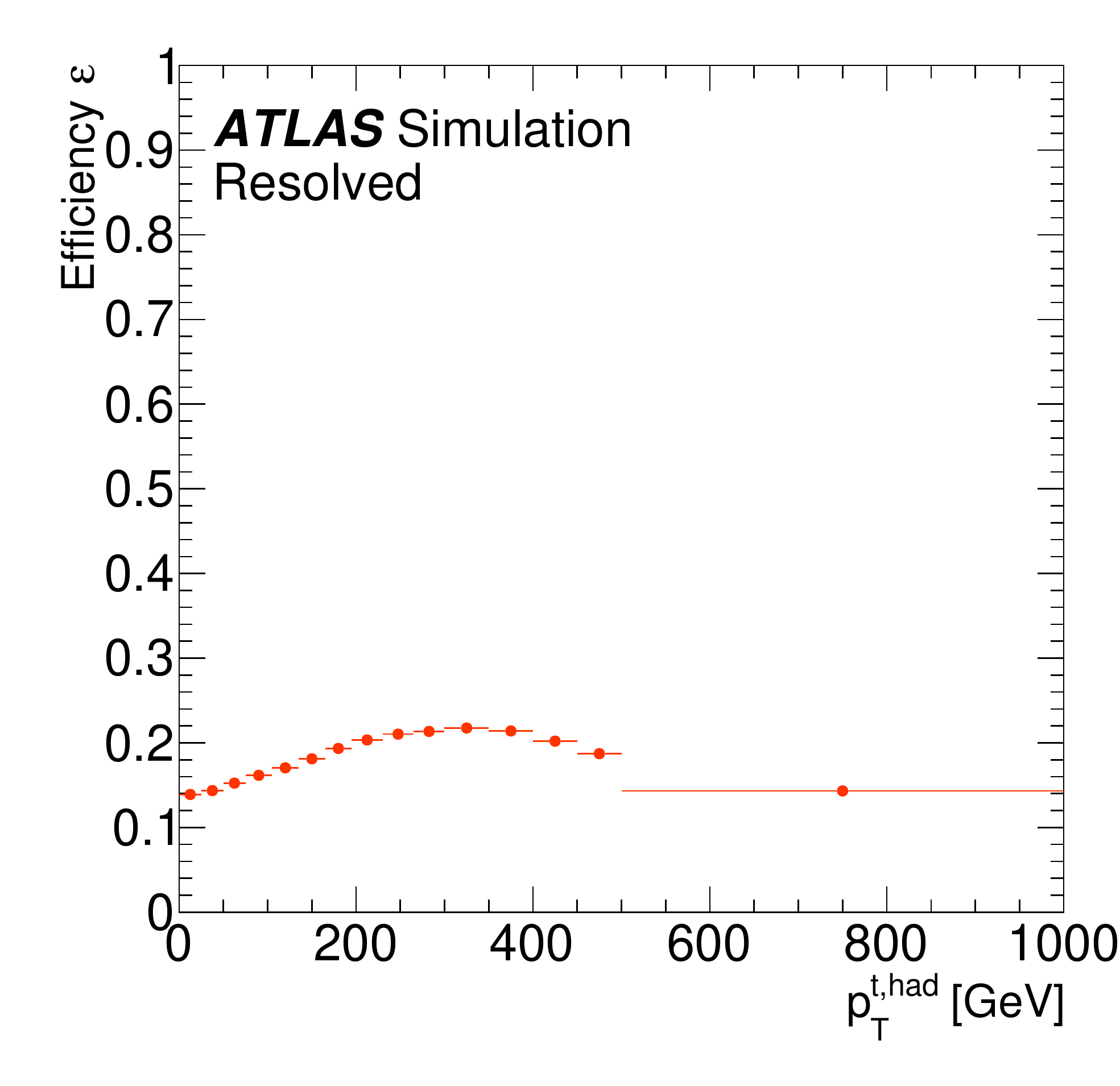}\label{fig:particle:eff:topH_pt} }
\subfigure[]{  \includegraphics[width=0.45\textwidth]{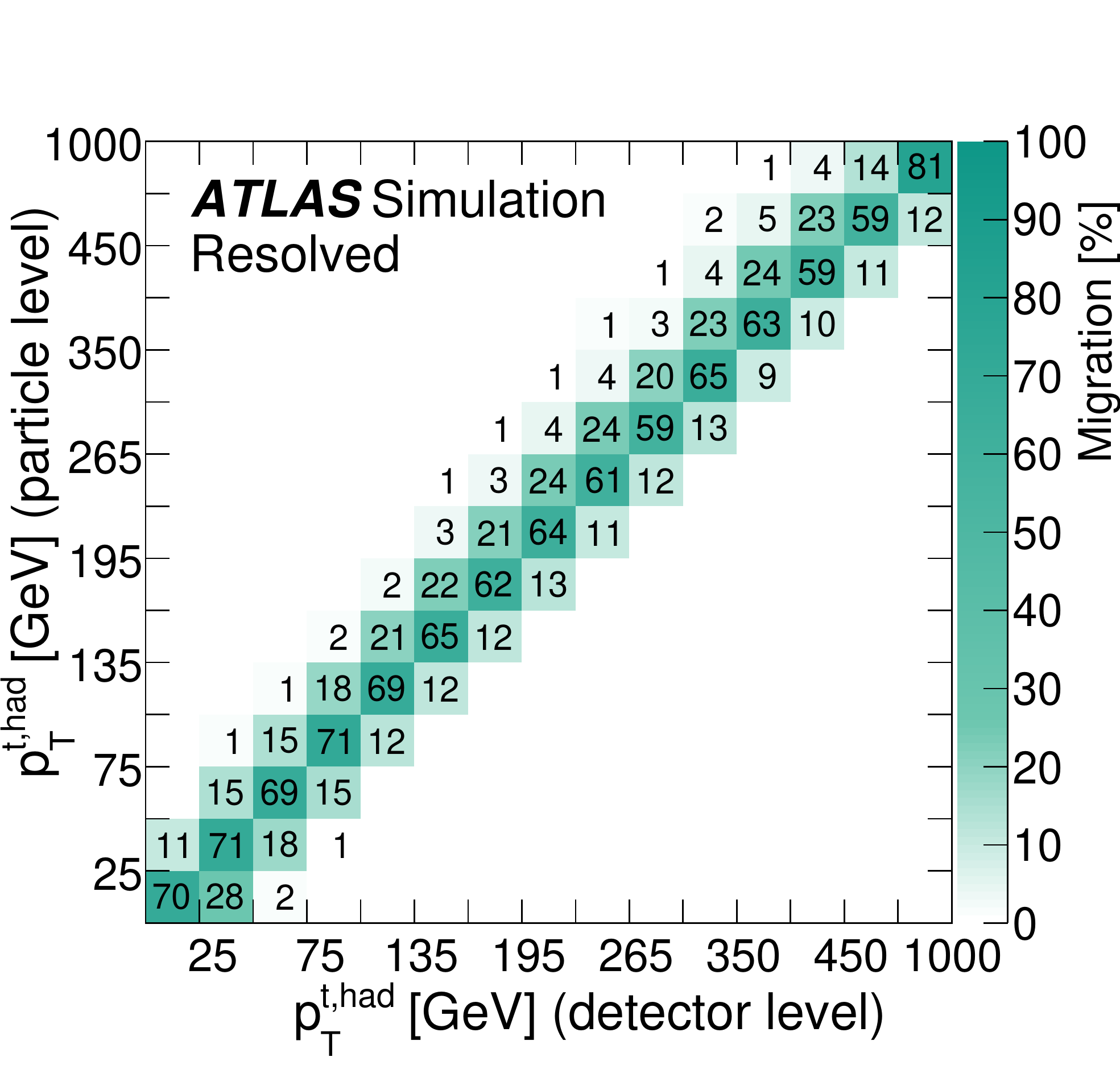}\label{fig:particle:migra:topH_pt} }
\caption{The \subref{fig:particle:acc:topH_pt} acceptance and \subref{fig:particle:match:topH_pt} matching corrections, \subref{fig:particle:eff:topH_pt} efficiency, and the \subref{fig:particle:migra:topH_pt} particle-to-detector-level migration matrix for the hadronic top-quark transverse momentum in the resolved topology evaluated with the \Powheg{}+\PythiaSix{} simulation sample with \HDampMT{} and using CT10 PDF.  In Figure \subref{fig:particle:migra:topH_pt}, the empty bins either contain no events or the fraction of events is less than 0.5\%. Following Section~\ref{sec:unfolding}, the acceptance and matching corrections are binned according to detector-level quantities, while the efficiency is binned according to particle-level quantities.}
\label{fig:corrs:fiducial:topH}
\end{figure*}

\begin{figure*}
\centering
\subfigure[]{\includegraphics[width=0.417\textwidth]{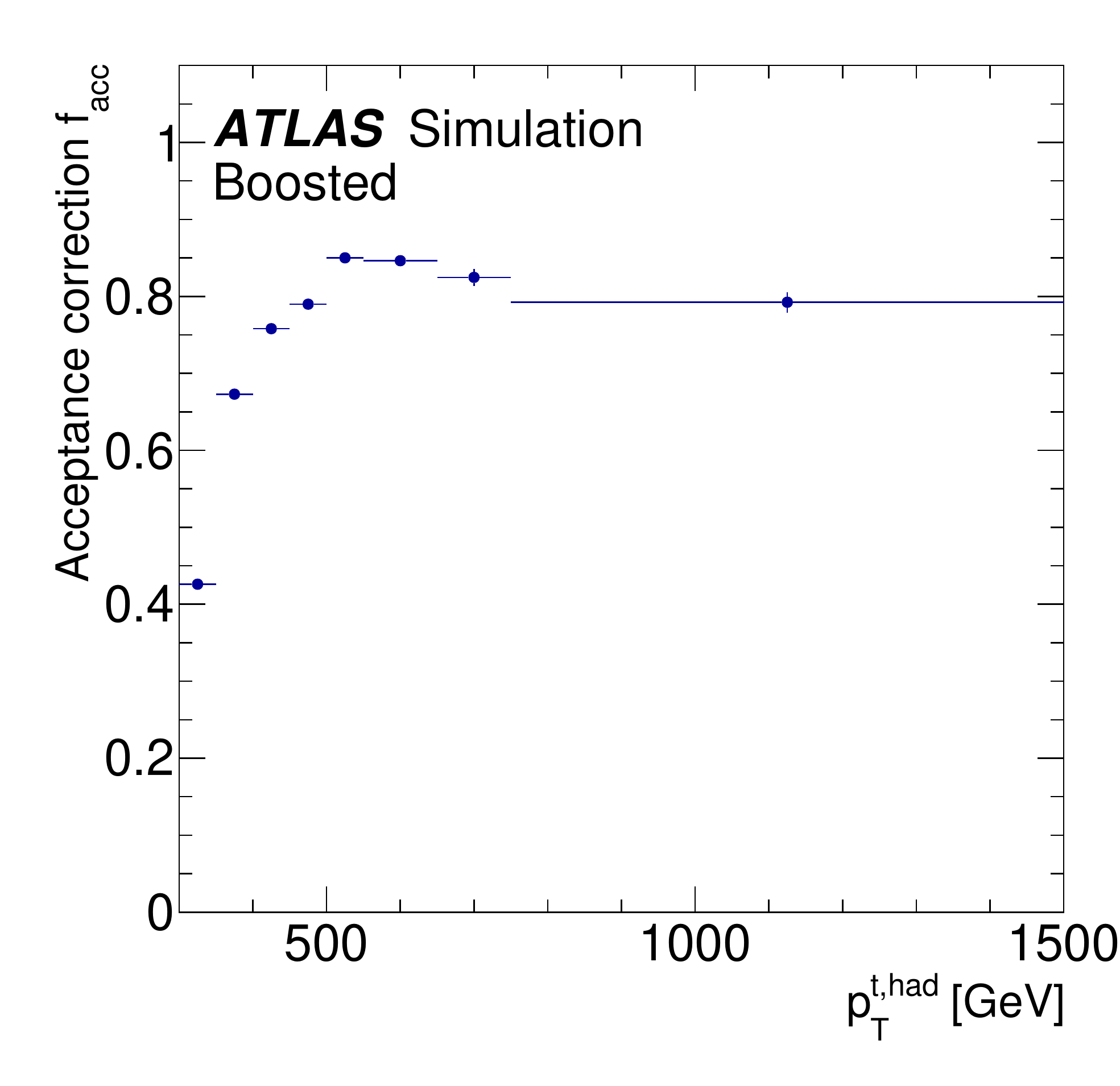}\label{Fig:corrsParticleBoosted1:topH_pt_acc} }
\subfigure[]{\includegraphics[width=0.417\textwidth]{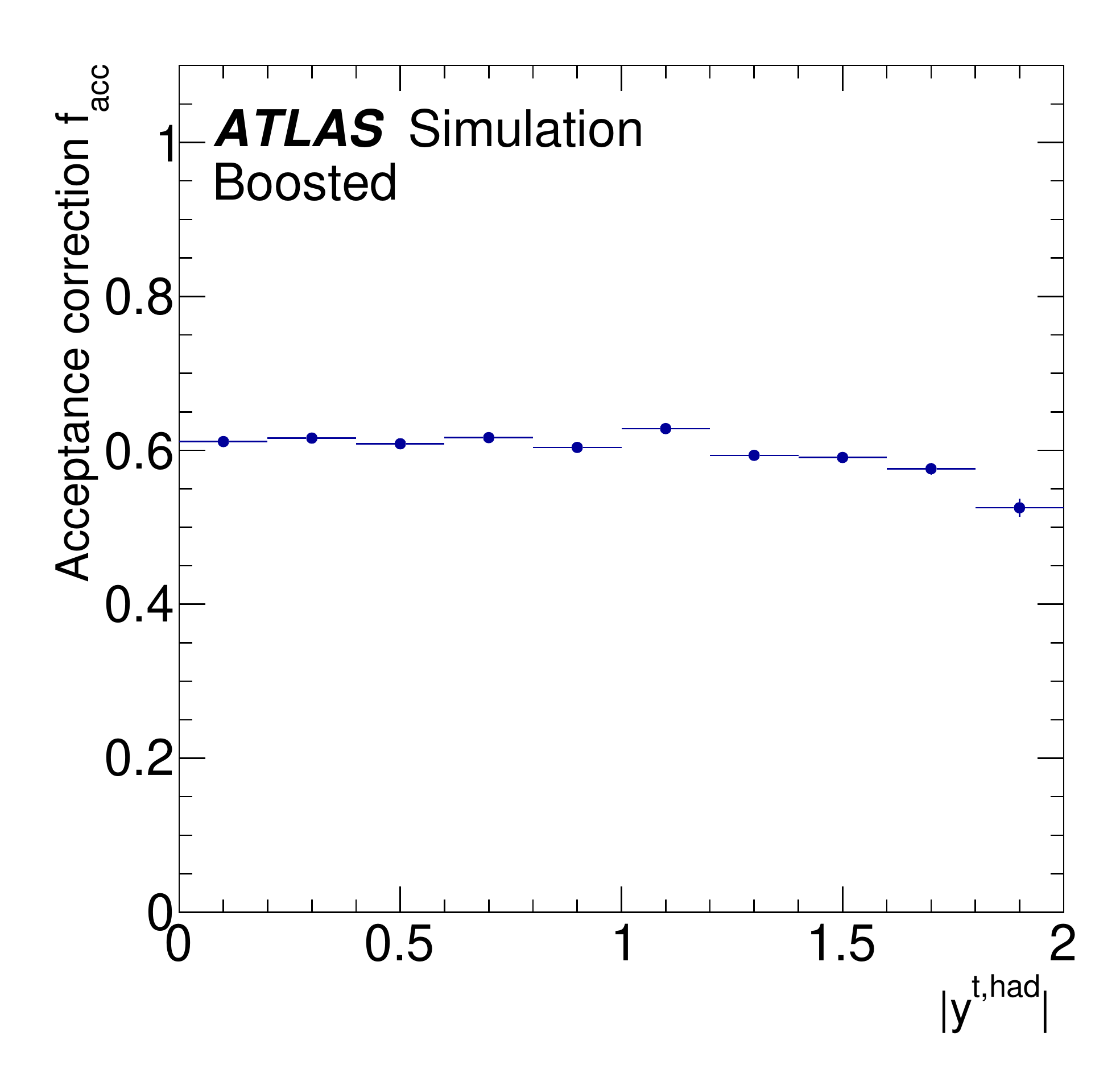}\label{Fig:corrsParticleBoosted1:topH_absrap_acc} }
\subfigure[]{\includegraphics[width=0.417\textwidth]{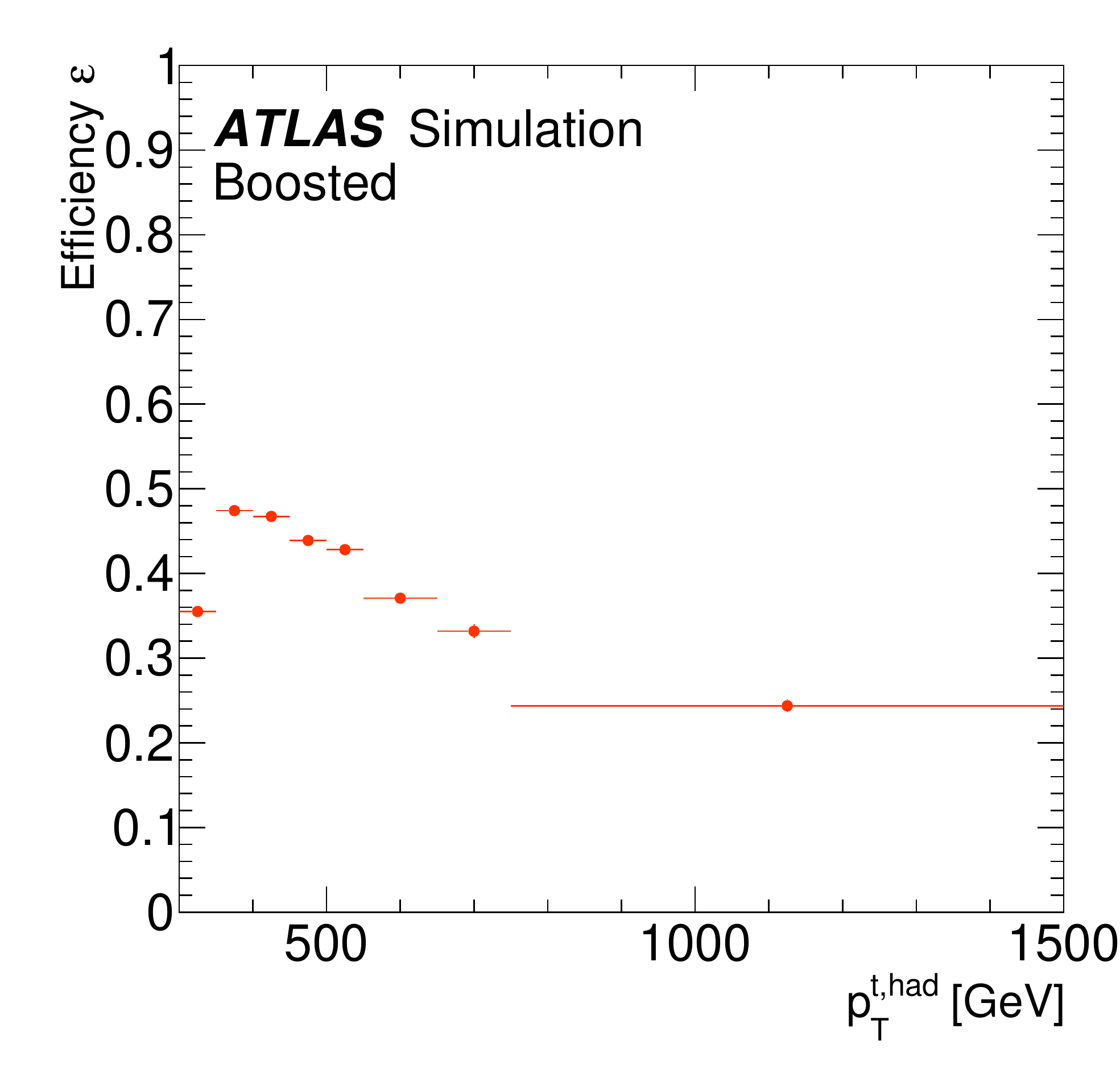}\label{Fig:corrsParticleBoosted1:topH_pt_eff} }
\subfigure[]{\includegraphics[width=0.417\textwidth]{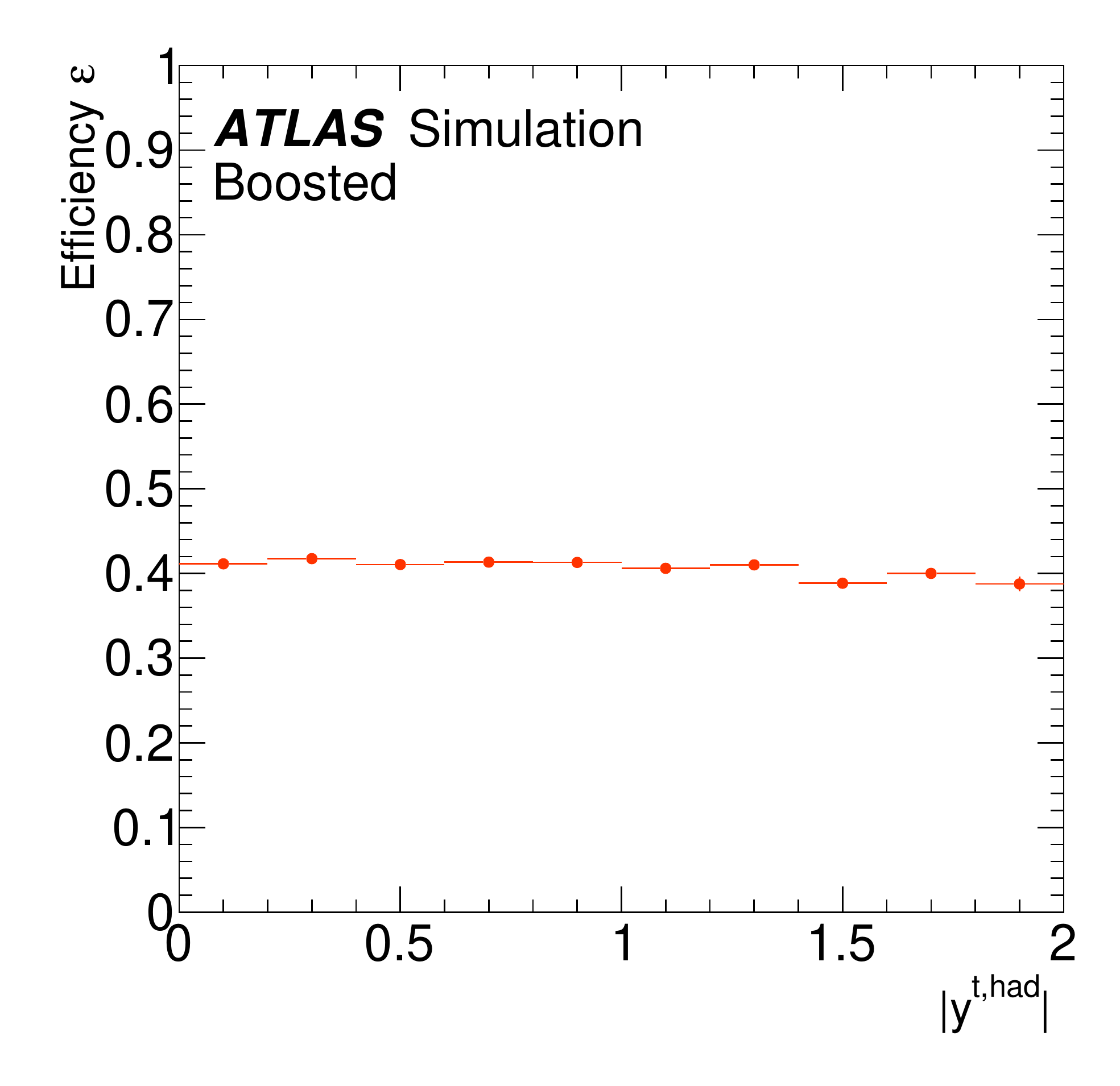}\label{Fig:corrsParticleBoosted1:topH_absrap_eff} }
\caption{The acceptance correction for \subref{Fig:corrsParticleBoosted1:topH_pt_acc} the hadronic top-quark transverse momentum $\ptthad$ and \subref{Fig:corrsParticleBoosted1:topH_absrap_acc} the absolute value of the rapidity $\absythad$, and the efficiency correction for \subref{Fig:corrsParticleBoosted1:topH_pt_eff} the hadronic top-quark transverse momentum $\ptthad$ and \subref{Fig:corrsParticleBoosted1:topH_absrap_eff} the absolute value of the rapidity $\absythad$ in the boosted topology, evaluated with the \Powheg{}+\PythiaSix{} simulation sample with \HDampMT{} and using CT10 PDF. Following Section~\ref{sec:unfolding}, the acceptance and matching corrections are binned according to detector-level quantities, while the efficiency is binned according to particle-level quantities.}
\label{Fig:corrsParticleBoosted1}
\end{figure*}

\begin{figure*}
\centering
\subfigure[]{\includegraphics[width=0.45\textwidth]{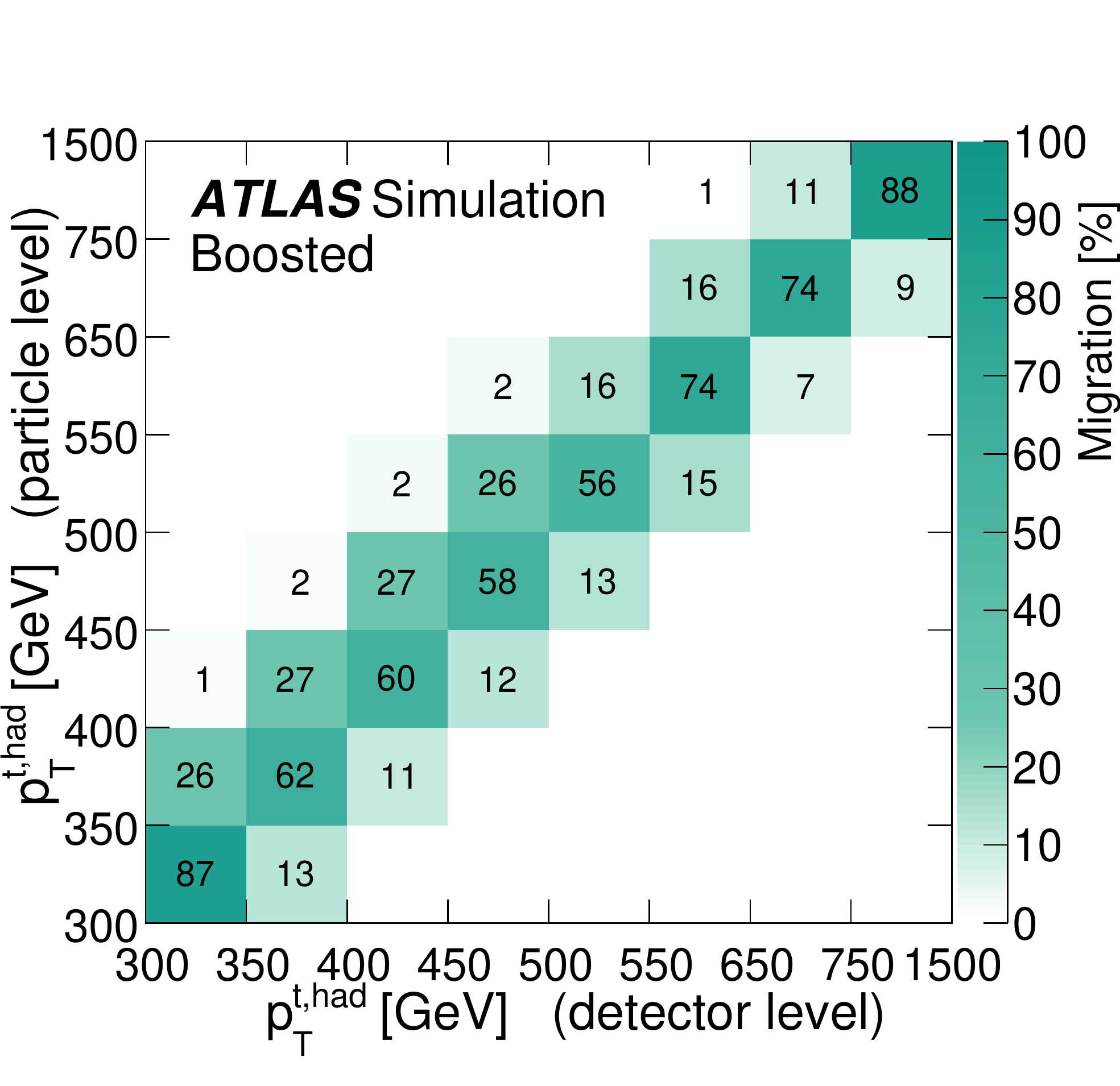}\label{Fig:corrsParticleBoosted2:topH_pt_acc} }
\subfigure[]{\includegraphics[width=0.45\textwidth]{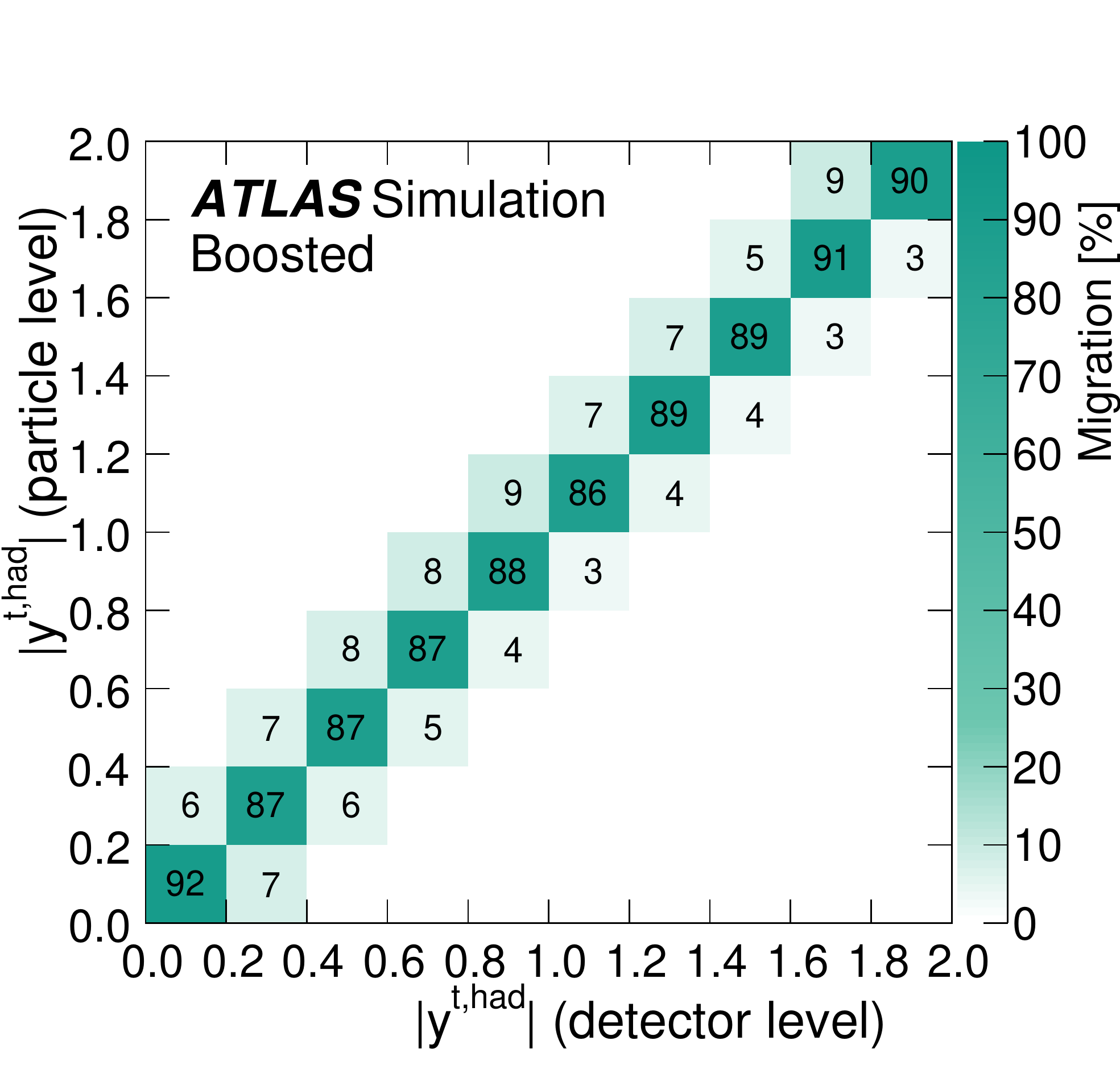}\label{Fig:corrsParticleBoosted2:topH_absrap_acc} }
\caption{Particle-to-detector-level migration matrices for \subref{Fig:corrsParticleBoosted2:topH_pt_acc} the hadronic top-quark transverse momentum and \subref{Fig:corrsParticleBoosted2:topH_absrap_acc} the absolute value of its rapidity, in the boosted topology. \Powheg{}+\PythiaSix{} is used to model the \ttb{} process and matrices are normalised so that the sum over the detector level yields 100\%. The empty bins either contain no events or the fraction of events is less than 0.5\%.}
\label{Fig:corrsParticleBoosted2}
\end{figure*}

\section{Systematic uncertainties determination} \label{sec:Uncertainties}

This section describes the estimation of systematic uncertainties related to object reconstruction and calibration, MC generator modelling and background estimation. 

To evaluate the impact of each uncertainty after the unfolding, the reconstructed distribution in simulation is varied, unfolded using corrections from the nominal \Powheg{}+\PythiaSix{} signal sample, and the unfolded varied distribution is compared to the corresponding particle-level distribution. All detector- and background-related systematic uncertainties  are evaluated using the same generator, while alternative generators and generator setups are employed to assess modelling systematic uncertainties.
In these cases, the corrections, derived from one generator, are used to unfold the detector-level spectra of the alternative generator.

The covariance matrices due to the statistical and systematic uncertainties are obtained for each observable by evaluating the covariance between the kinematic bins using pseudo-experiments. In particular, the correlations due to statistical fluctuations from the size of both data and simulated signal samples are evaluated by varying the event counts independently in every bin before unfolding, and then propagating the resulting variations through the unfolding.

\subsection{Object reconstruction and calibration}\label{sec:DetMod}
The small-$R$ jet energy scale (JES) uncertainty is derived using a~combination of simulations,
test beam data and \textit{in situ} measurements~\cite{Objects_jet_calibration,jer_2,Aad:2012vm}. Additional contributions
from jet flavour composition, $\eta$-intercalibration, punch-through, single-particle response, calorimeter response to different jet flavours and pile-up are taken into
account, resulting in 19 eigenvector systematic uncertainty subcomponents, including the uncertainties in the jet energy resolution 
obtained with an \textit{in situ} measurement of the jet response 
in dijet events~\cite{jer:2013}. 

The uncertainties in the large-$R$ JES, the jet mass scale (JMS) and the $\tau_{32}$ subjettiness ratio are obtained using a~data-driven method, which compares the ratio of each large-$R$ jet kinematic variable reconstructed from clusters in the calorimeter to that from inner-detector tracks between data and MC simulation \cite{PERF-2012-02}.  
The uncertainties in large-$R$ JES and JMS are assumed to be fully correlated and they result in a~global JES uncertainty split into three components representing the contributions from the baseline difference between data and simulation, the modelling of parton showers and hadronisation and the description of track reconstruction efficiency and impact parameter resolution.
The uncertainty in $\tau_{32}$ is considered uncorrelated with those in JES and JMS and consists of two components~\cite{ATL-PHYS-PUB-2015-053}
where an uncertainty obtained by applying the above procedure to $\sqrt{s}=8\,\TeV{}$
data is followed by applying an uncertainty in   
a~cross-calibration contribution derived by simulating the different data-taking conditions for 8~\TeV{} and 13~\TeV{} LHC $pp$ collisions in terms of reconstruction settings for topological clusters in the calorimeter, LHC bunch spacing and nuclear interaction modelling.
The uncertainty in the large-$R$ jet mass resolution (JMR) is determined by smearing the jet mass such that its mass resolution is degraded by 20\%~\cite{ATLAS:2012am,CONF-2016-008}.
The JES uncertainty for the large-$R$ jets is the dominant contribution to the total uncertainty of the measurements in the boosted topology.

The efficiency to tag jets containing $b$-hadrons is corrected in
simulated events by applying $b$-tagging scale factors, extracted from a~$\ttbar$ dilepton sample, in order to
account for the residual difference between data and simulation. Scale factors are also applied for jets originating from light quarks that are misidentified as $b$-jets.
The associated flavour-tagging systematic uncertainties, split into eigenvector components, are computed by varying the scale factors 
within their uncertainties~\cite{ATLAS-CONF-2014-004,btag,CONF-2012-040,ATL-PHYS-PUB-2017-003}. 

The lepton reconstruction efficiency in simulated events is corrected by scale factors 
derived from measurements of these efficiencies in data using a control region enriched in ~$Z \to \ell^+ \ell^-$ 
events. 
The lepton trigger and reconstruction efficiency scale factors, energy scale and resolution are varied 
within their uncertainties \cite{atlasElecPerf,ATLAS-CONF-2016-024,ATL-PHYS-PUB-2016-015,Aad:2016jkr} derived using the same sample.

The uncertainty associated with $\Etmiss$ is calculated by propagating the 
energy scale and resolution systematic uncertainties to all jets and leptons in the 
$\Etmiss$ calculation. Additional $\Etmiss$ uncertainties arising from energy deposits not associated with any reconstructed objects are also included \cite{atlasEtmisPerf}.

\subsection{Signal modelling}\label{sec:SigMod}
Uncertainties in the signal modelling
affect the kinematic properties of simulated $\ttbar$ events as well as detector- and particle-level efficiencies. 

In order to assess the uncertainty related to the matrix-element model used in the MC generator for the $\ttbar$ signal process, events simulated with \mgamcatnlo{}+\herwigpp{} are unfolded using the migration matrix and correction factors derived from an alternative \PowHeg{}+\herwigpp{} sample. The symmetrised full difference between the unfolded distribution and the known particle-level distribution of the \mgamcatnlo{}+\herwigpp{} sample is assigned as the relative uncertainty for the fiducial distributions. This uncertainty is found to be in the range $1$--$6$\%, depending on the variable, increasing up to $15$\% at large \ptthad, \mttbar, \ptttbar ~and \absyttbar. The observable that is most affected by these uncertainties is~\mttbar{}.

To assess the impact of different parton shower models, 
unfolded results using events simulated with \PowHeg interfaced 
to the \PythiaSix{} parton shower model are compared to events simulated with \PowHeg interfaced 
to the \herwigpp{} parton shower model, using the same procedure as described above to evaluate the uncertainty related to the \ttbar generator.
The resulting systematic uncertainties, taken as the symmetrised full difference, are found to be  typically at the $3$--$6$\% ($6$--$9$\%) level for 
the absolute spectra in the resolved (boosted) topology.

In order to evaluate the uncertainty related to
the modelling of initial- and final-state QCD radiation (ISR/FSR), two $\ttbar$ MC samples with modified ISR/FSR modelling are used.
The MC samples used for the evaluation of this uncertainty are generated using the 
\PowHeg generator interfaced to the \Pythia{} shower model,
where the parameters are 
varied as described in Section~\ref{sec:DataSimSamples}.
This uncertainty is found to be in the range $3$--$6$\% for the absolute spectra in both the resolved and boosted topology.

The impact of the uncertainty related to the PDF is assessed using the $\ttbar$  sample generated with a\textsc{MC@NLO} interfaced to \herwigpp. PDF-varied corrections for the unfolding procedure are obtained by reweighting the central PDF4LHC15 PDF set to the full set of 30 eigenvectors. Using these corrections, the central a\textsc{MC@NLO}+\herwigpp{} distribution is unfolded, the relative difference is computed with respect to the expected central particle-level spectrum, and the total uncertainty is obtained by adding these relative differences in quadrature. In addition, an inter-PDF uncertainty between the central PDF4LHC15 and CT10 sets is evaluated in a~similar way and added in quadrature.
The total PDF uncertainty is found to be less than $1$\% in most of the kinematic bins.

\subsection{Background modelling}\label{sec:BkgMod}

Systematic uncertainties affecting the backgrounds are evaluated by adding to the signal spectrum the difference between the varied and nominal backgrounds.
The shift between the resulting unfolded distribution and the nominal one is used to estimate the size of the uncertainty.

The single-top-quark background is assigned an uncertainty associated with its normalisation and the overall impact of this systematic uncertainty on the measured cross-section is less than  $0.5$\%. The ISR/FSR variations of the single-top sample were not considered since this would be at most a~$\sim\!\!5\%$ effect on a~$\sim\!\!5\%$ background.

The systematic uncertainties due to the 
overall normalisation and the heavy-flavour fractions of $W$+jets events 
are obtained by varying the data-driven scale factors.  The overall impact of these uncertainties is less than $0.5$\%. 
Each detector systematic uncertainty includes the impact of those on the $W$+jets estimate.

The uncertainty in the background from non-prompt and fake leptons is evaluated by
changing the selection used to form the control region
and propagating the statistical uncertainty of
parameterisations of the efficiency to pass the tighter lepton requirements for real and fake leptons.
The varied control regions are defined by inverting the $\Etmiss$ and $m_{\textrm T}^W$ requirements in the case of electrons and inverting the requirement on impact parameters of the associated track in the case of muons.
In addition, in the resolved-topology, an extra 50\% uncertainty is assigned to this background to account for the remaining mismodelling observed in various control regions. This systematic uncertainty, in the resolved topology, also includes the impact of this normalisation on extracting the $W+$jets estimate.
 In the case of the boosted topology, the mismodelling of this background is present only at large values of $m_{\textrm T}^W$. Consequently, for events satisfying $m_{\textrm T}^W$ > 150 \GeV, an extra 100\% uncertainty is included in the fake-leptons background estimate.
 Finally, in order to take into account the effect on the $W+$jets sample due to a~different non-prompt and fake leptons background normalization also in the boosted-topology, an extra systematic is added which reflects the difference in the $W+$jet estimate obtained by varying the non-prompt and fake leptons background  normalization by 30\%.
The combination of all these components also affects the shape of this background and the overall impact of these systematic uncertainties is at the 5\% level in both topologies. 

In the case of the $Z$+jets and diboson backgrounds, the uncertainties include a~contribution from the overall cross-section normalisation as well as an additional 24\% uncertainty 
added in quadrature for each reconstructed jet, not counting those from the boson decays~\cite{Alwall:2007fs}. The overall impact of these uncertainties is less than $1$\%, and the largest contribution is due to the $Z$+jets background.

\subsection{Finite size of the simulated samples and luminosity uncertainty}
In order to account for the finite size of the simulated samples, test distributions based on total predictions are varied in each bin according to their statistical uncertainty, excluding the data-driven fake-leptons background.
The effect on the absolute spectra is at most 1-2\% in the resolved case, 
while in the boosted case the effect is about 5\%, peaking at 12\% in the last top-quark $\pt$ bins. The uncertainty in the integrated luminosity of 2.1\% is not a~dominant uncertainty for the absolute differential cross-section results and it mostly cancels for the relative differential cross-section results. 

\subsection{Systematic uncertainties summary}

Figures~\ref{fig:unc_results:fiducial:topH:abs}--\ref{fig:unc_results:fiducial:tt:rel} present the uncertainties in the absolute and relative \ttbar{} fiducial phase-space differential cross-sections as a~function of the different observables. In particular, 
Figures~\ref{fig:unc_results:fiducial:topH:abs} and~\ref{fig:unc_results:fiducial:topH:rel} show uncertainties in the absolute and relative cross-sections as a function of the hadronic top-quark transverse momentum and of the absolute value of the rapidity in resolved and boosted topologies.
Figure~\ref{fig:unc_results:fiducial:tt:abs} presents the uncertainties in the absolute differential cross-sections as a function of the \ttb{} system invariant mass, transverse momentum, and absolute value of the rapidity in the resolved topology, with corresponding uncertainties in the relative cross-sections displayed in Figure~\ref{fig:unc_results:fiducial:tt:rel}.

The dominant systematic uncertainties are from the JES and flavour tagging for the resolved topology, while the large-$R$ jet uncertainties dominate the uncertainties for the boosted topology.
Other significant uncertainties include those from the signal modelling with, depending on the observable, either the generator modelling, parton shower or the ISR/FSR being the most dominant.
The uncertainties are smaller for the relative cross-section results. 

The measurements presented here exhibit, for most distributions in the resolved topology and in large parts of the phase-space, a~precision of the order of 10--15\% for the absolute spectra and 5--10\% for the relative differential cross-sections, while for the boosted topology the precision obtained varies from 20\% to about 50\%.

\begin{figure*}[htbp]
\centering
\subfigure[]{  \includegraphics[width=0.45\textwidth]{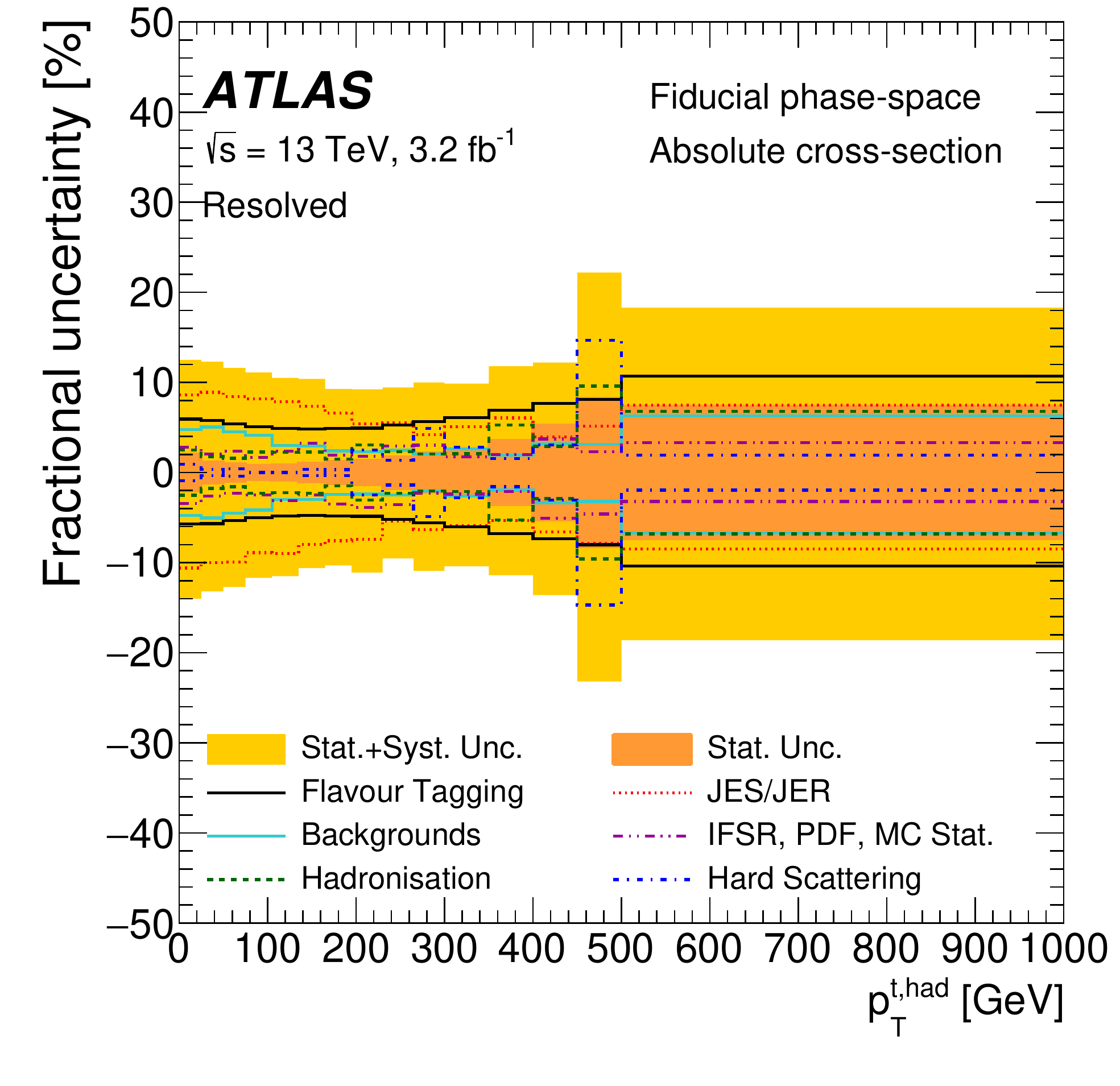}\label{fig:unc_particle:topH_pt:abs}}
\subfigure[]{  \includegraphics[width=0.45\textwidth]{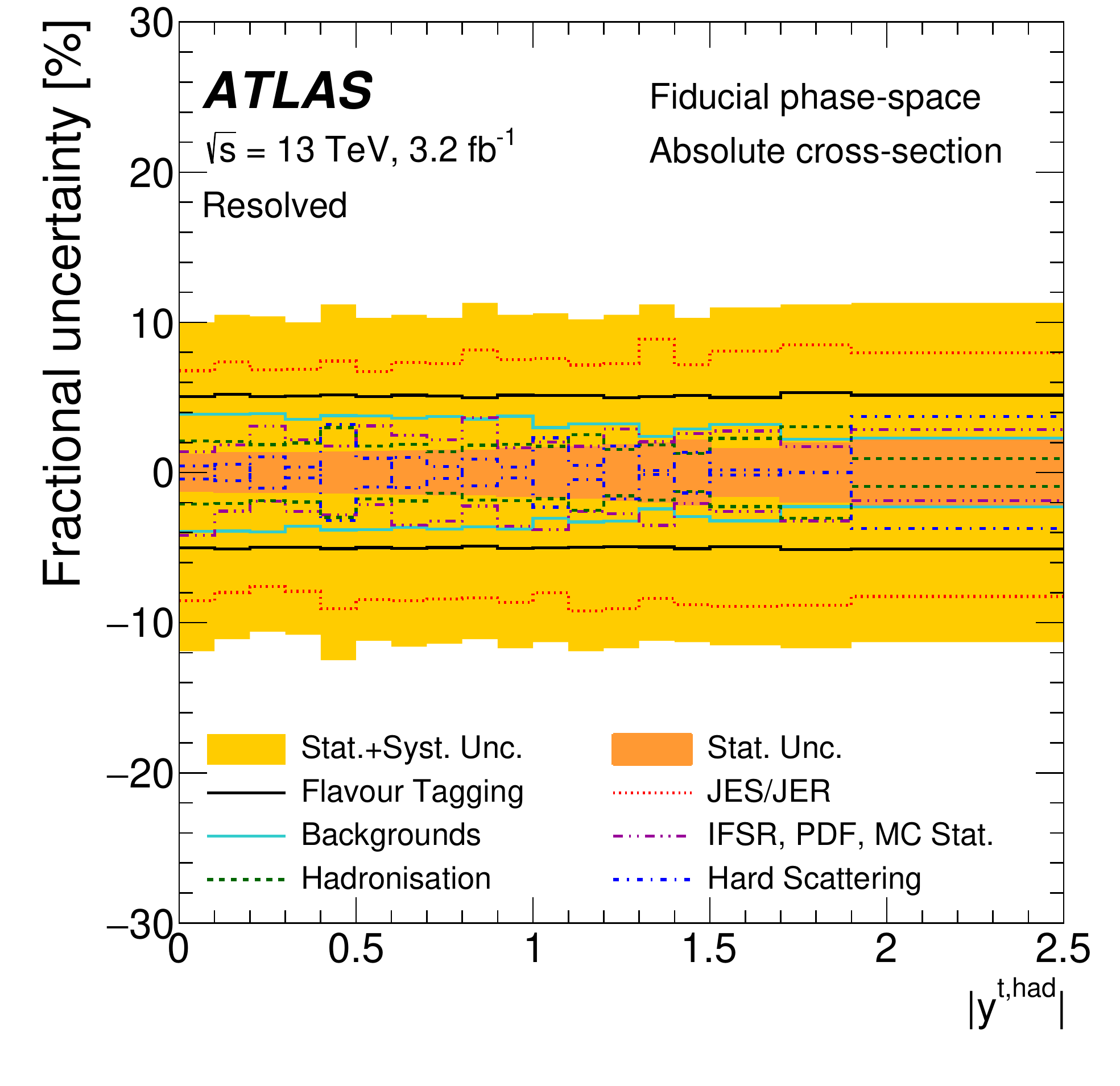}\label{fig:unc_particle:topH_absrap:abs}}
\subfigure[]{  \includegraphics[width=0.45\textwidth]{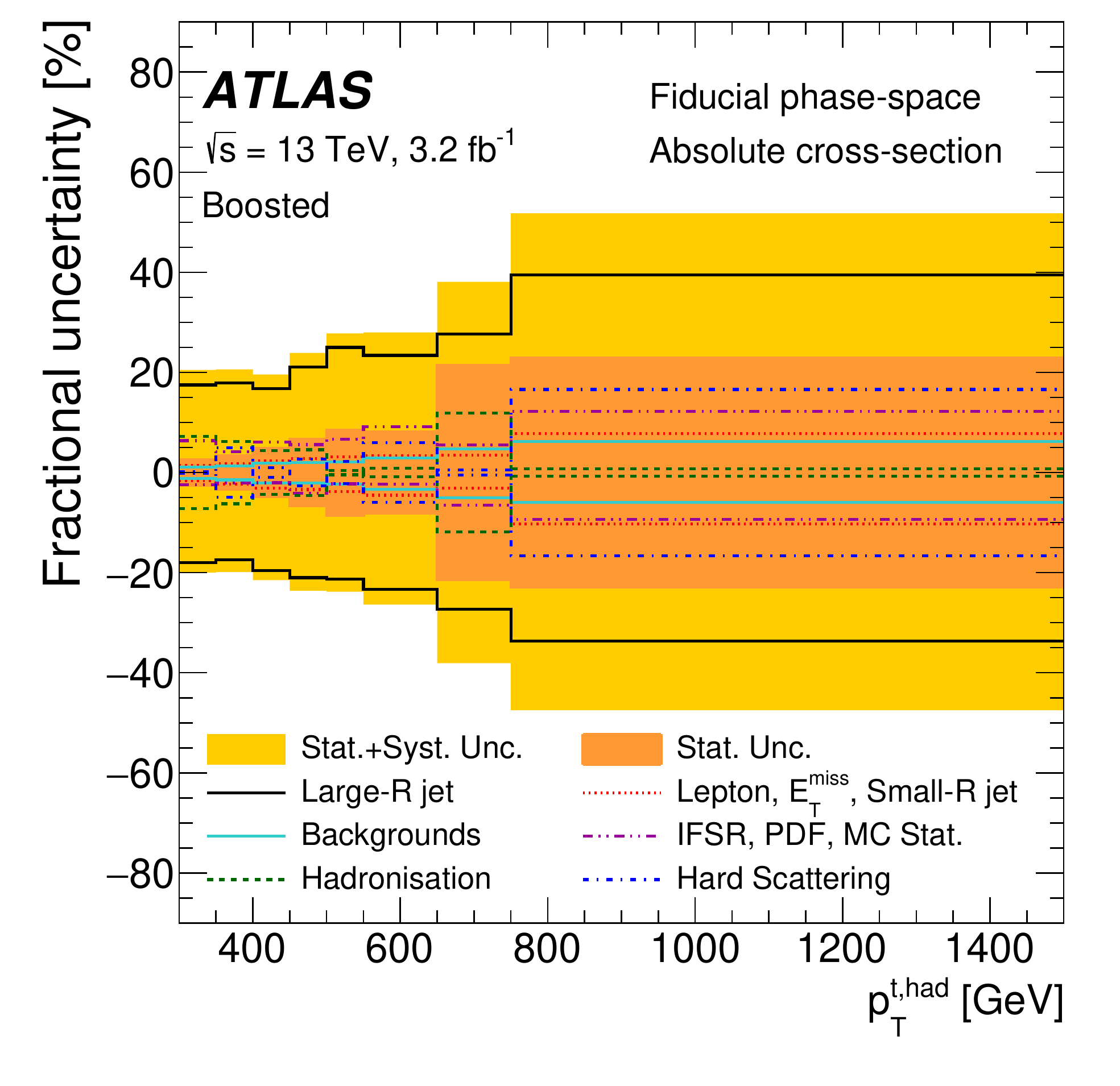}\label{fig:unc_particle:topH_pt:abs_boosted}}
\subfigure[]{  \includegraphics[width=0.45\textwidth]{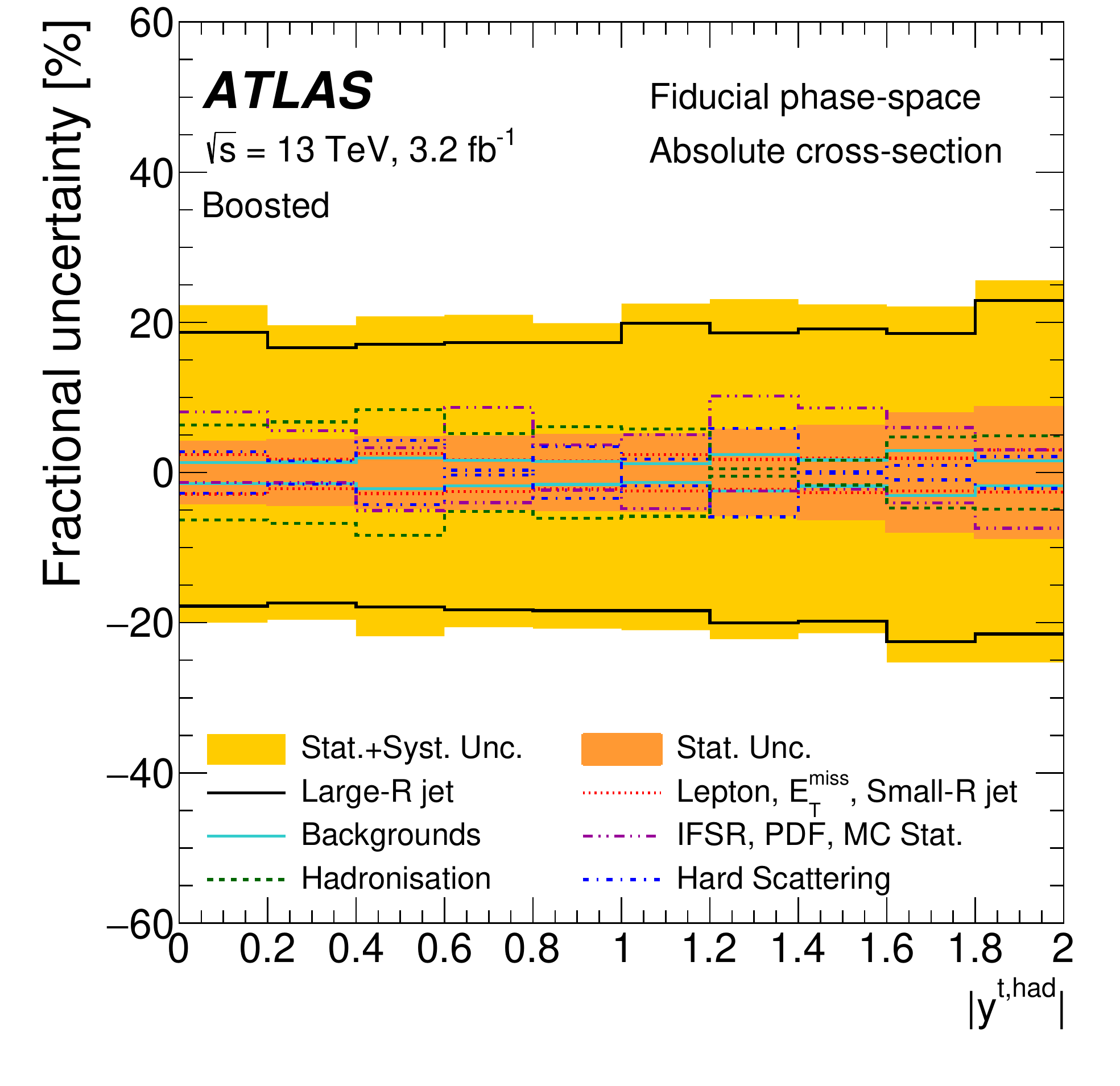}\label{fig:unc_particle:topH_absrap:abs_boosted}}
\caption{Uncertainties in the fiducial phase-space differential cross-sections as a~function of  \subref{fig:unc_particle:topH_pt:abs}~the transverse momentum (\ptthad{}) and \subref{fig:unc_particle:topH_absrap:abs} the absolute value of the rapidity (\absythad) of the hadronic top quark in the resolved topology and corresponding results in the boosted topology \subref{fig:unc_particle:topH_pt:abs_boosted}, \subref{fig:unc_particle:topH_absrap:abs_boosted}. The yellow bands indicate the total uncertainty of the data in each bin. The \PowHeg{}+\PythiaSix{} generator with \HDampMT~and the CT10 PDF is used as the nominal prediction to correct for detector effects.}
\label{fig:unc_results:fiducial:topH:abs}
\end{figure*}

\begin{figure*}[htbp]
\centering
\subfigure[]{\includegraphics[width=0.45\textwidth]{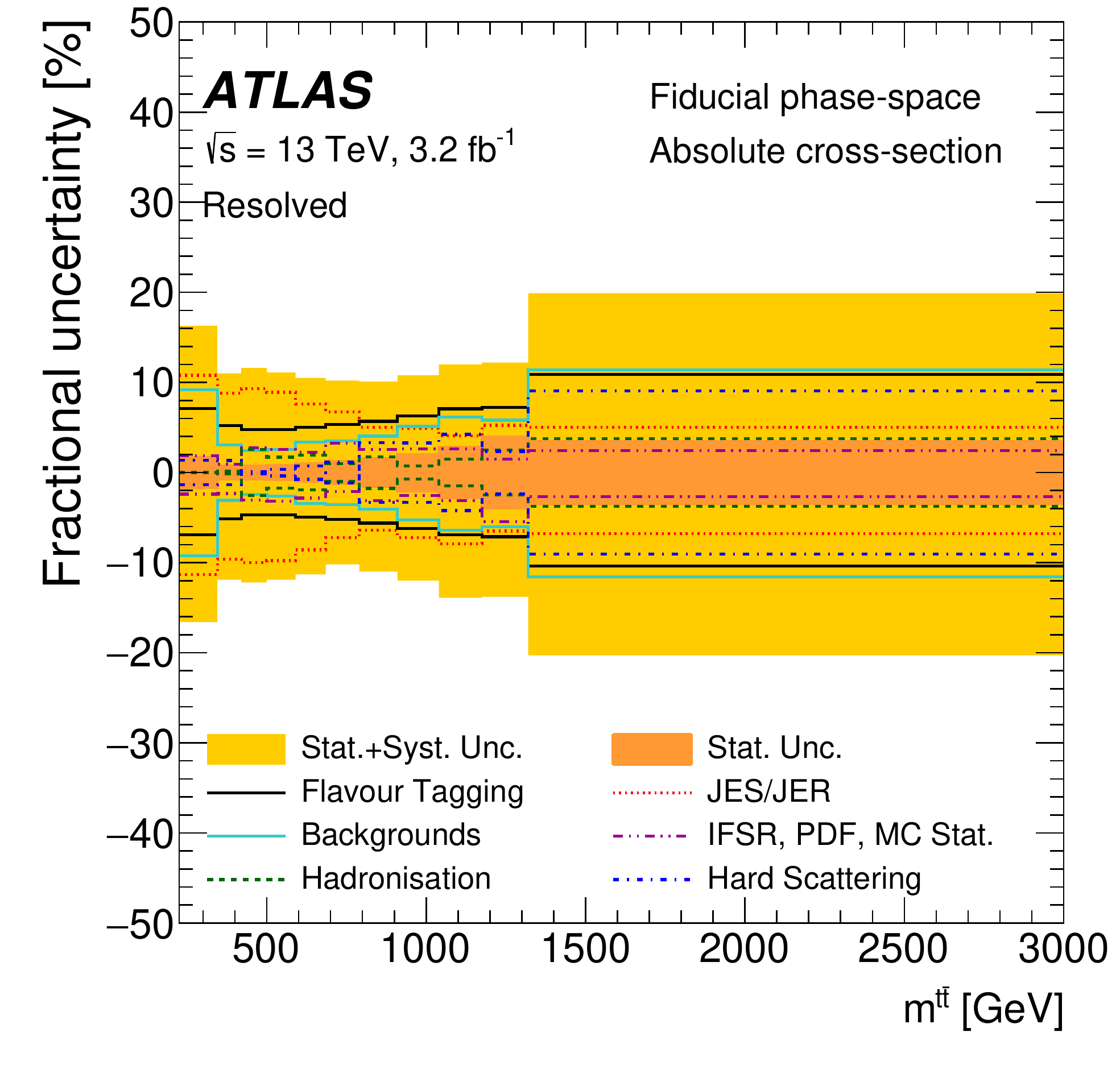}\label{fig:unc_particle:tt_m:abs}}
\subfigure[]{\includegraphics[width=0.45\textwidth]{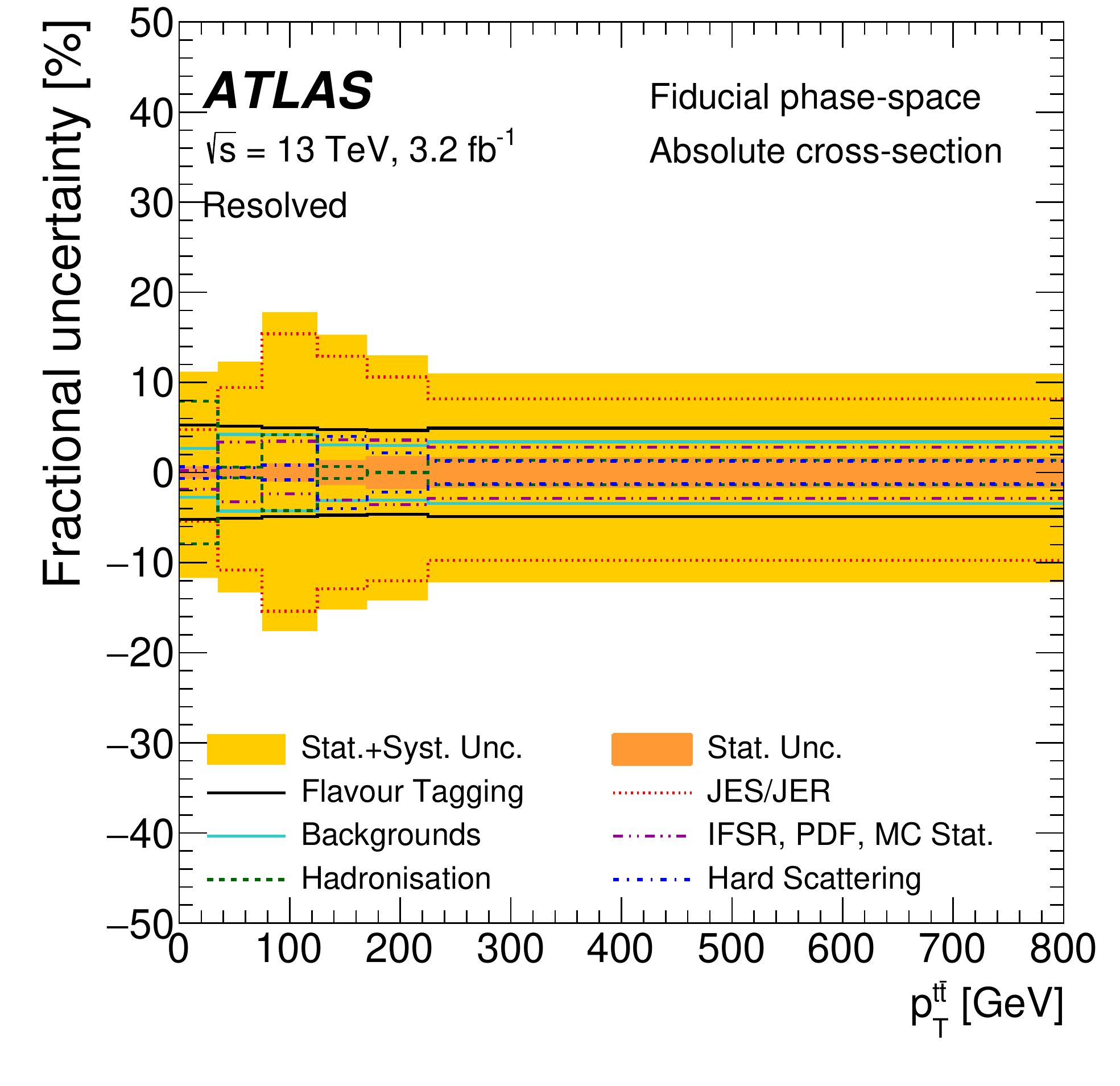}\label{fig:unc_particle:tt_pt:abs}}
\subfigure[]{\includegraphics[width=0.45\textwidth]{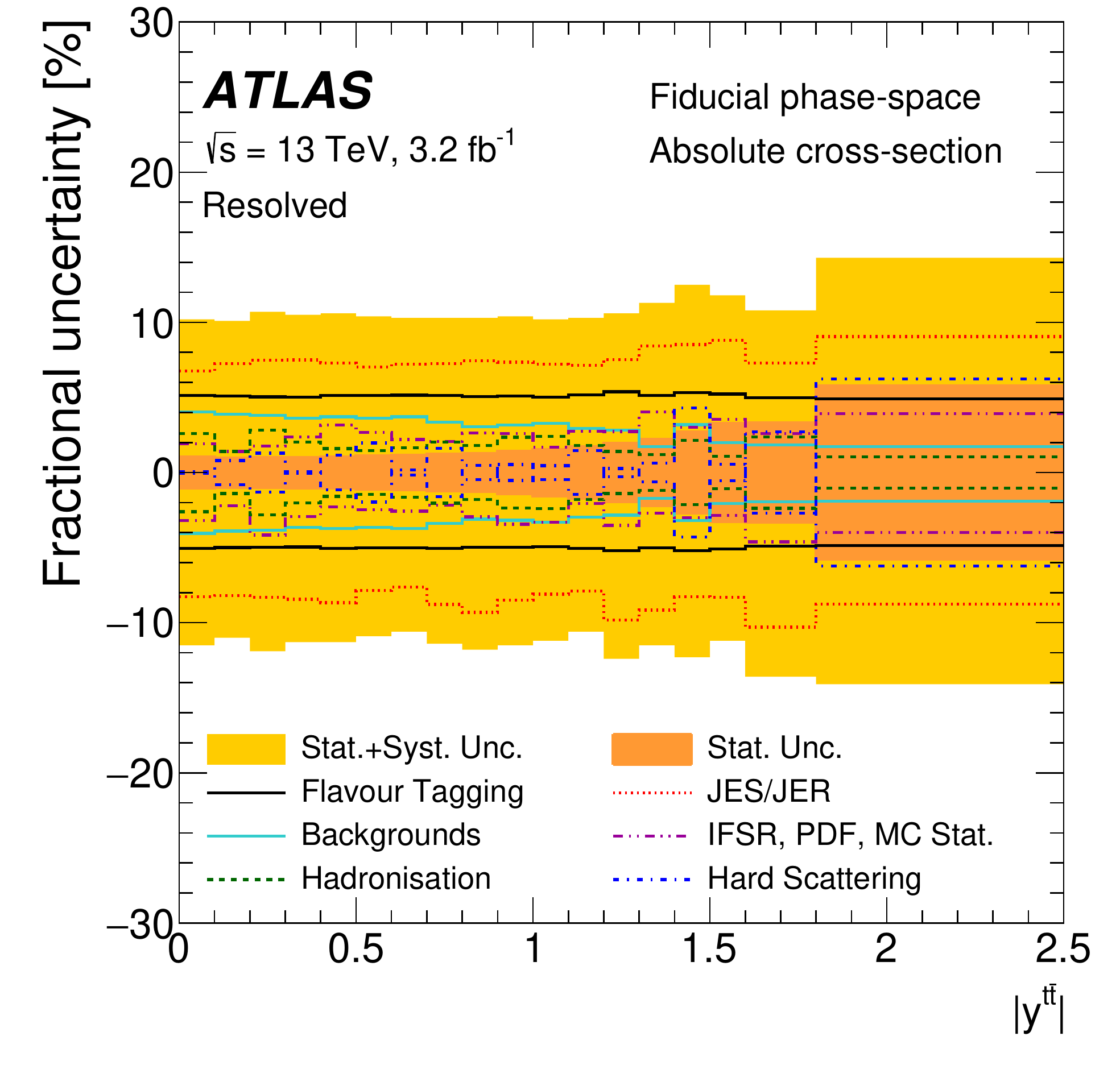}\label{fig:unc_particle:tt_absrap:abs}}
\caption{Uncertainties in the fiducial phase-space differential cross-sections as a~function of the \subref{fig:unc_particle:tt_m:abs}~invariant mass (\mttbar{}), \subref{fig:unc_particle:tt_pt:abs}~transverse momentum (\ptttbar{}) and \subref{fig:unc_particle:tt_absrap:abs} the absolute value of the rapidity (\absyttbar{}) of the \ttb{} system in the resolved topology. The yellow bands indicate the total uncertainty of the data in each bin. The \PowHeg{}+\PythiaSix{} generator with \HDampMT{} and the CT10 PDF is used as the nominal prediction to correct for detector effects.}
\label{fig:unc_results:fiducial:tt:abs}
\end{figure*}

\begin{figure*}[htbp]
\centering
\subfigure[]{  \includegraphics[width=0.45\textwidth]{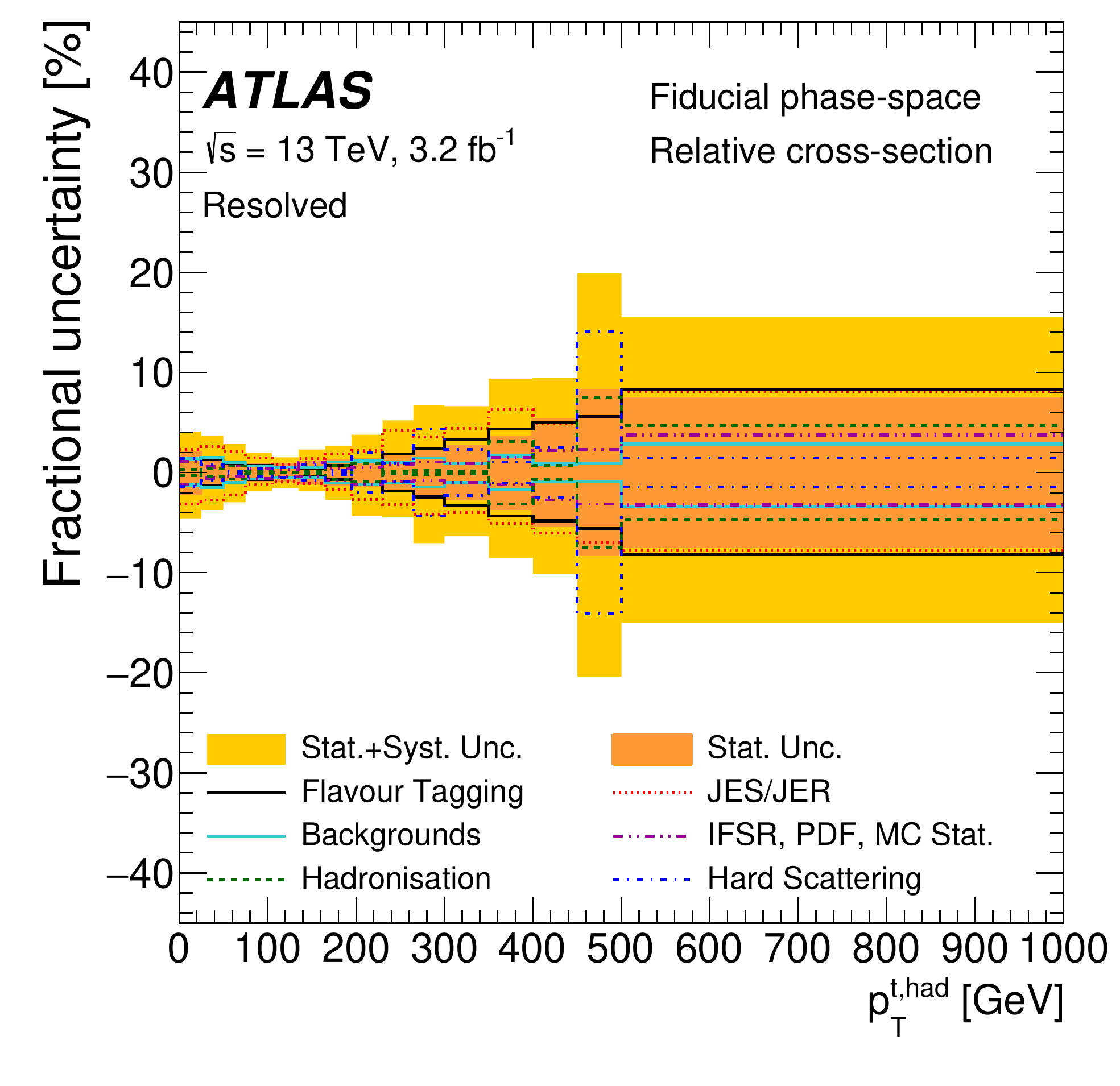}\label{fig:unc_particle:topH_pt:rel}}
\subfigure[]{ \includegraphics[width=0.45\textwidth]{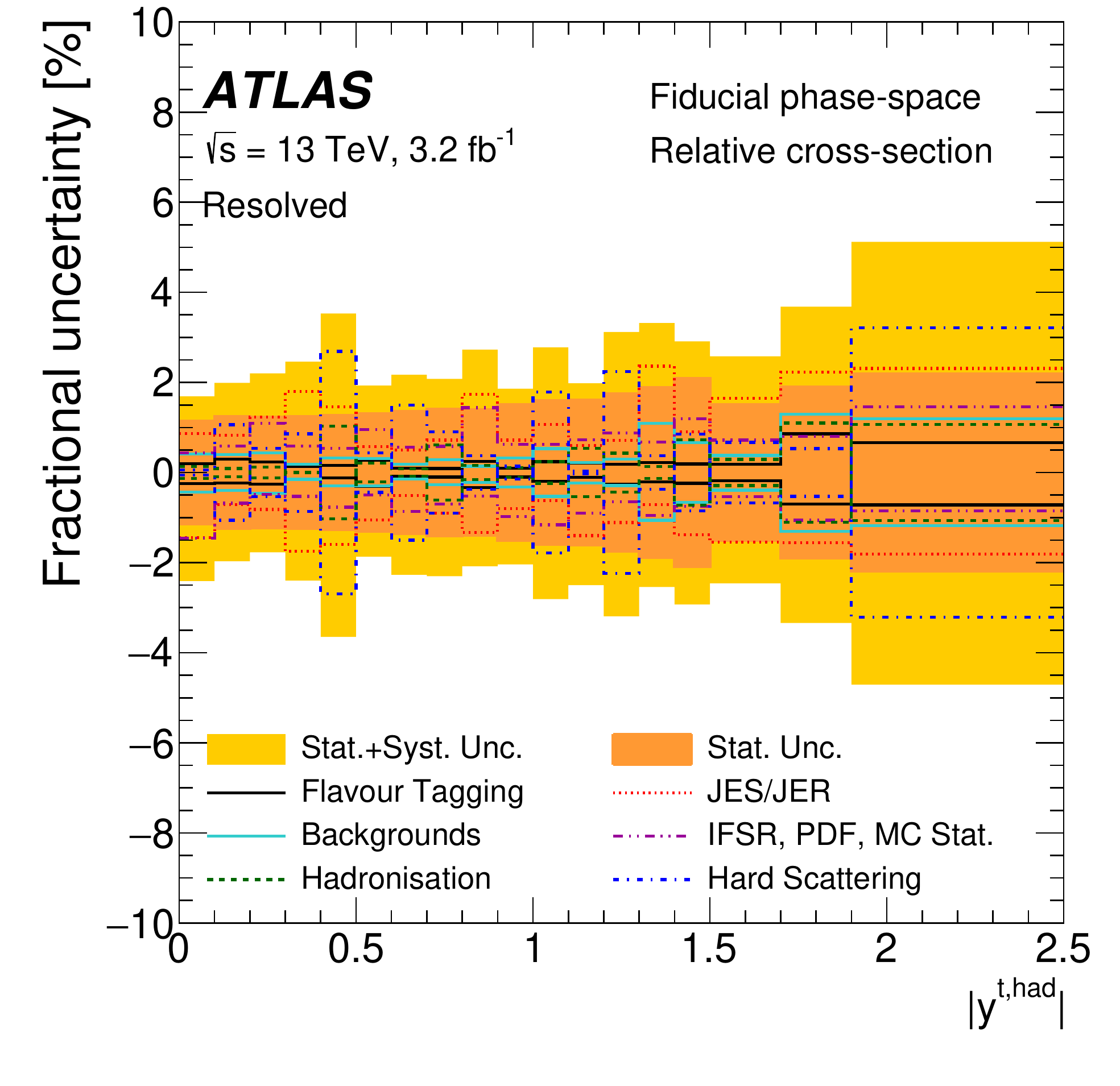}\label{fig:unc_particle:topH_absrap:rel}}
\subfigure[]{\includegraphics[width=0.45\textwidth]{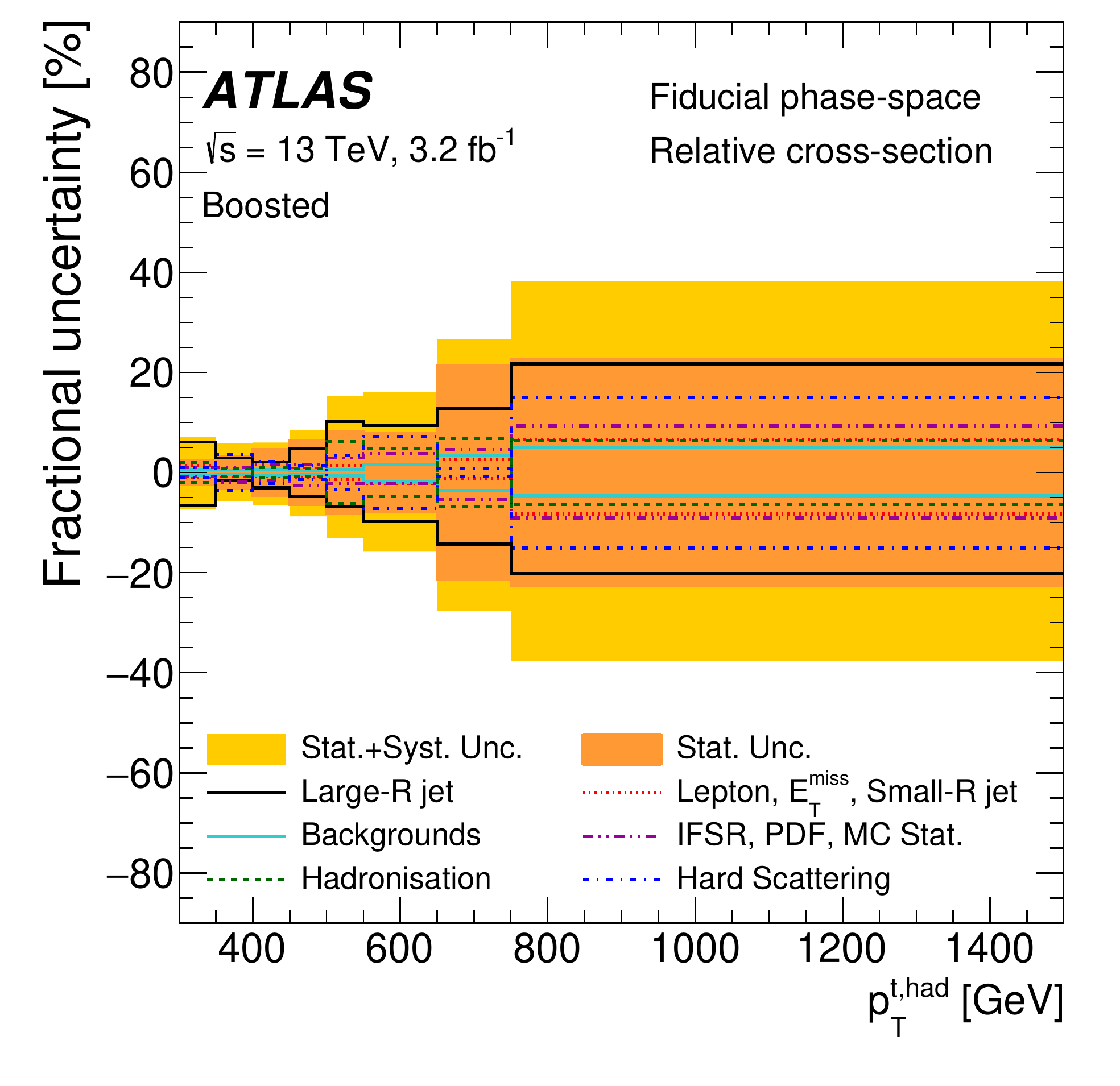}\label{fig:unc_particle:topH_pt:rel_boosted}}
\subfigure[]{\includegraphics[width=0.45\textwidth]{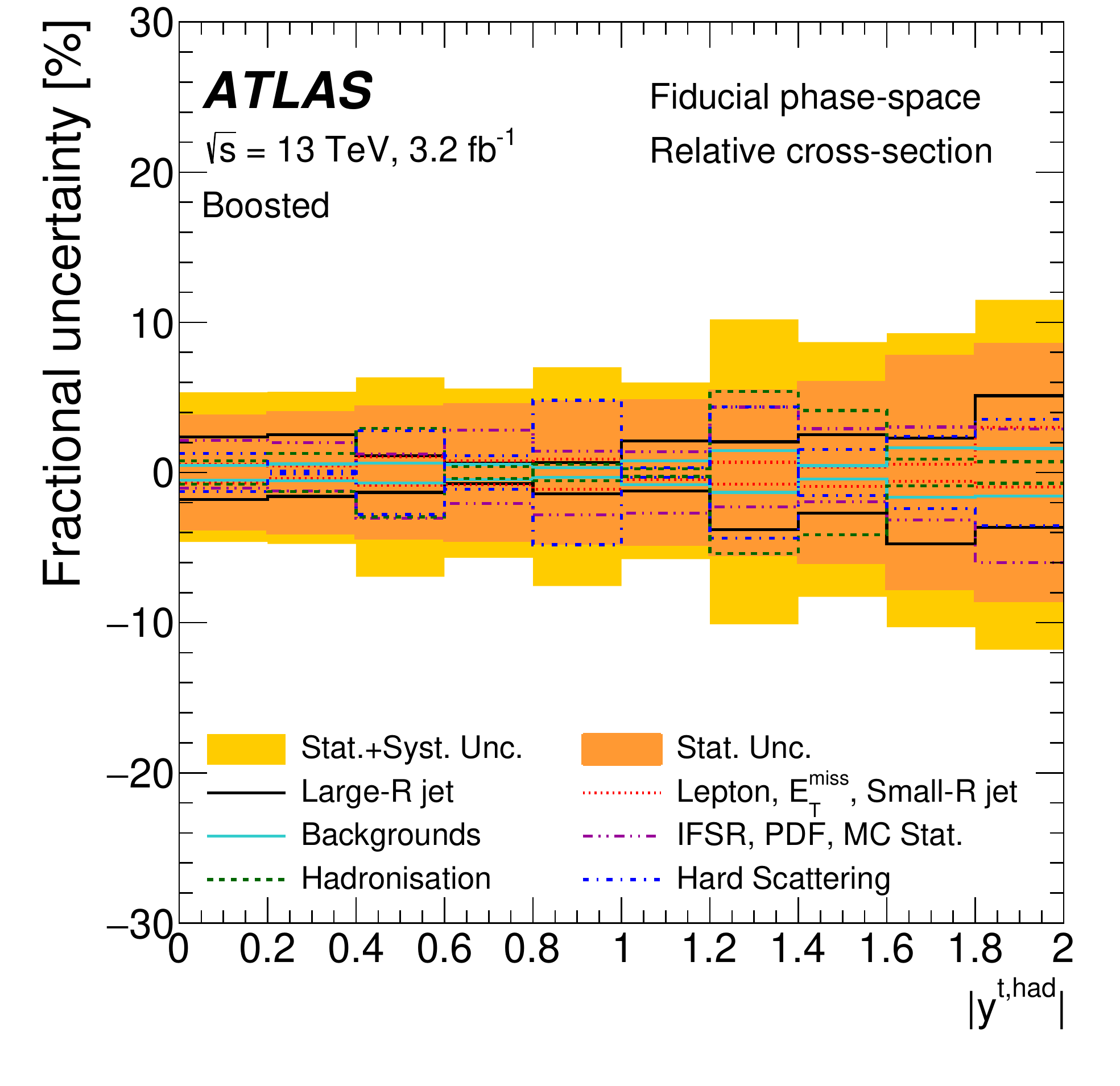}\label{fig:unc_particle:topH_absrap:rel_boosted}}
\caption{Uncertainties in the fiducial phase-space relative differential cross-sections as a~function of the \subref{fig:unc_particle:topH_pt:rel}~transverse momentum (\ptthad{}) and \subref{fig:unc_particle:topH_absrap:rel} the absolute value of the rapidity (\absythad) of the hadronic top quark in the resolved topology, and corresponding results in the boosted topology \subref{fig:unc_particle:topH_pt:rel_boosted}, \subref{fig:unc_particle:topH_absrap:rel_boosted}. The yellow bands indicate the total uncertainty of the data in each bin. The \PowHeg{}+\PythiaSix{} generator with \HDampMT{} and the CT10 PDF is used as the nominal prediction to correct for detector effects.}
\label{fig:unc_results:fiducial:topH:rel}
\end{figure*}

\begin{figure*}[htbp]
\centering
\subfigure[]{\includegraphics[width=0.45\textwidth]{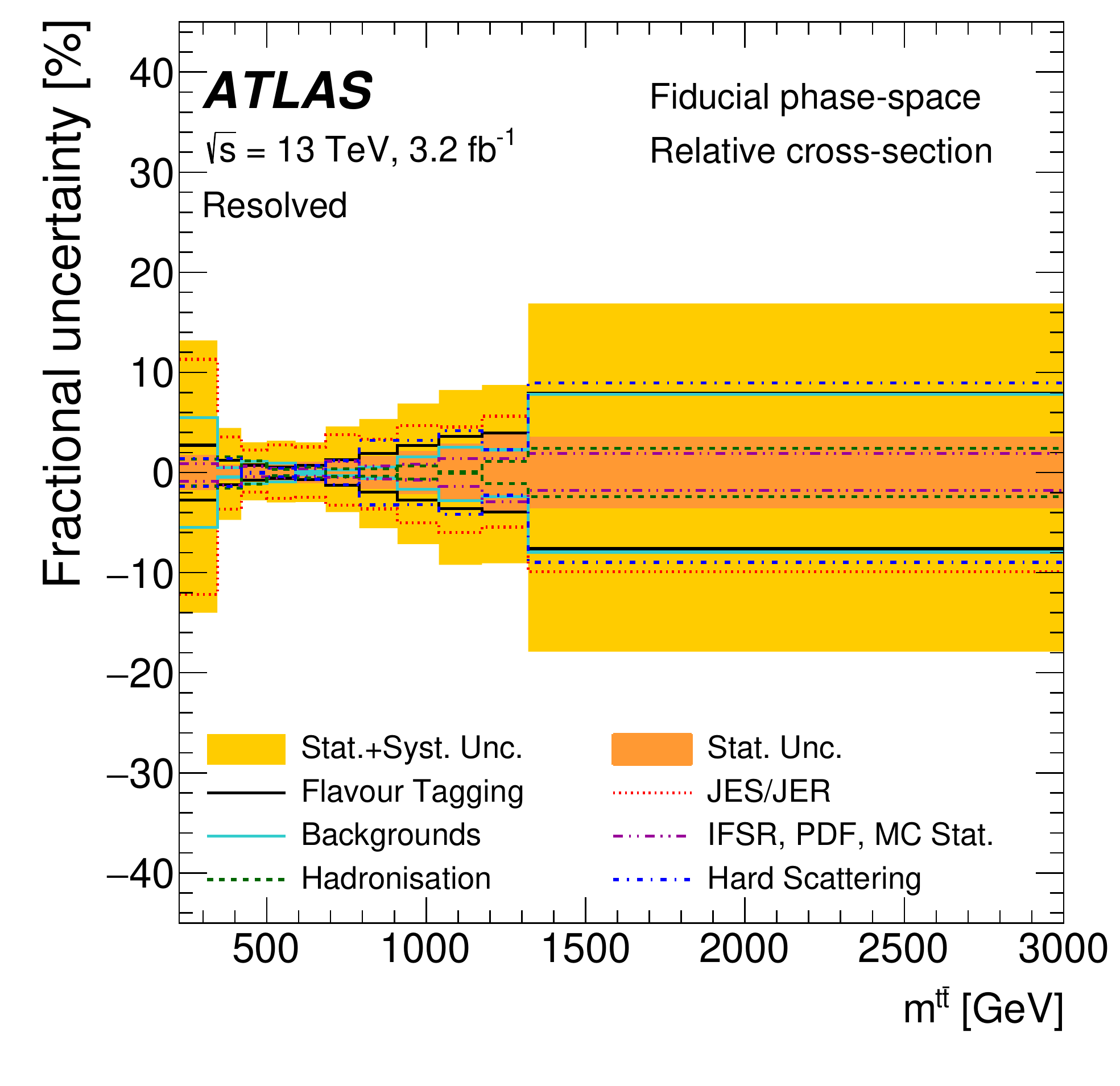}\label{fig:unc_particle:tt_m:rel}}
\subfigure[]{\includegraphics[width=0.45\textwidth]{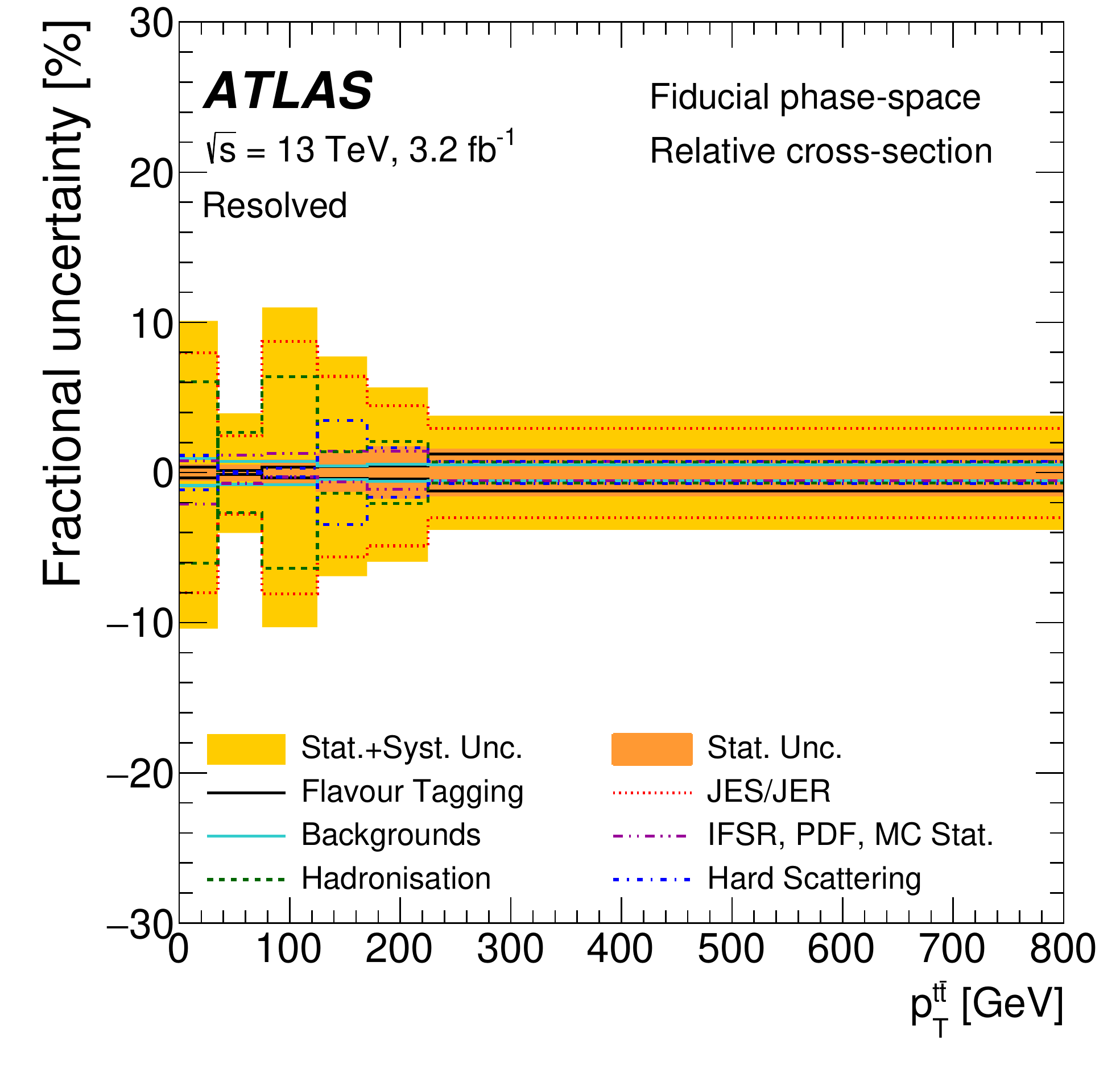}\label{fig:unc_particle:tt_pt:rel}}
\subfigure[]{\includegraphics[width=0.45\textwidth]{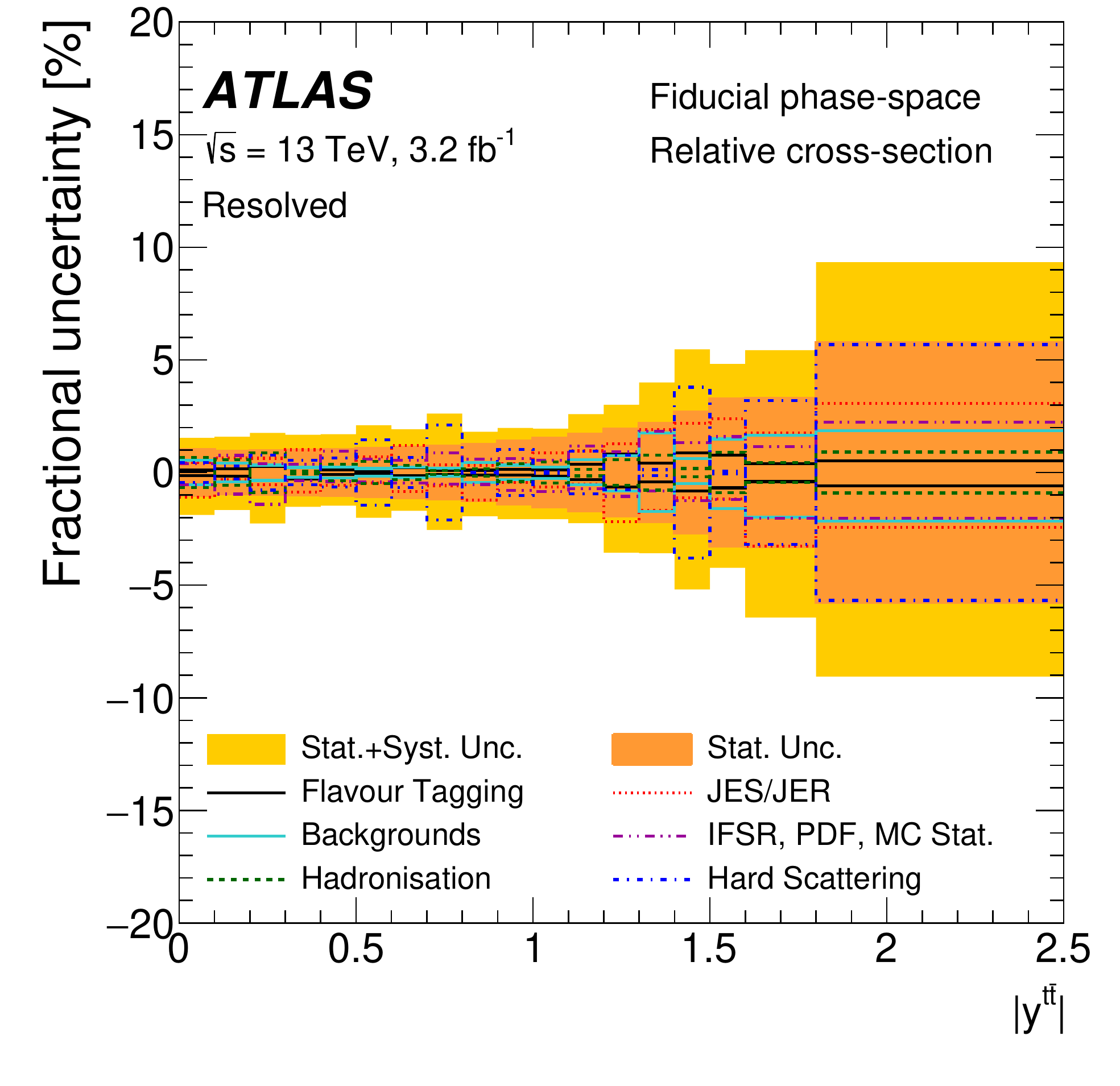}\label{fig:unc_particle:tt_absrap:rel}}
\caption{Uncertainties in the fiducial phase-space relative differential cross-sections as a~function of the \subref{fig:unc_particle:tt_m:rel}~invariant mass (\mttbar{}), \subref{fig:unc_particle:tt_pt:rel}~transverse momentum (\ptttbar{}) and \subref{fig:unc_particle:tt_absrap:rel} the absolute value of the rapidity (\absyttbar{}) of the \ttb{} system in the resolved topology. The yellow bands indicate the total uncertainty of the data in each bin. The \PowHeg{}+\PythiaSix{} generator with \HDampMT{} and the CT10 PDF is used as the nominal prediction to correct for detector effects.}
\label{fig:unc_results:fiducial:tt:rel}
\end{figure*}

\clearpage
\section{Results and comparisons with predictions} \label{sec:Results}

In this section, comparisons between unfolded data distributions and several SM predictions are presented for the observables discussed in Section~\ref{sec:YieldsAndPlots}, for both the resolved and boosted topologies. 
In addition to the absolute cross-sections, relative differential cross-sections are also studied in order to exploit the reduction of systematic uncertainties that are highly correlated across the kinematic bins.

The SM predictions are obtained using different MC generators. The \PowHegBox generator, denoted ``PWG'' in the figures, is employed with three different parton-shower models, namely \PythiaSix{}, \PythiaEight{} and \herwigpp{}, as well as two extra settings for radiation modelling (radHi, radLo). Finally, another \NLO{} generator is compared to the data, namely \mgamcatnlo{}+\herwigpp{}. All of these samples are described in detail in Section~\ref{sec:DataSimSamples}.

In order to quantify the level of agreement between the measured distributions and simulations with different theoretical predictions, $\chi^2$ values are evaluated employing the full covariance matrices of the experimental uncertainties but not including the uncertainties in the theoretical predictions. The $p$-values (probabilities that the $\chi^2$ is larger than or equal to the observed value) are then evaluated from the $\chi^2$ and the number of degrees of freedom (NDF). The normalisation constraint used to derive the relative differential cross-sections lowers the NDF and the rank of the $N_{\textrm b} \times N_{\textrm b}$ covariance matrix by one unit, where $N_{\textrm b}$ is the number of bins of the spectrum under consideration. In order to evaluate the $\chi^2$ for the normalised spectra, the following relation is used

\begin{equation*}
\chi^2 = V_{N_{\textrm b}-1}^{\textrm T} \cdot {\textrm{Cov}}_{N_{\textrm b}-1}^{-1} \cdot V_{N_{\textrm b}-1} \,,
\end{equation*} 

where $V_{N_{\textrm b}-1}$ is the vector of differences between data and prediction obtained by discarding one of the $N_{\textrm b}$ elements and ${\textrm{Cov}}_{N_{\textrm b}-1}$ is the $(N_{\textrm b}-1) \times (N_{\textrm b}-1)$ sub-matrix derived from the full covariance matrix discarding the corresponding row and column. 
The sub-matrix obtained in this way is invertible and allows the $\chi^2$ to be computed. The $\chi^2$ value does not depend on the choice of element discarded for the vector $V_{N_{\textrm b}-1}$ and the corresponding sub-matrix ${\textrm{Cov}}_{N_{\textrm b}-1}$. 

The total covariance matrix including the effect of all uncertainties is calculated for each distribution at particle level in order to quantitatively compare with theoretical predictions. This matrix is obtained by summing two covariance matrices.

The first covariance matrix incorporates the statistical uncertainty and the systematic uncertainties from detector and background modelling. It is obtained by performing pseudo-experiments, where, in each pseudo-experiment, each bin of the data distribution is varied following a~Poisson distribution. Gaussian-distributed shifts are coherently added for each systematic uncertainty by scaling each Poisson-fluctuated bin with the relative variation from the associated systematic uncertainty effect. Differential cross-sections are obtained by unfolding each varied reconstruction distribution with the nominal corrections, and the results are used to compute the first covariance matrix.

The second covariance matrix is obtained by summing four separate theory-model covariance matrices corresponding to the effects of the \ttbar{} generator, parton shower, ISR/FSR and PDF uncertainties. Elements of these covariance matrices are computed by multiplying the relative systematic uncertainties scaled by the measured cross-section in each bin. The bin-to-bin correlation value is set to unity for each contribution. This procedure is needed for the signal modelling uncertainties  because they cannot be represented as a~smooth variation at detector level, and so cannot be included in the pseudo-experiment formalism used for the first covariance matrix.

If the number of events in a~given bin of a~pseudo-experiment
becomes negative due to the effect of the combined systematic shifts,
this value is set to zero before the unfolding stage.  
This is the case for the \ptthad{} distribution in the boosted topology where the total uncertainty is about 50$\%$ in the last two bins and the negative fluctuations appeared in 2$\%$ of pseudo-experiments in the seventh bin and in 7$\%$ for the last bin.
The expected effect is thus only a few per cent decorrelation of the last two bins from the others.

Figures~\ref{fig:results:fiducial:topH:abs}--\ref{fig:results:fiducial:tt:rel} present the absolute and relative \ttbar{} fiducial phase-space differential cross-sections as functions of the different observables.
In particular, Figure~\ref{fig:results:fiducial:topH:abs} shows the absolute differential cross section as
a function of the hadronic top-quark transverse momentum and the absolute value of the
rapidity in the resolved topology in the top row and the boosted topology in the bottom row.
Figure~\ref{fig:results:fiducial:tt:abs} presents the absolute cross section as a function of the \ttbar{} system invariant
mass, transverse momentum and absolute value of the rapidity in the resolved topology.
Figures~\ref{fig:results:fiducial:topH:rel} and~\ref{fig:results:fiducial:tt:rel} show the corresponding relative cross-sections.

In Tables~\ref{tab:cor_topH_pt_rel_resol} and \ref{tab:cor_topH_pt_rel_boosted},
correlation matrices are presented for the
relative differential cross-section measurements as a function of the
hadronic top-quark transverse momentum for the resolved and boosted topologies. 
Large correlations across the bins are present for the absolute cross-section results due to highly correlated systematic uncertainties which change the overall cross-section. For the relative cross-section results, there is typically a~strong correlation between a few neighbouring bins, and an anti-correlation with distant bins due to the normalisation condition.

For the hadronic top-quark transverse momentum, the values of the absolute
differential cross-sections are shown in Table~\ref{Tab:XsecValues} along with their uncertainties for both the resolved and boosted topologies. In addition, the inclusive fiducial cross-section in each of the resolved and boosted topology is presented in Table~\ref{tab:fid_xsects}
alongside those from different models for comparison.
The inclusive cross-section is extracted in a~single bin, i.e. not by integrating a~particular differential cross-section. Most of the systematic uncertainties associated with this fiducial measurement are uncorrelated with the fiducial measurement in the dilepton channel~\cite{atlasInclusive13} and the results agree at the level of about one standard deviation.

Most predictions do not describe well all the distributions, as also witnessed by the $\chi^2$ values and the $p$-values listed in Tables~\ref{Tab:chi2_res_abs}--\ref{Tab:chi2_boo_rel}. 
In particular, tension between data and most predictions is observed in the case of the differential cross-sections as a function of the hadronic top-quark transverse momentum distribution (Figures~\ref{fig:particle:topH_pt:abs}, \ref{fig:particle:topH_pt:abs_boosted}, \ref{fig:particle:topH_pt:rel}, \ref{fig:particle:topH_pt:rel_boosted}). 

No electroweak corrections \cite{Kuhn:2005it,Kuhn:2006vh,Bernreuther:2008aw,Kuhn:2006vh,Manohar:2012rs,Kuhn:2013zoa} are included in these predictions. 
Although these have been shown to have a~measurable impact at very high values of the top-quark transverse momentum~\cite{CON-2014-057},
the electroweak correction of $10$--$15$\% \cite{Kuhn:2013zoa} for values of the top-quark transverse momentum of about 1~\TeV{} 
is not large enough to remove the discrepancy observed in the differential cross-section as a function of the boosted \ptthad{} distribution as shown in Figure~\ref{fig:particle:topH_pt:rel_boosted}. For the case of the differential cross-sections as a function of the \ptthad{} distribution in both the resolved and boosted topologies the \PowHeg{}+\HerwigSeven{} generator gives the best $\chi^2$ value. It was shown that, at 8~\TeV{}, the agreement at parton level improves when using the full \NNLO{} calculations~\cite{Czakon:2016dgf,Czakon:2015owf, atlasDiff4, cmsDiff2}. The shape of the differential cross-sections as a function of the \absythad distributions  (Figures~\ref{fig:particle:topH_absrap:abs} and~\ref{fig:particle:topH_absrap:abs_boosted}) show good agreement for all the generators for both the resolved and boosted topologies. 
 
For the differential cross-section as a function of the \mttbar{} distribution (Figures~\ref{fig:particle:tt_m:abs} and~\ref{fig:particle:tt_m:rel}), all the predictions agree reasonably well with the data except for the two \herwigpp{} samples. As shown in the differential cross-section as a function of \ptttbar{} distributions (Figures~\ref{fig:particle:tt_pt:abs} and~\ref{fig:particle:tt_pt:rel}), the radHi and radLo samples bracket the nominal \PowHeg{}+\PythiaSix{} prediction. As illustrated by the $\chi^2$ values of the \ptttbar{} spectrum, for the case of the absolute differential cross-sections, none of the predictions agree well with the data, while for the case of the relative differential cross-sections only the radLo and the \PowHeg{}+\herwigpp predictions disagree with the data. 

There is an indication (Figure~\ref{fig:particle:tt_absrap:rel}) that the data at high values of \ttb{} rapidity for the relative differential cross-sections may not be adequately described by many of the generators considered. These distributions are especially sensitive to different choices of PDF sets, as was observed at 8~\TeV{}~\cite{atlasDiff4}.
The \PowHeg{}+\herwigpp prediction gives the worst $\chi^2$ value for this observable.

Overall, it can be seen
that the \PowHeg{}+\herwigpp prediction disagrees the most with data, having a~$p$-value of less than
$1\%$~for four of the five observables studied in the resolved channel, while the \Powheg{}+\HerwigSeven{} prediction agrees adequately with the data for all five observables. 

Since the definitions of the phase space and the particle-level hadronic top quark differ between the resolved and boosted topologies, a~direct comparison of the measured differential cross-sections is not possible. However, it can be seen in~Figure~\ref{fig:results:fiducial:topH:abs_resolved_boosted} that the ratio of data to prediction is consistent between the two topologies in the overlap region.
Also, the trend observed in~Figure~\ref{fig:results:fiducial:topH:abs_resolved_boosted} explains the difference in the overall data/prediction normalisation in~Figure~\ref{fig:particle:topH_pt:abs} and Figure~\ref{fig:particle:topH_pt:abs_boosted}.

About 50\% of the selected data events that satisfy the boosted selection also satisfy the resolved selection.
This fraction depends on the kinematic properties of the events and decreases to about 30\% at a~top-quark $\pt$ of $1\,$\TeV. Only 0.3\% of the events that satisfy the resolved event selection also satisfy the boosted selection requirements.

\clearpage

\begin{table}[!p]
\centering 
\resizebox*{\textwidth}{!}{
\begin{tabular}{r|rrrrrrrrrrrrrrr}
Bin \GeV{}  &  0--25  &  25--50  &  50--75  &  75--105  &  105--135  &  135--165  &  165--195  &  195--230  &  230--265  &  265--300  &  300--350  &  350--400  &  400--450  &450--500&500--1000\\ \hline
 0--25  & \textbf{1.00} & 0.70  & 0.61  & 0.59  & 0.08  & $-$0.23  & $-$0.49  & $-$0.30  & $-$0.52  & $-$0.22  & $-$0.39  & $-$0.42  & $-$0.23  & 0.07  & $-$0.17  \\ 
 25--50  & 0.70  & \textbf{1.00} & 0.77  & 0.69  & $-$0.01  & $-$0.39  & $-$0.65  & $-$0.45  & $-$0.66  & $-$0.35  & $-$0.50  & $-$0.53  & $-$0.31  & $-$0.01  & $-$0.21  \\ 
 50--75  & 0.61  & 0.77  & \textbf{1.00} & 0.67  & $-$0.01  & $-$0.40  & $-$0.58  & $-$0.46  & $-$0.70  & $-$0.49  & $-$0.60  & $-$0.56  & $-$0.41  & $-$0.12  & $-$0.28  \\ 
 75--105  & 0.59  & 0.69  & 0.67  & \textbf{1.00} & 0.06  & $-$0.21  & $-$0.53  & $-$0.38  & $-$0.56  & $-$0.35  & $-$0.50  & $-$0.52  & $-$0.37  & $-$0.08  & $-$0.32  \\ 
 105--135  & 0.08  & $-$0.01  & $-$0.01  & 0.06  & \textbf{1.00} & 0.35  & 0.09  & 0.15  & $-$0.12  & 0.05  & $-$0.12  & $-$0.16  & $-$0.19  & 0.11  & $-$0.27  \\ 
 135--165  & $-$0.23  & $-$0.39  & $-$0.40  & $-$0.21  & 0.35  & \textbf{1.00} & 0.57  & 0.54  & 0.50  & 0.47  & 0.40  & 0.37  & 0.25  & 0.35  & 0.12  \\ 
 165--195  & $-$0.49  & $-$0.65  & $-$0.58  & $-$0.53  & 0.09  & 0.57  & \textbf{1.00} & 0.66  & 0.64  & 0.50  & 0.53  & 0.61  & 0.38  & 0.33  & 0.29  \\ 
 195--230  & $-$0.30  & $-$0.45  & $-$0.46  & $-$0.38  & 0.15  & 0.54  & 0.66  & \textbf{1.00} & 0.67  & 0.76  & 0.71  & 0.62  & 0.56  & 0.66  & 0.39  \\ 
 230--265  & $-$0.52  & $-$0.66  & $-$0.70  & $-$0.56  & $-$0.12  & 0.50  & 0.64  & 0.67  & \textbf{1.00} & 0.68  & 0.73  & 0.73  & 0.58  & 0.42  & 0.45  \\ 
 265--300  & $-$0.22  & $-$0.35  & $-$0.49  & $-$0.35  & 0.05  & 0.47  & 0.50  & 0.76  & 0.68  & \textbf{1.00} & 0.77  & 0.60  & 0.65  & 0.71  & 0.45  \\ 
 300--350  & $-$0.39  & $-$0.50  & $-$0.60  & $-$0.50  & $-$0.12  & 0.40  & 0.53  & 0.71  & 0.73  & 0.77  & \textbf{1.00} & 0.71  & 0.66  & 0.59  & 0.57  \\ 
 350--400  & $-$0.42  & $-$0.53  & $-$0.56  & $-$0.52  & $-$0.16  & 0.37  & 0.61  & 0.62  & 0.73  & 0.60  & 0.71  & \textbf{1.00} & 0.59  & 0.49  & 0.62  \\ 
 400--450  & $-$0.23  & $-$0.31  & $-$0.41  & $-$0.37  & $-$0.19  & 0.25  & 0.38  & 0.56  & 0.58  & 0.65  & 0.66  & 0.59  & \textbf{1.00} & 0.54  & 0.57  \\ 
 450--500  & 0.07  & $-$0.01  & $-$0.12  & $-$0.08  & 0.11  & 0.35  & 0.33  & 0.66  & 0.42  & 0.71  & 0.59  & 0.49  & 0.54  & \textbf{1.00} & 0.46  \\ 
 500--1000  & $-$0.17  & $-$0.21  & $-$0.28  & $-$0.32  & $-$0.27  & 0.12  & 0.29  & 0.39  & 0.45  & 0.45  & 0.57  & 0.62  & 0.57  & 0.46  & \textbf{1.00} \\ 
\end{tabular}}

\caption{Correlation matrix of the relative cross-section as a function of the hadronic top-quark \pt{}, accounting for the statistical and systematic uncertainties in the resolved topology.}
\label{tab:cor_topH_pt_rel_resol}
\end{table}
\begin{table}[!p]

\footnotesize 
\centering 
\begin{tabular}{r|rrrrrrrr}
Bin \GeV{}  &  300--350  &  350--400  &  400--450  &  450--500  &  500--550  &  550--650  &  650--750  &  750--1500   \\ \hline
 300--350  & \textbf{1.00} & 0.36  & $-$0.42  & $-$0.57  & $-$0.46  & $-$0.47  & $-$0.53  & $-$0.52  \\ 
 350--400  & 0.36  & \textbf{1.00} & $-$0.01  & $-$0.22  & $-$0.03  & 0.04  & $-$0.23  & $-$0.11  \\ 
 400--450  & $-$0.42  & $-$0.01  & \textbf{1.00} & 0.34  & 0.30  & 0.50  & 0.27  & 0.37  \\ 
 450--500  & $-$0.57  & $-$0.22  & 0.34  & \textbf{1.00} & 0.51  & 0.45  & 0.48  & 0.49  \\ 
 500--550  & $-$0.46  & $-$0.03  & 0.30  & 0.51  & \textbf{1.00} & 0.59  & 0.44  & 0.51  \\ 
 550--650  & $-$0.47  & 0.04  & 0.50  & 0.45  & 0.59  & \textbf{1.00} & 0.43  & 0.54  \\ 
 650--750  & $-$0.53  & $-$0.23  & 0.27  & 0.48  & 0.44  & 0.43  & \textbf{1.00} & 0.44  \\ 
 750--1500  & $-$0.52  & $-$0.11  & 0.37  & 0.49  & 0.51  & 0.54  & 0.44  & \textbf{1.00} \\ 
\end{tabular} 

\caption{Correlation matrix for the relative cross-section as a function of the hadronic top-quark \pt{}, accounting for the statistical and systematic uncertainties in the boosted topology.}
\label{tab:cor_topH_pt_rel_boosted}
\end{table}

\begin{table}[t]
\centering
\begin{tabular}{r|r}
\hline 
\hline
 \multicolumn{1}{l|}{Resolved} &  \multicolumn{1}{l}{$\sigma$ in resolved}       \\
 \multicolumn{1}{l|}{particle-level \ptthad{} [\GeV{}]} & \multicolumn{1}{l}{fiducial phase-space [pb]}      \\
\hline
  0--25  &	$3.37 \pm 0.07 \pm 0.44$ \\
  25--50  &	$9.77 \pm 0.11 \pm 1.22$ \\
  50--75  &	$14.51 \pm 0.14 \pm 1.73$ \\
  75--105  &	$19.26 \pm 0.15 \pm 2.17$ \\
  105--135  &	$17.21 \pm 0.15 \pm 1.88$ \\
  135--165  &	$12.34 \pm 0.12 \pm 1.28$ \\
  165--195  &	$8.40 \pm 0.10 \pm 0.81$ \\
  195--230  &	$6.42 \pm 0.09 \pm 0.65$ \\
  230--265  &	$3.95 \pm 0.07 \pm 0.37$ \\
  265--300  &	$2.69 \pm 0.06 \pm 0.28$ \\
  300--350  &	$2.04 \pm 0.05 \pm 0.21$ \\
  350--400  &	$1.11 \pm 0.04 \pm 0.13$ \\
  400--450  &	$0.55 \pm 0.03 \pm 0.07$ \\
  450--500  &	$0.26 \pm 0.02 \pm 0.06$ \\
  500--1000  &	$0.36 \pm 0.03 \pm 0.07$ \\
\hline
\multicolumn{1}{l|}{Boosted} & \multicolumn{1}{l}{$\sigma$ in boosted}       \\
 \multicolumn{1}{l|}{particle-level \ptthad{} [\GeV{}]} & \multicolumn{1}{l}{fiducial phase-space [pb]}      \\
\hline
  300--350  &	$0.95 \pm 0.02 \pm 0.19$ \\
  350--400  &	$0.61 \pm 0.02 \pm 0.12$ \\
  400--450  &	$0.35 \pm 0.02 \pm 0.07$ \\
  450--500  &	$0.20 \pm 0.01 \pm 0.05$ \\
  500--550  &	$0.14 \pm 0.01 \pm 0.04$ \\
  550--650  &	$0.17 \pm 0.01 \pm 0.05$ \\
  650--750  &	$0.042 \pm 0.009 \pm 0.016$ \\
  750--1500  &	$0.043 \pm 0.010 \pm 0.023$ \\
\hline
\hline
\end{tabular}
\caption{Unfolded fiducial phase-space differential cross-section values in bins of hadronic top-quark transverse momentum for the resolved (top) and boosted (bottom) topologies. The first uncertainty is statistical and the second one is systematic. }
\label{Tab:XsecValues}
\end{table}

\begin{table}[!h]
  \centering
\begin{tabular}{l|ll}
  \hline\hline

Sample  &  \multicolumn{2}{l}{Fiducial cross-section $[\textrm{pb}]$} \\ \hline
          &   Resolved           & Boosted \\ \hline
\Powheg{}+\PythiaSix{}  &  92.0       &  2.96    \\ 
\Powheg{}+\Pythia{} radHi  &  90.9       &    3.10  \\ 
\Powheg{}+\Pythia{} radLo  &  94.2       &     2.89 \\ 
\aMCatNLO{}+\Herwig{}++  &  94.9       &      3.19 \\ 
\Powheg{}+\Herwig{}++  &  93.5       &   2.84   \\ 
\Powheg{}+\PythiaEight{}  & 97.5   & 3.07 \\
\Powheg{}+\HerwigSeven{} & 97.2 & 2.84 \\
\aMCatNLO{}+\PythiaEight{} & 98.5 & 2.88 
\\ \hline
Data & $110 {}^{+13}_{-14} $ (stat+syst)   &  $2.54 \pm 0.54$ (stat+syst)   \\ \hline\hline

\end{tabular}
\caption{Fiducial cross-sections in the resolved and boosted topologies for data and different models. Each model's cross-section is scaled to the NNLO+NNLL value from Refs.~\cite{Cacciari:2011hy,Beneke:2011mq,Baernreuther:2012ws,Czakon:2012zr,Czakon:2012pz,Czakon:2013goa}, hence the quoted fiducial cross-sections result from different kinematic regions and thus acceptance from each model.}
\label{tab:fid_xsects}
\end{table}

\clearpage

\begin{figure*}[htbp]
\centering
\subfigure[]{ \includegraphics[width=0.45\textwidth]{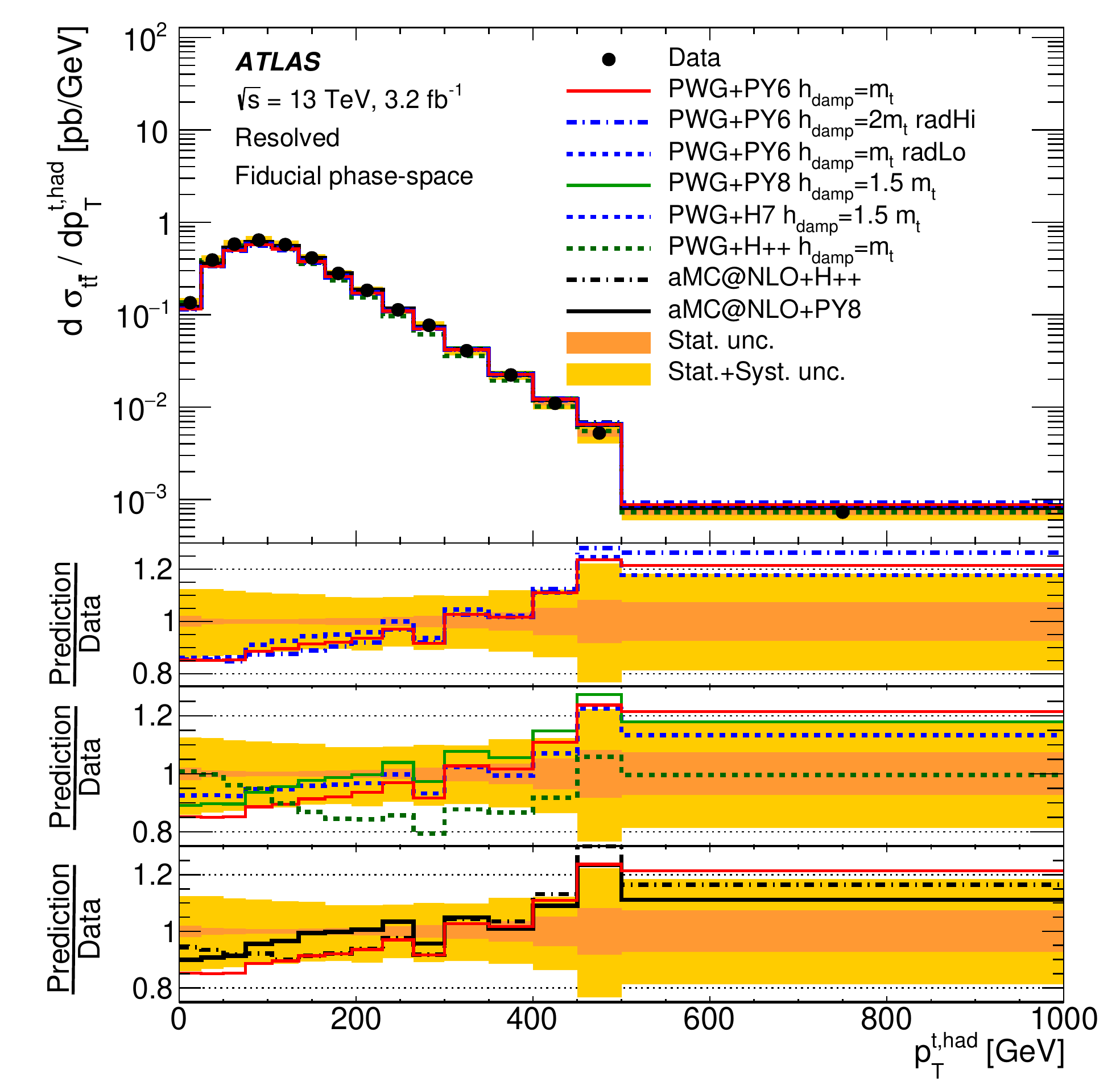}\label{fig:particle:topH_pt:abs}}
\subfigure[]{ \includegraphics[width=0.45\textwidth]{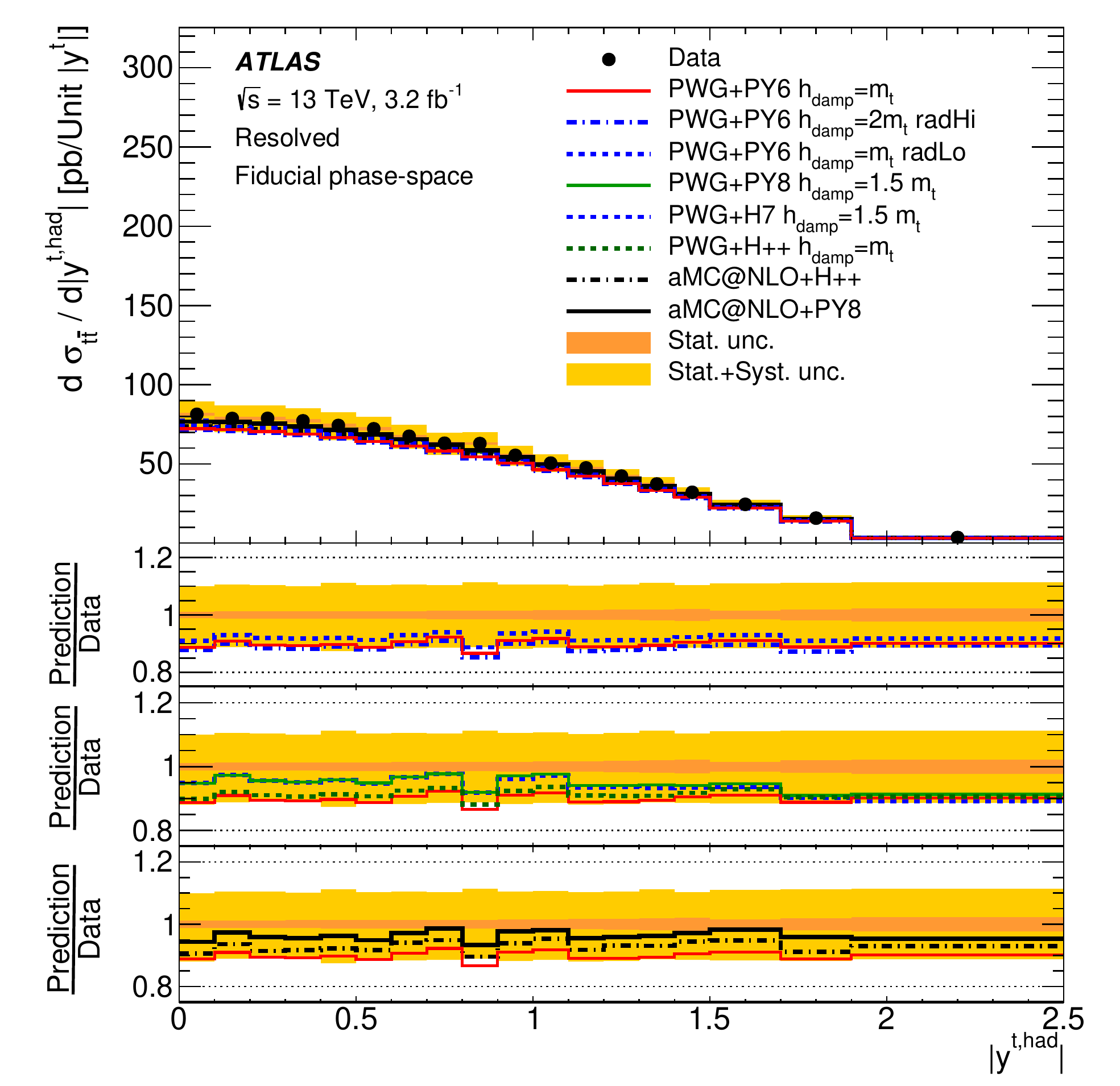}\label{fig:particle:topH_absrap:abs}}
\subfigure[]{ \includegraphics[width=0.45\textwidth]{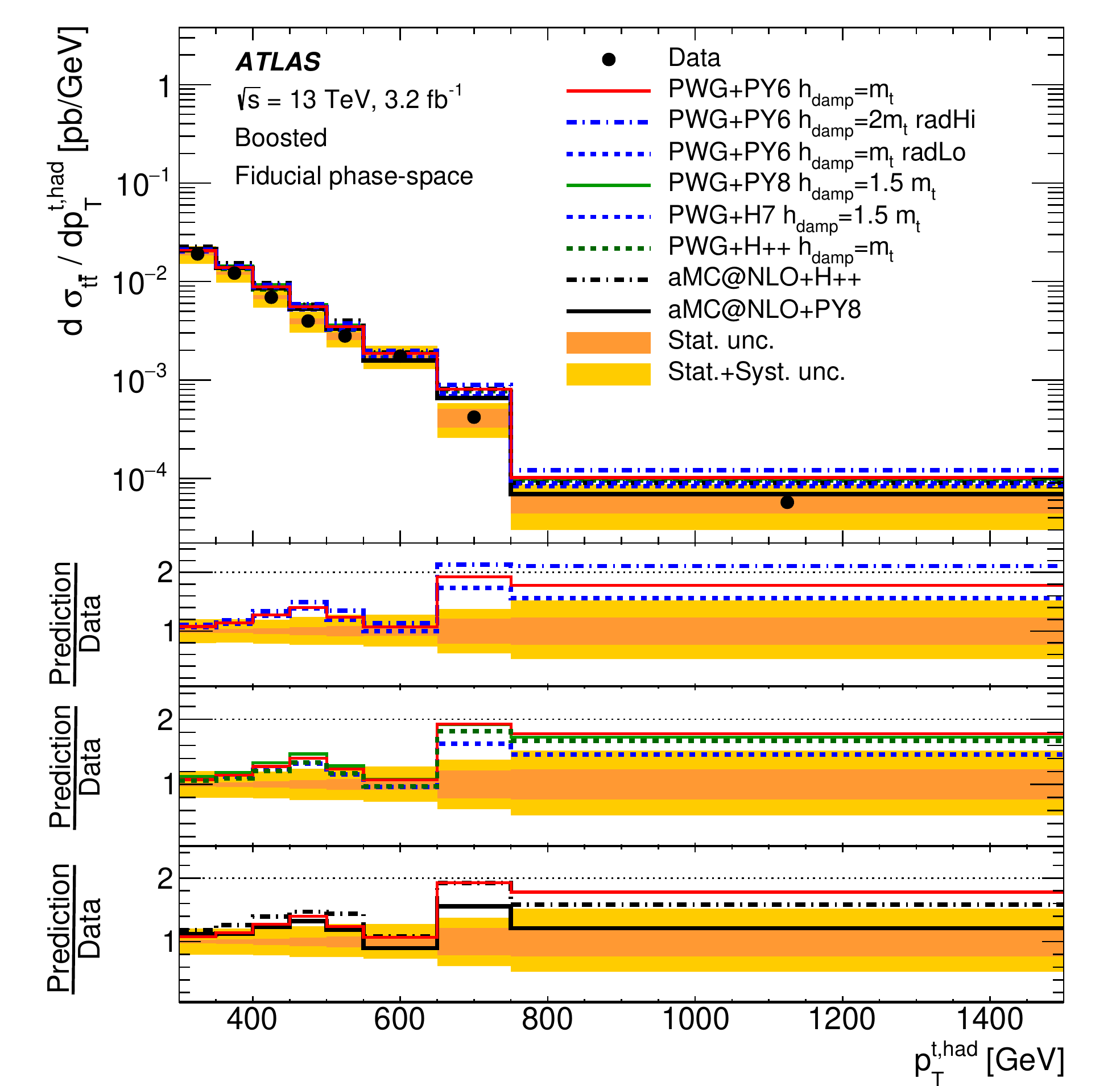}\label{fig:particle:topH_pt:abs_boosted}}
\subfigure[]{ \includegraphics[width=0.45\textwidth]{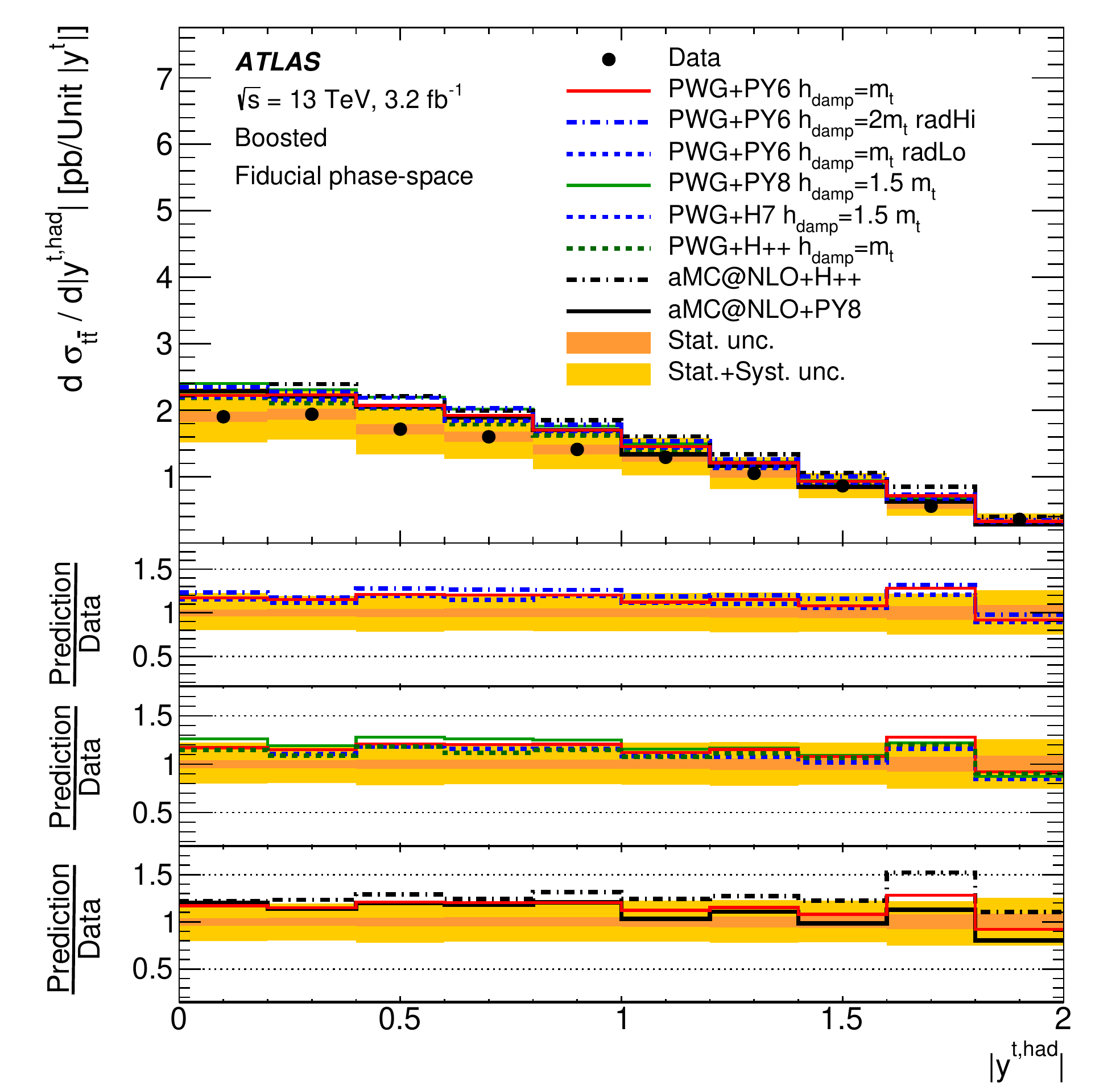}\label{fig:particle:topH_absrap:abs_boosted}}
\caption{Fiducial phase-space absolute differential cross-sections as a function of the \subref{fig:particle:topH_pt:abs}~transverse momentum (\ptthad{}) and \subref{fig:particle:topH_absrap:abs} the absolute value of the rapidity (\absythad) of the hadronic top quark in the resolved topology and corresponding results in the boosted topology \subref{fig:particle:topH_pt:abs_boosted}, \subref{fig:particle:topH_absrap:abs_boosted}. The yellow bands indicate the total uncertainty of the data in each bin. The \PowHeg{}+\PythiaSix{} generator with \HDampMT~and the CT10 PDF is used as the nominal prediction to correct for detector effects.
The lower three panels show the ratio of the predictions to the data. 
The first panel compares the three \Powheg{}+\PythiaSix{} samples with different settings for additional radiation, the second panel compares the nominal \Powheg{}+\PythiaSix{} sample with the other \Powheg{} samples and the third panel compares the nominal \Powheg{}+\PythiaSix{} sample with the \mgamcatnlo{} samples.
}
\label{fig:results:fiducial:topH:abs}
\end{figure*}

\begin{figure*}[htbp]
\centering
\subfigure[]{ \includegraphics[width=0.45\textwidth]{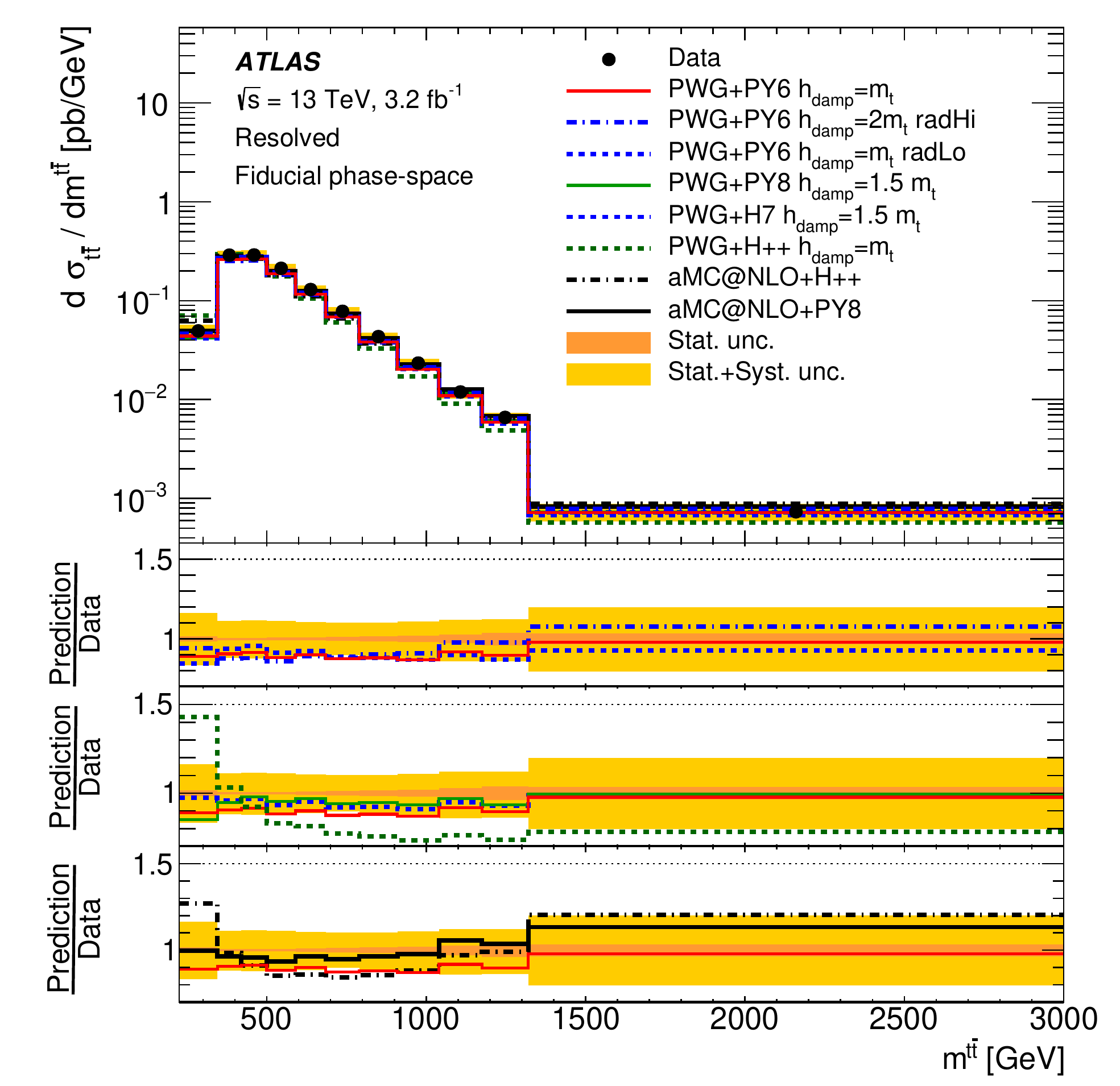}\label{fig:particle:tt_m:abs}}
\subfigure[]{ \includegraphics[width=0.45\textwidth]{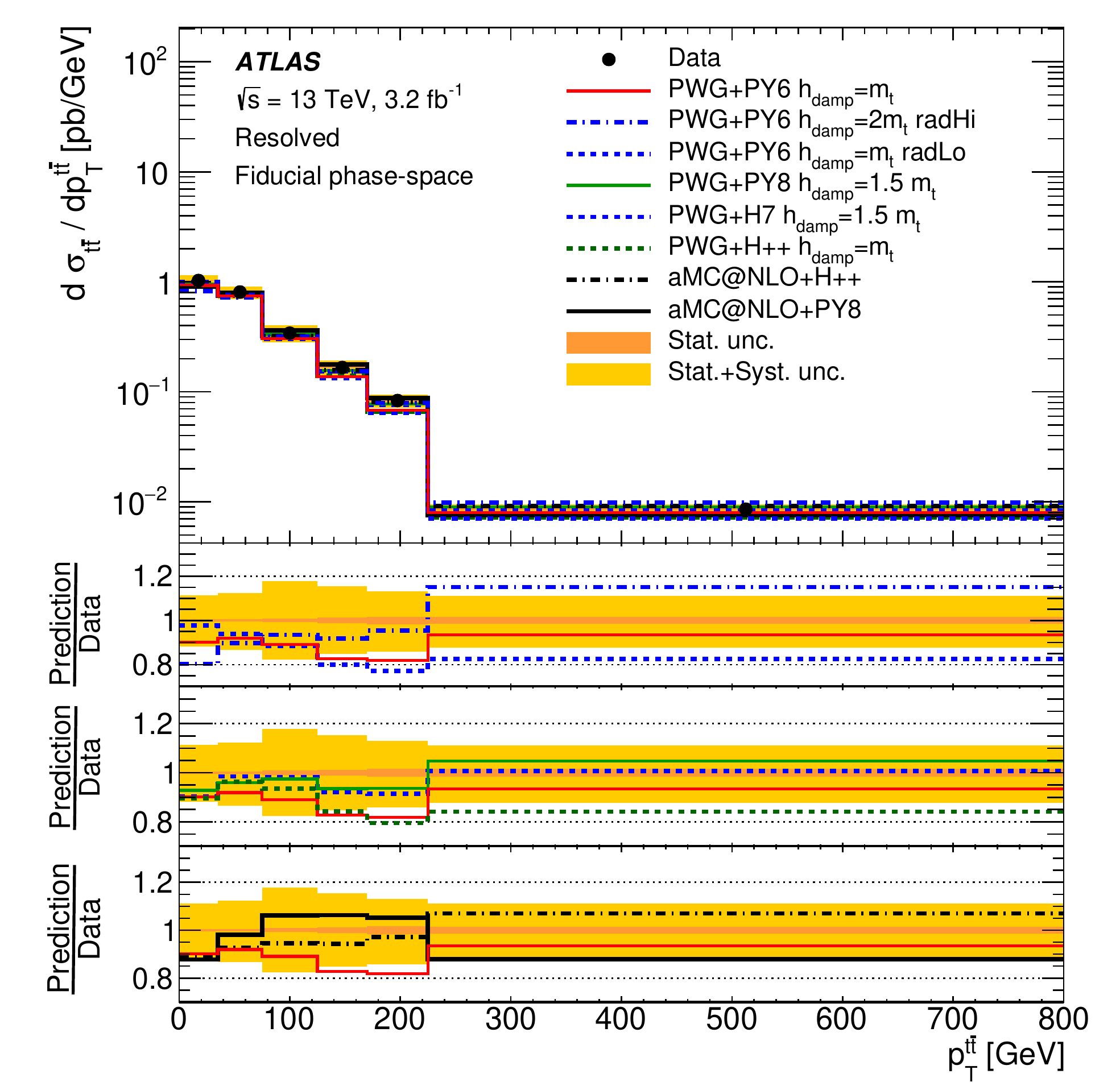}\label{fig:particle:tt_pt:abs}}
\subfigure[]{ \includegraphics[width=0.45\textwidth]{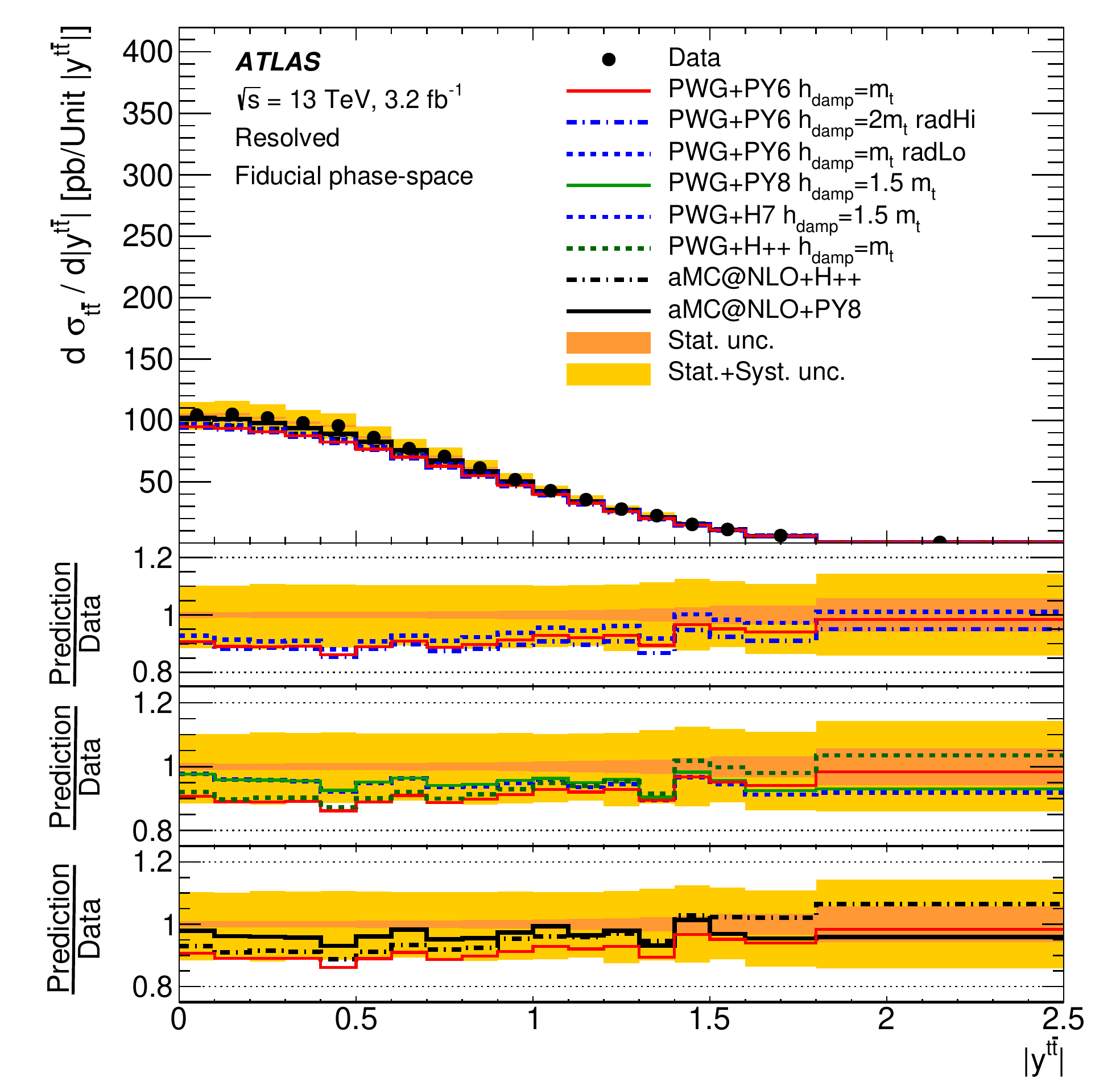}\label{fig:particle:tt_absrap:abs}}
\caption{Fiducial phase-space absolute differential cross-sections as a function of the \subref{fig:particle:tt_m:abs}~invariant mass (\mttbar{}), \subref{fig:particle:tt_pt:abs}~transverse momentum (\ptttbar{}) and \subref{fig:particle:tt_absrap:abs} the absolute value of the rapidity (\absyttbar{}) of the \ttb{} system in the resolved topology. The yellow bands indicate the total uncertainty of the data in each bin. The \PowHeg{}+\PythiaSix{} generator with \HDampMT{} and the CT10 PDF is used as the nominal prediction to correct for detector effects.
The lower three panels show the ratio of the predictions to the data. 
The first panel compares the three \Powheg{}+\PythiaSix{} samples with different settings for additional radiation, the second panel compares the nominal \Powheg{}+\PythiaSix{} sample with the other \Powheg{} samples and the third panel compares the nominal \Powheg{}+\PythiaSix{} sample with the \mgamcatnlo{} samples.
}
\label{fig:results:fiducial:tt:abs}
\end{figure*}

\begin{figure*}[htbp]
\centering
\subfigure[]{ \includegraphics[width=0.45\textwidth]{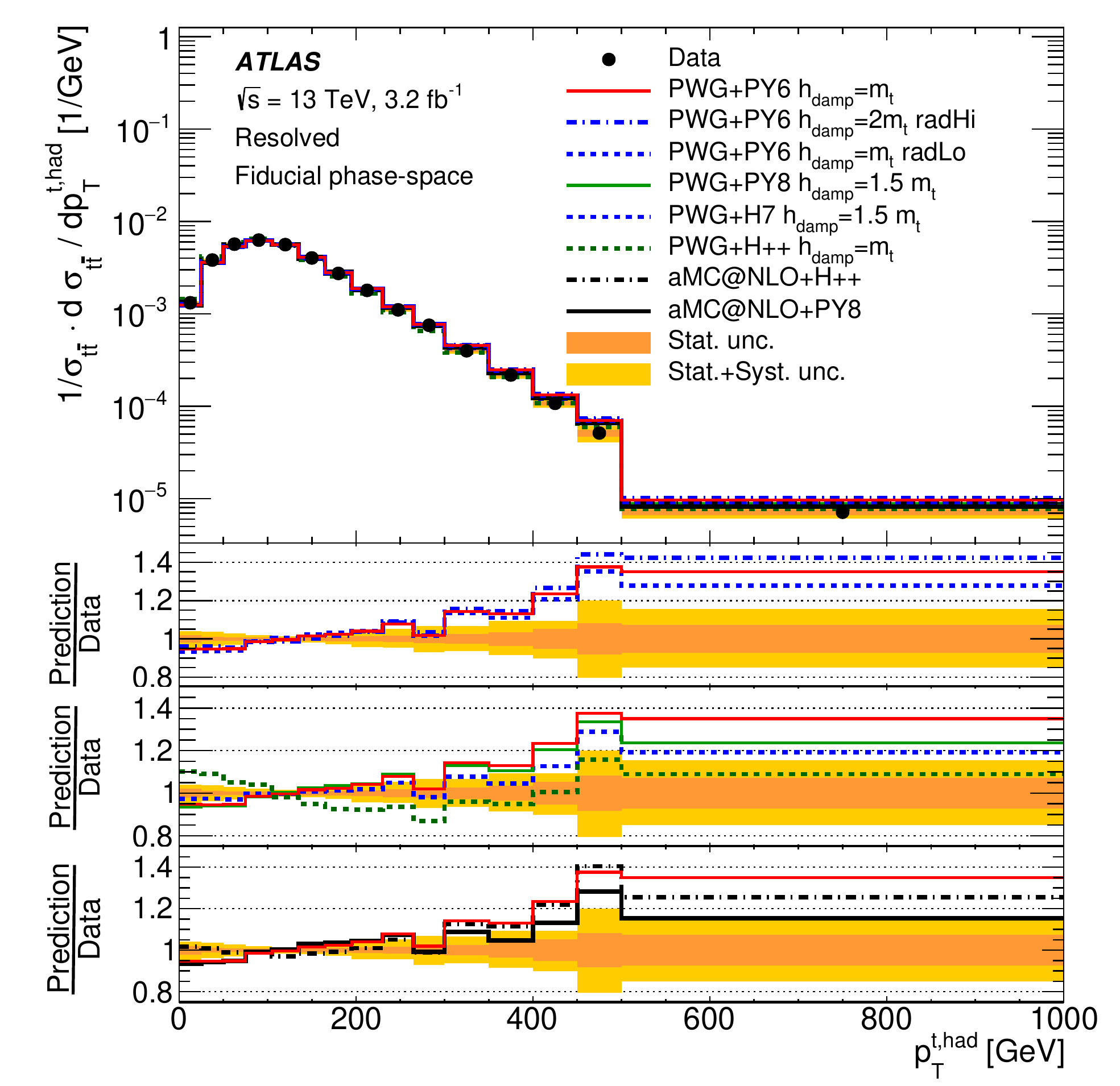}\label{fig:particle:topH_pt:rel}}
\subfigure[]{ \includegraphics[width=0.45\textwidth]{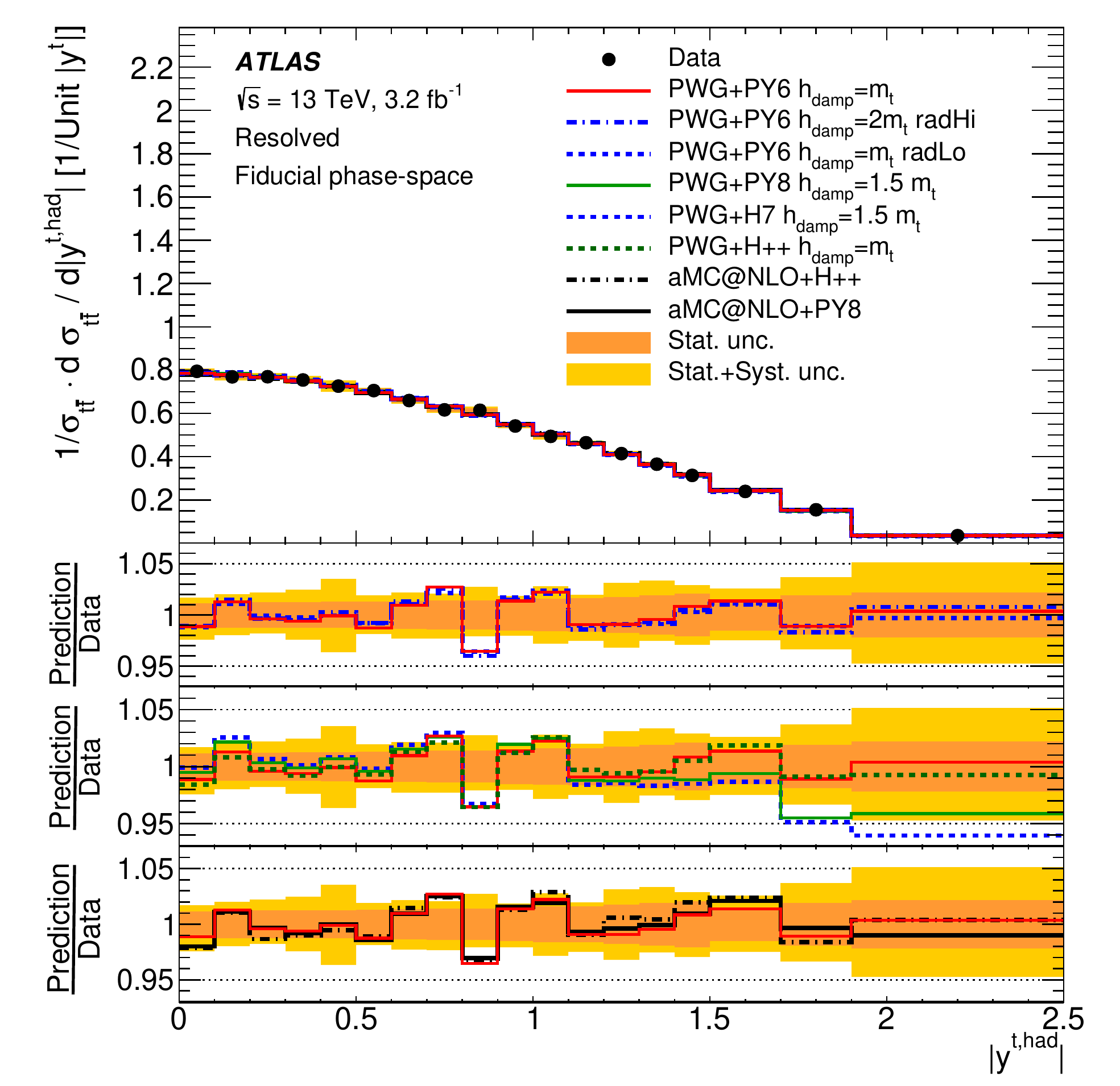}\label{fig:particle:topH_absrap:rel}}
\subfigure[]{ \includegraphics[width=0.45\textwidth]{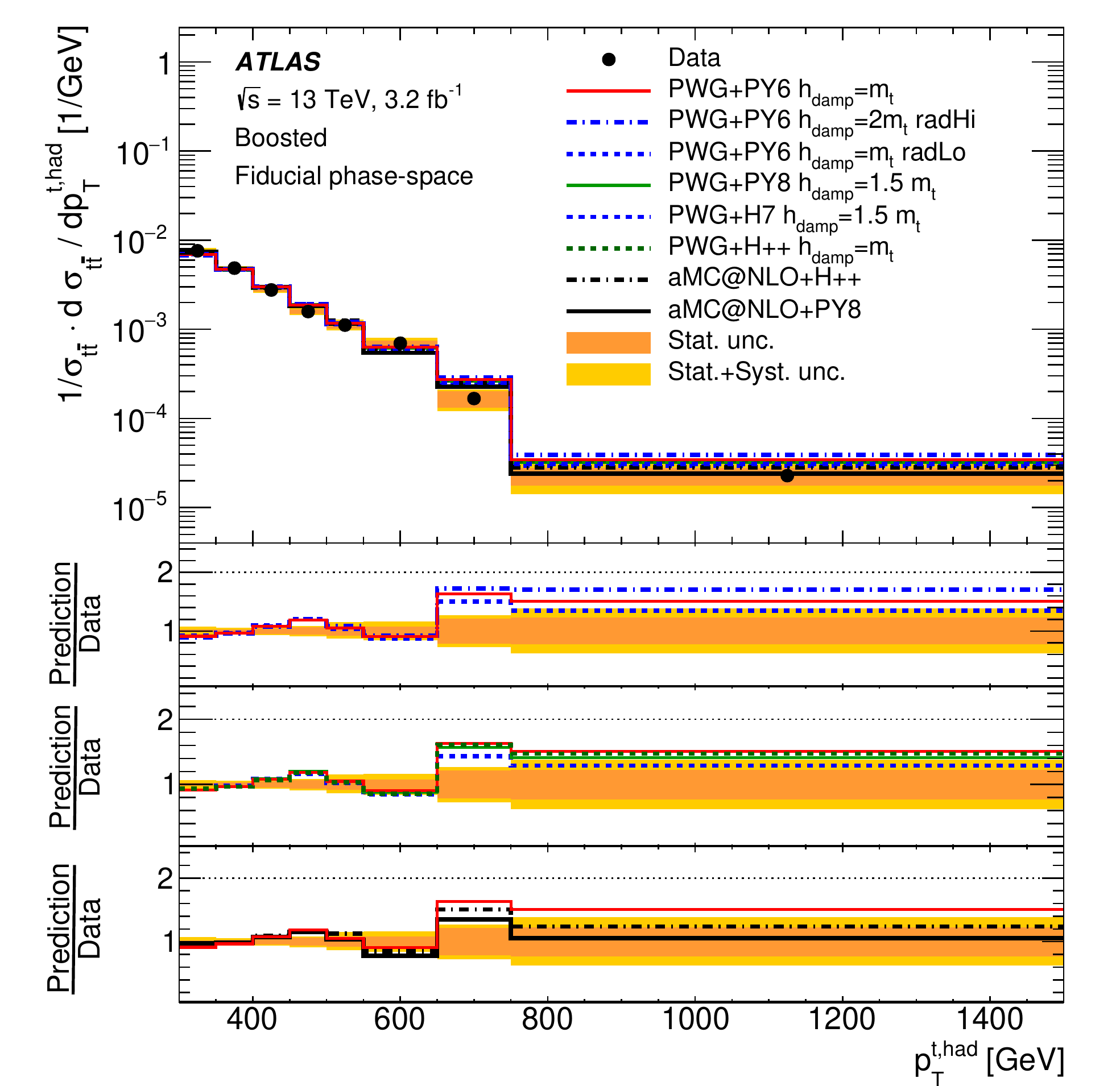}\label{fig:particle:topH_pt:rel_boosted}}
\subfigure[]{ \includegraphics[width=0.45\textwidth]{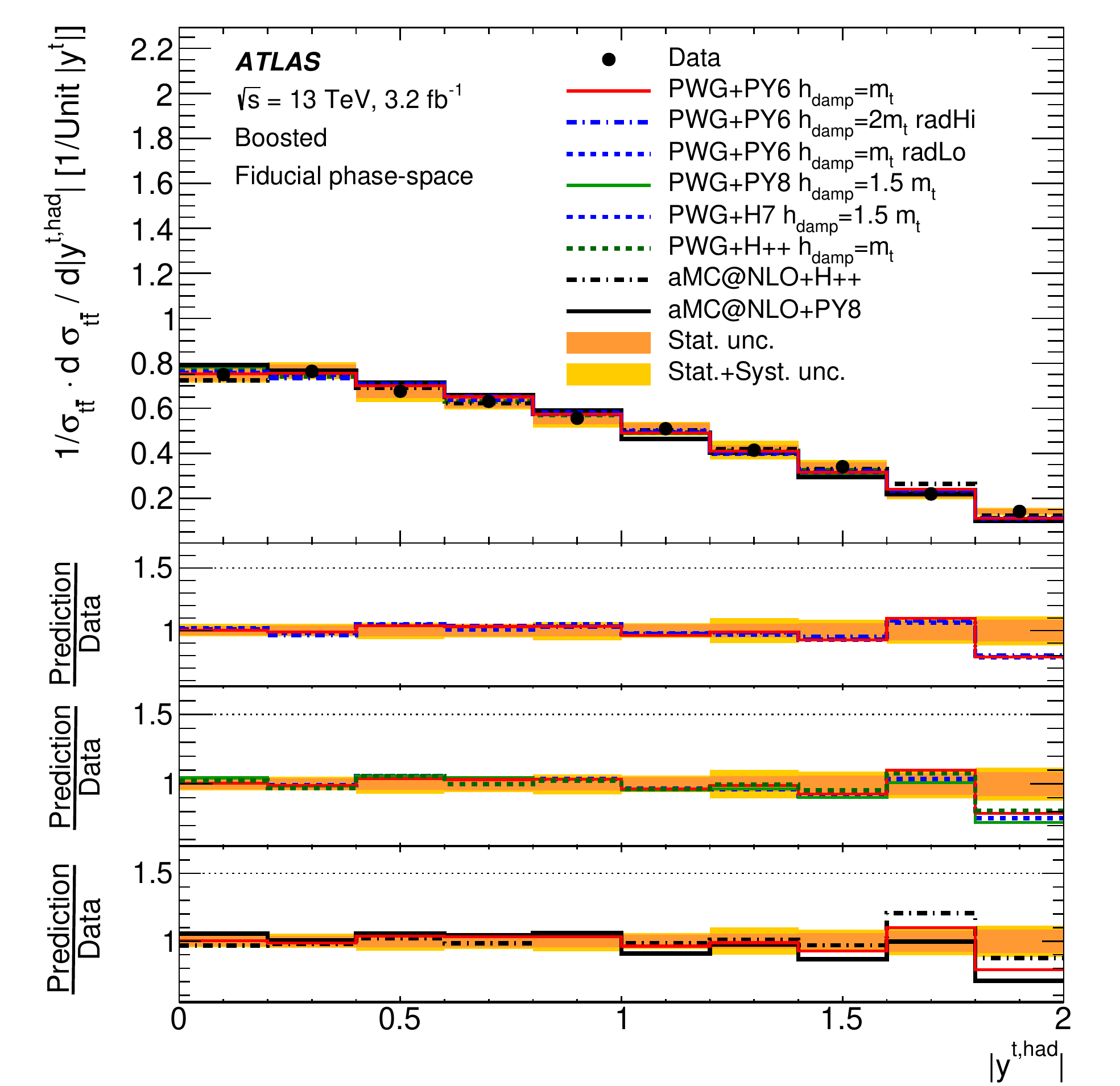}\label{fig:particle:topH_absrap:rel_boosted}}
\caption{Fiducial phase-space relative differential cross-sections as a function of the \subref{fig:particle:topH_pt:rel}~transverse momentum (\ptthad{}) and \subref{fig:particle:topH_absrap:rel} the absolute value of the rapidity (\absythad) of the hadronic top quark in the resolved topology, and corresponding results in the boosted topology \subref{fig:particle:topH_pt:rel_boosted}, \subref{fig:particle:topH_absrap:rel_boosted}. The yellow bands indicate the total uncertainty of the data in each bin. The \PowHeg{}+\PythiaSix{} generator with \HDampMT{} and the CT10 PDF is used as the nominal prediction to correct for detector effects.
The lower three panels show the ratio of the predictions to the data. 
The first panel compares the three \Powheg{}+\PythiaSix{} samples with different settings for additional radiation, the second panel compares the nominal \Powheg{}+\PythiaSix{} sample with the other \Powheg{} samples and the third panel compares the nominal \Powheg{}+\PythiaSix{} sample with the \mgamcatnlo{} samples.
}
\label{fig:results:fiducial:topH:rel}
\end{figure*}

\begin{figure*}[htbp]
\centering
\subfigure[]{ \includegraphics[width=0.45\textwidth]{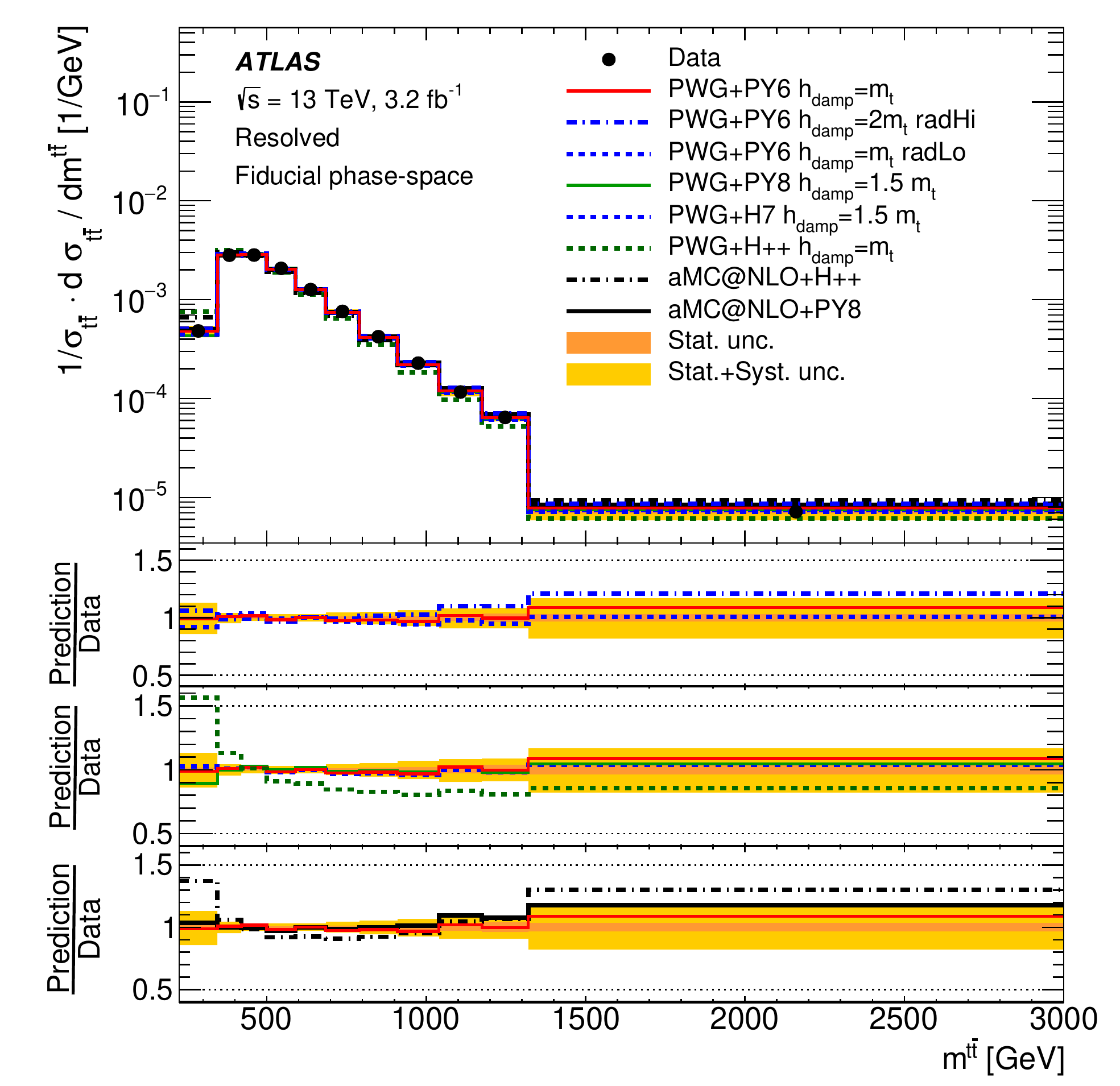}\label{fig:particle:tt_m:rel}}
\subfigure[]{ \includegraphics[width=0.45\textwidth]{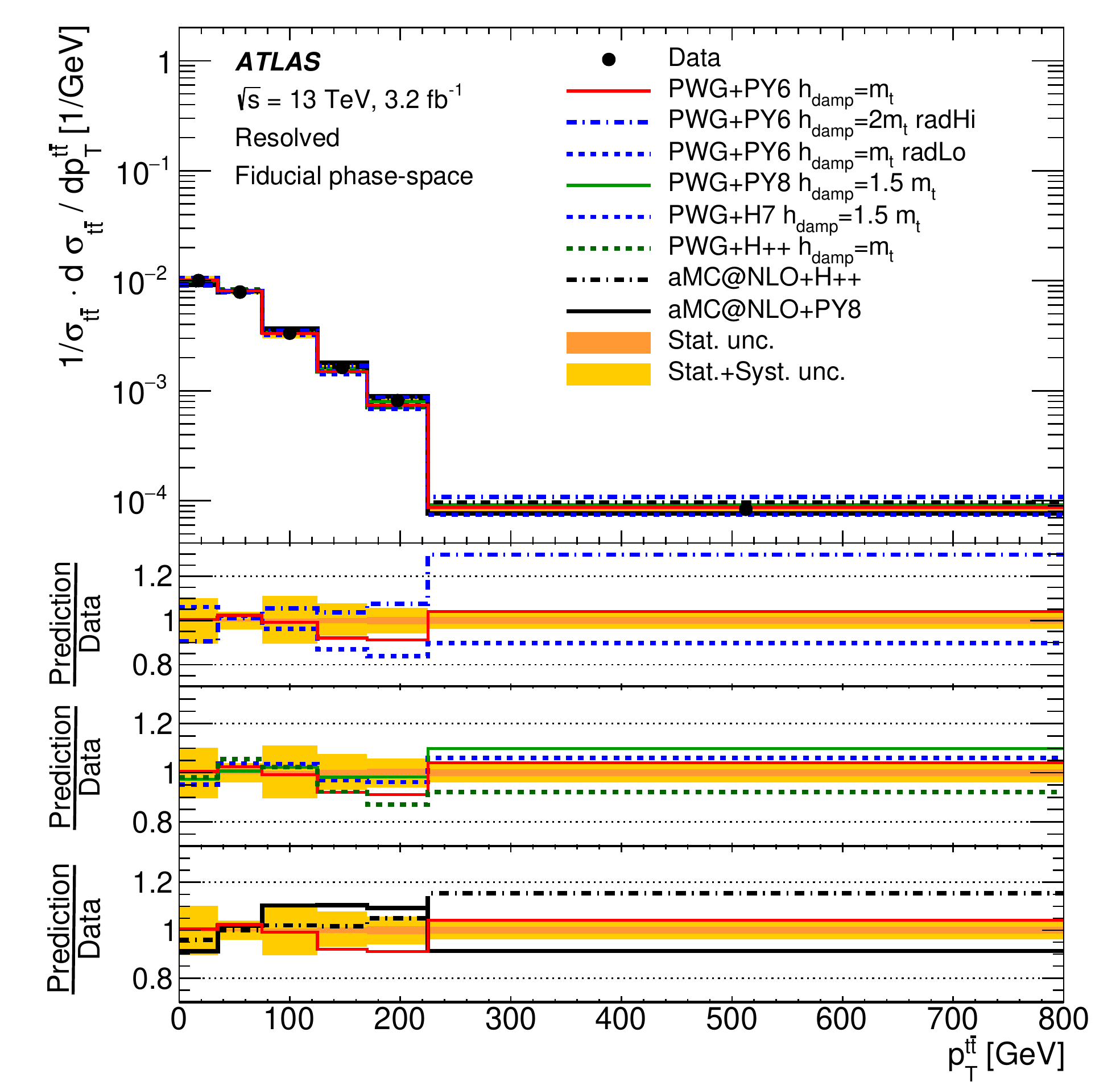}\label{fig:particle:tt_pt:rel}}
\subfigure[]{ \includegraphics[width=0.45\textwidth]{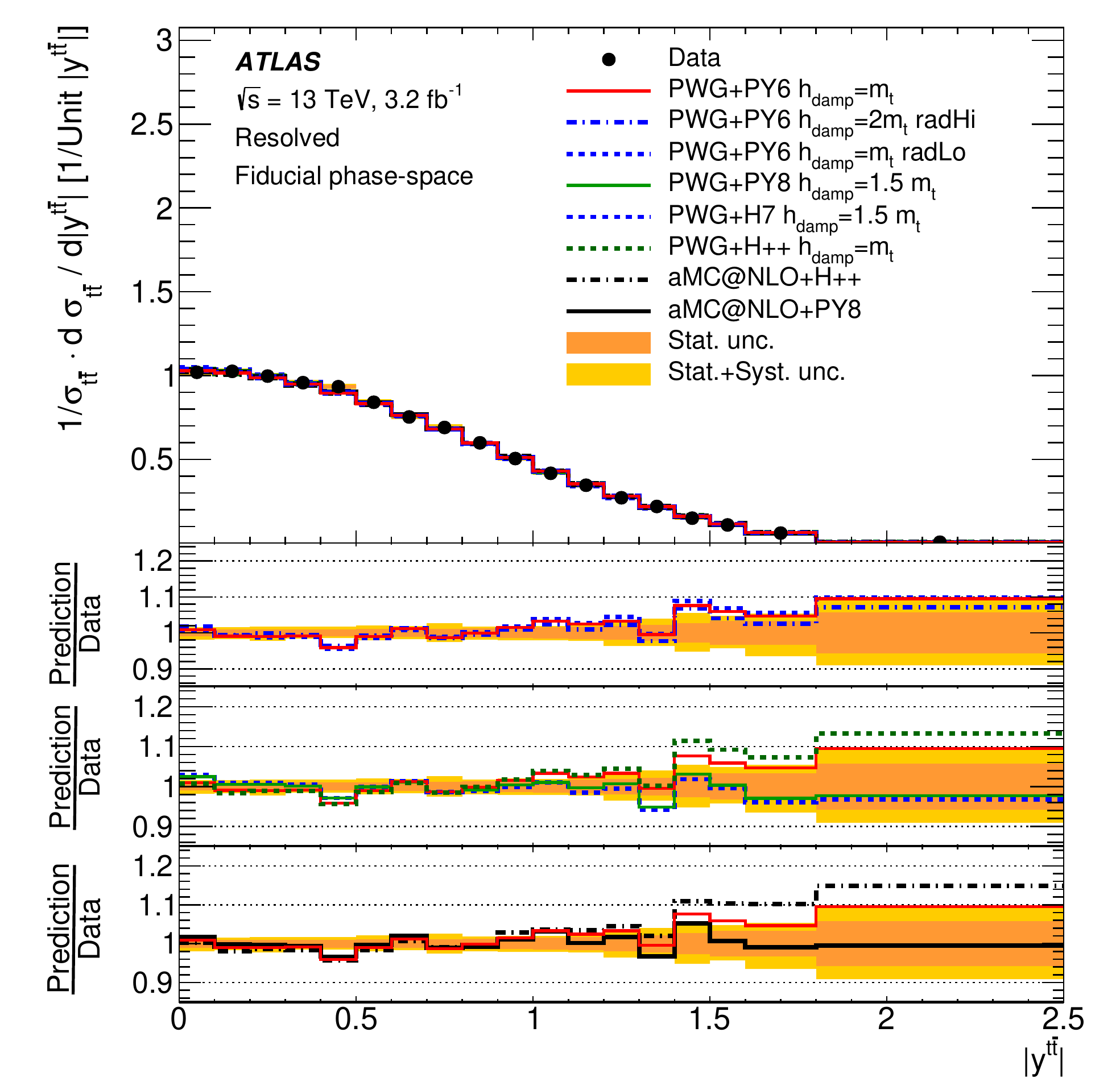}\label{fig:particle:tt_absrap:rel}}
\caption{Fiducial phase-space relative differential cross-sections as a function of the \subref{fig:particle:tt_m:rel}~invariant mass (\mttbar{}), \subref{fig:particle:tt_pt:rel}~transverse momentum (\ptttbar{}) and \subref{fig:particle:tt_absrap:rel} the absolute value of the rapidity (\absyttbar{}) of the \ttb{} system in the resolved topology. The yellow bands indicate the total uncertainty of the data in each bin. The \PowHeg{}+\PythiaSix{} generator with \HDampMT{} and the CT10 PDF is used as the nominal prediction to correct for detector effects.
The lower three panels show the ratio of the predictions to the data. 
The first panel compares the three \Powheg{}+\PythiaSix{} samples with different settings for additional radiation, the second panel compares the nominal \Powheg{}+\PythiaSix{} sample with the other \Powheg{} samples and the third panel compares the nominal \Powheg{}+\PythiaSix{} sample with the \mgamcatnlo{} samples.
}
\label{fig:results:fiducial:tt:rel}
\end{figure*}

\begin{figure*}[htbp]
\centering
\includegraphics[width=0.95\textwidth]{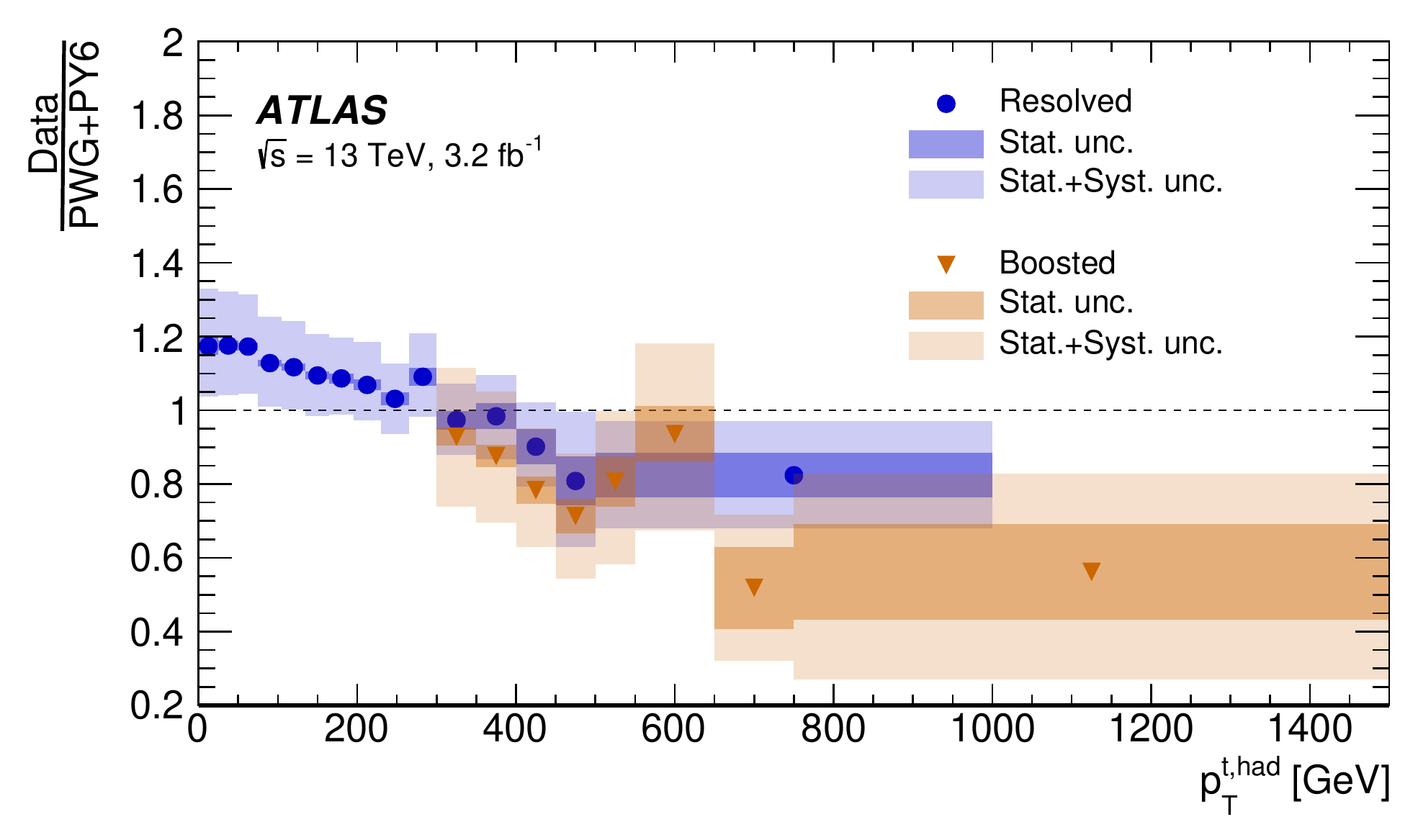}\label{fig:particle:topH_absrap:abs_resolved_boosted}

\caption{Ratios of the measured fiducial phase-space absolute differential cross-section to the prediction from \Powheg{}+\PythiaSix{} in the resolved and boosted topologies as a function of their respective transverse momentum of the hadronic top quark. The bands indicate the statistical and total uncertainties of the data in each bin.
The \PowHeg{}+\PythiaSix{} generator with \HDampMT~and the CT10 PDF is used as the nominal prediction to correct for detector effects.
}
\label{fig:results:fiducial:topH:abs_resolved_boosted}
\end{figure*}

\clearpage

\begin{landscape}
\begin{table}
\footnotesize
\centering
\noindent\makebox[\textwidth]{
\begin{tabular}{l | r @{/} l r  | r @{/} l r  | r @{/} l r  | r @{/} l r  | r @{/} l r }
\hline
& \multicolumn{3}{c|}{$p_{\textrm{T}}^{t,\textrm{had}}$} & \multicolumn{3}{|c|}{$  |y^{t,\textrm{had}}|$} & \multicolumn{3}{|c|}{$        m^{t\bar{t}}$}& \multicolumn{3}{|c|}{$p_{\textrm{T}}^{t\bar{t}}$} & \multicolumn{3}{|c}{$      |y^{t\bar{t}}|$} \\
& \multicolumn{2}{|c}{$\chi^{2}$/NDF} &  ~$p$-val  & \multicolumn{2}{|c}{$\chi^{2}$/NDF} &  ~$p$-val  & \multicolumn{2}{|c}{$\chi^{2}$/NDF} &  ~$p$-val  & \multicolumn{2}{|c}{$\chi^{2}$/NDF} &  ~$p$-val  & \multicolumn{2}{|c}{$\chi^{2}$/NDF} &  ~$p$-val  \\
\hline
\hline
\Powheg{}+\PythiaSix{}&
{\ }  19.0 & 15 & 0.22  & {\ } 7.8 & 18 & 0.98  &  {\ } 9.8 & 11 & 0.55  & {\ } 14.9 & 6 & 0.02  &  {\ } 20.0 & 18 & 0.33 \\
\Powheg{}+\PythiaSix{} (radHi)&
{\ } 20.9 & 15 & 0.14  & {\ } 8.5 & 18 & 0.97  &  {\ } 8.7 & 11 & 0.65  & {\ } 56.1 & 6 & $<$0.01  &  {\ } 17.3 & 18 & 0.51 \\
\Powheg{}+\PythiaSix{} (radLo)&
{\ } 20.8 & 15 & 0.14  & {\ } 7.4 & 18 & 0.99  &  {\ } 12.7 & 11 & 0.32  &  {\ } 22.1 & 6 & $<$0.01  &  {\ } 25.5 & 18 & 0.11 \\
\mgamcatnlo{}+\herwigpp{}&
{\ } 23.5 & 15 & 0.07  &   {\ } 10.7 & 18 & 0.91  &  {\ } 32.4 & 11 & $<$0.01& {\ } 16.4 & 6 & 0.01  &  {\ } 28.1 & 18 & 0.06 \\
\Powheg{}+\herwigpp{}&
{\ } 30.3 & 15 & 0.01  & {\ } 7.9 & 18 & 0.98  &  {\ } 34.8 & 11 & $<$0.01  & {\ } 28.0 & 6 & $<$0.01  &  {\ } 30.4 & 18 & 0.03 \\
\mgamcatnlo{}+\PythiaEight{} &
{\ } 19.1 & 15 & 0.21  & {\ } 8.4 & 18 & 0.97  &  {\ }  7.6 & 11 & 0.75  &{\ } 19.0 & 6 & $<$0.01  &  {\ } 16.1 & 18 & 0.59 \\

\Powheg{}+\PythiaEight{} &
{\ } 18.4 & 15 & 0.24  & {\ } 10.5 & 18 & 0.92  &  {\ } 7.7 & 11 & 0.74  & {\ } 11.7 & 6 & 0.07  &  {\ } 12.3 & 18 & 0.83 \\

\Powheg{}+\HerwigSeven{}&
{\ } 13.8 & 15 & 0.54  & {\ } 10.9 & 18 & 0.90  &  {\ }  7.0 & 11 & 0.80  & {\ } 11.6 & 6 & 0.07  &  {\ } 12.8 & 18 & 0.80 \\
\hline
\end{tabular}}
\caption{Comparison between the measured fiducial phase-space absolute
  differential cross-sections and the predictions from several MC
  generators in the resolved topology in terms of a~$\chi^2$ divided by the number of degrees of freedom (NDF) and $p$-values with NDF equal to $N_{\textrm b}$ where $N_{\textrm b}$ denotes the number of bins in the distribution. }
\label{Tab:chi2_res_abs}
\end{table}

\begin{table}
\footnotesize
\centering
\noindent\makebox[\textwidth]{
\begin{tabular}{l | r @{/} l r  | r @{/} l r  }
\hline
  & \multicolumn{3}{c|}{$ p_{\text{T}}^{t,\text{had}}$} & \multicolumn{3}{|c}{$ |y^{t,\text{had}}|$} \\
& \multicolumn{2}{|c}{$\chi^{2}$/NDF} &  ~$p$-val  & \multicolumn{2}{|c}{$\chi^{2}$/NDF} &  ~$p$-val  \\
\hline
\hline
\Powheg{}+\PythiaSix{}&
  {\ } 14.7 & 8 & 0.06 & {\ } 11.0 & 10 & 0.36   \\
\Powheg{}+\PythiaSix{} (radHi)&
 {\ } 19.5 & 8 & 0.01 & {\ } 12.3 & 10 & 0.27   \\
\Powheg{}+\PythiaSix{} (radLo)&
 {\ } 15.0 & 8 & 0.06 & {\ } 10.0 & 10 & 0.44   \\
\mgamcatnlo{}+\herwigpp{} &
 {\ } 17.9 & 8 & 0.02 & {\ } 12.8 & 10 & 0.24   \\
\Powheg{}+\herwigpp{} &
 {\ } 14.1 & 8 & 0.08 & {\ } 8.0 & 10 & 0.63  \\

\mgamcatnlo{}+\PythiaEight{} &
 {\ } 12.8 & 8 & 0.12  & {\ } 20.4 & 10 & 0.03   \\

\Powheg{}+\PythiaEight{} &
 {\ } 16.7 & 8 & 0.03  & {\ } 18.4 & 10 & 0.05   \\

\Powheg{}+\HerwigSeven{} &
 {\ } 11.9 & 8 & 0.15 & {\ } 11.7 & 10 & 0.30    \\

\hline
\end{tabular}}
\caption{Comparison between the measured fiducial phase-space absolute
  differential cross-sections and the predictions from several MC
  generators in the boosted topology in terms of a~$\chi^2$ divided by the number of degrees of freedom (NDF) and $p$-values with NDF equal to $N_{\textrm b}$ where $N_{\textrm b}$ denotes the number of bins in the distribution. }
\label{Tab:chi2_boo_abs}
\end{table}
\end{landscape}

\begin{landscape}
\begin{table}
\footnotesize
\centering
\noindent\makebox[\textwidth]{
\begin{tabular}{l | r @{/} l r  | r @{/} l r  | r @{/} l r  | r @{/} l r  | r @{/} l r  }
\hline
Observable  & \multicolumn{3}{c|}{$p_{\textrm{T}}^{t,\textrm{had}}$} & \multicolumn{3}{|c|}{$  |y^{t,\textrm{had}}|$} & \multicolumn{3}{|c|}{$        m^{t\bar{t}}$} & \multicolumn{3}{c|}{$p_{\textrm{T}}^{t\bar{t}}$} & \multicolumn{3}{|c}{$      |y^{t\bar{t}}|$} \\
& \multicolumn{2}{|c}{$\chi^{2}$/NDF} &  ~$p$-val  & \multicolumn{2}{|c}{$\chi^{2}$/NDF} &  ~$p$-val  & \multicolumn{2}{|c}{$\chi^{2}$/NDF} &  ~$p$-val  & \multicolumn{2}{|c}{$\chi^{2}$/NDF} &  ~$p$-val  & \multicolumn{2}{|c}{$\chi^{2}$/NDF} &  ~$p$-val  \\
\hline
\hline
\Powheg{}+\PythiaSix{}&
{\ } 23.0 & 14 & 0.06 &  {\ } 8.1 & 17 & 0.96  &  {\ }  6.3 & 10 & 0.79  & {\ } 7.7 & 5 & 0.17  &  {\ } 22.5 & 17 & 0.17 \\
\Powheg{}+\PythiaSix{} (radHi)&
{\ } 23.8 & 14 & 0.05  & {\ } 8.5 & 17 & 0.95  &  {\ } 7.7 & 10 & 0.66  & {\ } 5.1 & 5 & 0.41  &  {\ } 19.3 & 17 & 0.31 \\
\Powheg{}+\PythiaSix{} (radLo)&
{\ } 25.9 & 14 & 0.03  &{\ } 7.5 & 17 & 0.98  &  {\ } 8.2 & 10 & 0.61  &  {\ } 20.4 & 5 & $<$0.01  &  {\ } 28.0 & 17 & 0.04 \\
\mgamcatnlo{}+\herwigpp{} &
{\ } 24.4 & 14 & 0.04  & {\ } 10.8 & 17 & 0.87  &  {\ } 23.6 & 10 & $<$0.01  & {\ } 2.6 & 5 & 0.76  &  {\ } 30.0 & 17 & 0.03 \\
\Powheg{}+\herwigpp{}&
{\ }24.0 & 14 & 0.05  &   {\ } 7.4 & 17 & 0.98  &  {\ } 37.9 & 10 & $<$0.01  & {\ } 25.0 & 5 & $<$0.01  &  {\ } 32.8 & 17 & 0.01 \\

\mgamcatnlo{}+\PythiaEight{}  &
{\ } 21.8 & 14 & 0.08  &{\ } 7.8 & 17 & 0.97  &  {\ } 6.8 & 10 & 0.75  &  {\ } 3.3 & 5 & 0.66  &  {\ } 18.0 & 17 & 0.39 \\

\Powheg{}+\PythiaEight{} &
{\ } 21.5 & 14 & 0.09  &{\ } 9.6 & 17 & 0.92  &  {\ } 6.5 & 10 & 0.77  &   {\ } 1.1 & 5 & 0.96  &  {\ } 14.0 & 17 & 0.67 \\

\Powheg{}+\HerwigSeven{}&
{\ }15.4 & 14 & 0.35  &  {\ } 9.3 & 17 & 0.93  &  {\ }  6.7 & 10 & 0.76  & {\ } 5.4 & 5 & 0.37  &  {\ } 15.1 & 17 & 0.59 \\
\hline
\end{tabular}}
\caption{Comparison between the measured fiducial phase-space relative
  differential cross-sections and the predictions from several MC
  generators in the resolved topology in terms of a~$\chi^2$ divided by the number of degrees of freedom (NDF) and $p$-values with NDF equal to $N_{\textrm b}-1$ where $N_{\textrm b}$ denotes the number of bins in the distribution. }
\label{Tab:chi2_res_rel}
\end{table}

\begin{table}
\footnotesize
\centering
\noindent\makebox[\textwidth]{
\begin{tabular}{l | r @{/} l r  | r @{/} l r }
\hline
  & \multicolumn{3}{c|}{$p_{\text{T}}^{t,\text{had}} $} & \multicolumn{3}{|c}{$ |y^{t,\text{had}}|$} \\
& \multicolumn{2}{|c}{$\chi^{2}$/NDF} &  ~$p$-val  & \multicolumn{2}{|c}{$\chi^{2}$/NDF} &  ~$p$-val  \\
\hline
\hline
\Powheg{}+\PythiaSix{}&
 {\ } 10.2 & 7 & 0.18 & {\ } 2.9 & 9 & 0.97    \\
\Powheg{}+\PythiaSix{} (radHi)&
 {\ } 11.3 & 7 & 0.12 & {\ } 2.9 & 9 & 0.97    \\
\Powheg{}+\PythiaSix{} (radLo)&
 {\ } 11.5 & 7 & 0.12 & {\ } 2.8 & 9 & 0.97   \\
\mgamcatnlo{}+\herwigpp{} &
 {\ } 11.1 & 7 & 0.13 & {\ } 4.6 & 9 & 0.87   \\
\Powheg{}+\herwigpp{} &
 {\ } 10.7 & 7 & 0.15 & {\ } 2.5 & 9 & 0.98   \\
\mgamcatnlo{}+\PythiaEight{}  &
 {\ } 10.9 & 7 & 0.14 & {\ } 7.2 & 9 & 0.62   \\

\Powheg{}+\PythiaEight{} &
 {\ } 11.3 & 7 & 0.13 & {\ } 4.3 & 9 & 0.89   \\
 
\Powheg{}+\HerwigSeven{}&
 {\ } 9.9 & 7 & 0.20 & {\ } 3.6 & 9 & 0.94   \\
\hline
\end{tabular}}
\caption{Comparison between the measured fiducial phase-space relative
  differential cross-sections and the predictions from several MC
  generators in the boosted topology in terms of a~$\chi^2$ divided by the number of degrees of freedom (NDF) and $p$-values with NDF equal to $N_{\textrm b}-1$ where $N_{\textrm b}$ denotes the number of bins in the distribution. }
\label{Tab:chi2_boo_rel}
\end{table}
\end{landscape}

\clearpage
\section{Conclusions} \label{sec:Conclusion}

Kinematic distributions of hadronically decaying top quarks in both resolved and boosted topologies, and of top-quark pairs in the resolved topology are measured in a fiducial phase-space, using events from the lepton+jets channel using data from 13~\TeV{} proton--proton collisions collected by the ATLAS detector at the CERN Large Hadron Collider, corresponding to an integrated luminosity of \lumitot{}. Absolute as well as relative differential cross-sections are measured as a~function of the hadronic top-quark transverse momentum and rapidity. For the resolved topology, the differential cross-sections are also measured as a~function of the mass, transverse momentum and rapidity of the \ttbar{} system.

In general, the Monte Carlo predictions agree with data in a~wide kinematic region. However, the shape of the transverse momentum distribution of hadronically decaying top quarks is poorly modelled by all NLO+PS predictions, where the disagreement is largest at high transverse momentum. This behaviour is consistent between the resolved and boosted topologies, and also with the results at lower centre-of-mass energies.

In the resolved topology, the precision of the measurement of the transverse momentum of the \ttbar{} system makes it possible to distinguish between different settings in the NLO+PS calculations, indicating that the data have discriminating power sufficient to allow parameter values for these generators to be improved. For the relative differential cross-section results, the transverse momentum of hadronically decaying top quarks is the most poorly modelled observable. 


\section*{Acknowledgements}
We thank CERN for the very successful operation of the LHC, as well as the
support staff from our institutions without whom ATLAS could not be
operated efficiently.

We acknowledge the support of ANPCyT, Argentina; YerPhI, Armenia; ARC, Australia; BMWFW and FWF, Austria; ANAS, Azerbaijan; SSTC, Belarus; CNPq and FAPESP, Brazil; NSERC, NRC and CFI, Canada; CERN; CONICYT, Chile; CAS, MOST and NSFC, China; COLCIENCIAS, Colombia; MSMT CR, MPO CR and VSC CR, Czech Republic; DNRF and DNSRC, Denmark; IN2P3-CNRS, CEA-DSM/IRFU, France; SRNSF, Georgia; BMBF, HGF, and MPG, Germany; GSRT, Greece; RGC, Hong Kong SAR, China; ISF, I-CORE and Benoziyo Center, Israel; INFN, Italy; MEXT and JSPS, Japan; CNRST, Morocco; NWO, Netherlands; RCN, Norway; MNiSW and NCN, Poland; FCT, Portugal; MNE/IFA, Romania; MES of Russia and NRC KI, Russian Federation; JINR; MESTD, Serbia; MSSR, Slovakia; ARRS and MIZ\v{S}, Slovenia; DST/NRF, South Africa; MINECO, Spain; SRC and Wallenberg Foundation, Sweden; SERI, SNSF and Cantons of Bern and Geneva, Switzerland; MOST, Taiwan; TAEK, Turkey; STFC, United Kingdom; DOE and NSF, United States of America. In addition, individual groups and members have received support from BCKDF, the Canada Council, CANARIE, CRC, Compute Canada, FQRNT, and the Ontario Innovation Trust, Canada; EPLANET, ERC, ERDF, FP7, Horizon 2020 and Marie Sk{\l}odowska-Curie Actions, European Union; Investissements d'Avenir Labex and Idex, ANR, R{\'e}gion Auvergne and Fondation Partager le Savoir, France; DFG and AvH Foundation, Germany; Herakleitos, Thales and Aristeia programmes co-financed by EU-ESF and the Greek NSRF; BSF, GIF and Minerva, Israel; BRF, Norway; CERCA Programme Generalitat de Catalunya, Generalitat Valenciana, Spain; the Royal Society and Leverhulme Trust, United Kingdom.

The crucial computing support from all WLCG partners is acknowledged gratefully, in particular from CERN, the ATLAS Tier-1 facilities at TRIUMF (Canada), NDGF (Denmark, Norway, Sweden), CC-IN2P3 (France), KIT/GridKA (Germany), INFN-CNAF (Italy), NL-T1 (Netherlands), PIC (Spain), ASGC (Taiwan), RAL (UK) and BNL (USA), the Tier-2 facilities worldwide and large non-WLCG resource providers. Major contributors of computing resources are listed in Ref.~\cite{ATL-GEN-PUB-2016-002}.

\appendix

\clearpage
\label{app:References}
\printbibliography

\clearpage

\newpage 
\begin{flushleft}
{\Large The ATLAS Collaboration}

\bigskip

M.~Aaboud$^\textrm{\scriptsize 137d}$,
G.~Aad$^\textrm{\scriptsize 88}$,
B.~Abbott$^\textrm{\scriptsize 115}$,
J.~Abdallah$^\textrm{\scriptsize 8}$,
O.~Abdinov$^\textrm{\scriptsize 12}$$^{,*}$,
B.~Abeloos$^\textrm{\scriptsize 119}$,
S.H.~Abidi$^\textrm{\scriptsize 161}$,
O.S.~AbouZeid$^\textrm{\scriptsize 139}$,
N.L.~Abraham$^\textrm{\scriptsize 151}$,
H.~Abramowicz$^\textrm{\scriptsize 155}$
% ...
\bigskip
\\
% ...
$^{8}$ Department of Physics, The University of Texas at Arlington, Arlington TX, United States of America\\
% ...
$^{12}$ Institute of Physics, Azerbaijan Academy of Sciences, Baku, Azerbaijan\\
% ...
$^{88}$ CPPM, Aix-Marseille Universit{\'e} and CNRS/IN2P3, Marseille, France\\
% ...
$^{115}$ Homer L. Dodge Department of Physics and Astronomy, University of Oklahoma, Norman OK, United States of America\\
% ...
$^{119}$ LAL, Univ. Paris-Sud, CNRS/IN2P3, Universit{\'e} Paris-Saclay, Orsay, France\\
% ...
$^{137}$ $^{(a)}$ Facult{\'e} des Sciences Ain Chock, R{\'e}seau Universitaire de Physique des Hautes Energies - Universit{\'e} Hassan II, Casablanca; $^{(b)}$ Centre National de l'Energie des Sciences Techniques Nucleaires, Rabat; $^{(c)}$ Facult{\'e} des Sciences Semlalia, Universit{\'e} Cadi Ayyad, LPHEA-Marrakech; $^{(d)}$ Facult{\'e} des Sciences, Universit{\'e} Mohamed Premier and LPTPM, Oujda; $^{(e)}$ Facult{\'e} des sciences, Universit{\'e} Mohammed V, Rabat, Morocco\\
% ...
$^{139}$ Santa Cruz Institute for Particle Physics, University of California Santa Cruz, Santa Cruz CA, United States of America\\
% ...
$^{151}$ Department of Physics and Astronomy, University of Sussex, Brighton, United Kingdom\\
% ...
$^{155}$ Raymond and Beverly Sackler School of Physics and Astronomy, Tel Aviv University, Tel Aviv, Israel\\
% ...
$^{161}$ Department of Physics, University of Toronto, Toronto ON, Canada\\
% ...
$^{a}$ Also at Department of Physics, King's College London, London, United Kingdom\\
$^{b}$ Also at Institute of Physics, Azerbaijan Academy of Sciences, Baku, Azerbaijan\\
$^{c}$ Also at Novosibirsk State University, Novosibirsk, Russia\\
$^{d}$ Also at TRIUMF, Vancouver BC, Canada\\
$^{e}$ Also at Department of Physics {\&} Astronomy, University of Louisville, Louisville, KY, United States of America\\
% ...
$^{*}$ Deceased
\end{flushleft}

\clearpage

\end{document}